\documentclass[12pt]{iopart}

\usepackage{iopams}
\usepackage[T1]{fontenc}
\usepackage{amsmath}
\usepackage{amsfonts}
\usepackage[numbers]{natbib}
\usepackage{graphicx}
\usepackage{color}
\usepackage[caption = false]{subfig}
\usepackage[utf8]{}
\usepackage{float}
\usepackage{tikz}

\begin{document}

\title[]{Perturbative Field-Theoretical Analysis of Three-Species Cyclic Predator-Prey Models}

\author{Louie Hong Yao$^1$, Mohamed Swailem$^1$, Ulrich Dobramysl$^2$ \&  Uwe C. Täuber$^{1,3}$}

\address{1 Department of Physics \& Center for Soft Matter and Biological Physics, MC 0435,
Robeson Hall, 850 West Campus Drive, Virginia Tech, Blacksburg, VA 24061, USA}
\address{2 Peter Medawar Building for Pathogen Research, University of Oxford, \\ Oxford OX1 3SY, 
United Kingdom}
\address{3 Faculty of Health Sciences, Virginia Tech, Blacksburg, VA 24061, USA}
\eads{\mailto{hyao0731@vt.edu}, \mailto{mswailem@vt.edu}, \mailto{ulrich.dobramysl@gmail.com}, \mailto{tauber@vt.edu}} 

\vspace{10pt}
\begin{indented}
\item[] \today
\end{indented}

\begin{abstract}
We apply a perturbative Doi--Peliti field-theoretical analysis to the stochastic spatially extended symmetric 
Rock-Paper-Scissors (RPS) and May--Leonard (ML) models, in which three species compete cyclically. 
Compared to the two-species Lotka--Volterra predator-prey (LV) model, according to numerical 
simulations, these cyclical models appear to be less affected by intrinsic stochastic fluctuations. 
Indeed, we demonstrate that the qualitative features of the ML model are insensitive to intrinsic 
reaction noise. 
In contrast, and although not yet observed in numerical simulations, we find that the RPS model 
acquires significant fluctuation-induced renormalizations in the perturbative regime, similar to the 
LV model. 
We also study the formation of spatio-temporal structures in the framework of stability analysis and 
provide a clearcut explanation for the absence of spatial patterns in the RPS system, whereas the 
spontaneous emergence of spatio-temporal structures features prominently in the LV and the ML 
models.
\end{abstract}

\noindent{\it Keywords\/}:   
predator-prey model, cyclic competition, field-theoretical analysis, pattern formation, 
fluctuation-induced behavior

\section{Introduction}

Population dynamics has been and continues to be an extremely active field of research since about forty 
years \cite{may2019stability,maynard1974models,murray2002mathematical,hofbauer1998evolutionary,horsthemke1984noise, nisbet2003modelling, samanta2021deterministic}. 
Steady progress in the development of mathematical and computational tools as well as the application of 
methods from statistical physics have allowed qualitative and quantitative insight into the behavior of 
interacting species. 
Various simplified models have been invoked to address prototypical situations in real ecosystems: 
The paradigmatic two-species Lotka--Volterra (LV) predator-prey model \cite{lotka1920undamped,
volterra1926variazioni} was originally introduced to study fish population oscillations in the Adriatic sea, 
as well as to explain auto-catalytic chemical reaction cycles. The Rock-Paper-Scissors (RPS) model
\cite{hofbauer1998evolutionary, dawkins1983john, tainaka1994vortices,reichenbach2007mobility,
reichenbach2008self,he2010spatial} addresses the case of three cyclically interacting species with a 
conserved total number of individuals, whereas the May--Leonard (ML) model \cite{may2019stability, may1975nonlinear,frey2010evolutionary,he2011coexistence,rulands2013global,rulands2011threefold,
serrao2017stochastic} describes a more general, non-conserved situation. 
These models are obviously and necessarily rather simplified and lack many of the details of ecological
 neighborhoods. 
However, recent efforts aim at the realization and experimental implementation of such systems 
\cite{elton1942ten,utida1957cyclic,mclaren1994wolves,kerr2002local,kirkup2004antibiotic,
muhlbauer2020gauser}. 
Furthermore, it is reasonable to assume that simplified constructs such as the LV, ML, and RPS systems 
should be useful as elementary motifs and building blocks of models for more extended ecosystems. 
It is therefore imperative to investigate which of their features are qualitatively and/or quantitatively
robust and remain important when multiple interacting species are coupled to environments with richer 
structures.

Traditionally, species dynamics in ecosystems are modelled via coupled non-linear ordinary differential 
equations. 
In the case of spatially extended systems, this approach is generalized by using partial differential 
equations that represent species dispersion through simple diffusion, i.e., coupled reaction-diffusion 
equations. 
However, this mean-field or mass action approach fails to take into account the inherent randomness and 
stochastic nature of the underlying processes stemming from fluctuations in the discrete number of 
individuals, and neglects spatio-temporal correlations. 
Yet fluctuations and correlations can lead to dramatically different behavior than predicted by mean-field 
theory \cite{durrett1999stochastic}. 
For example, the classical LV mean-field rate equations predict neutral cycles and hence non-linear 
oscillations around a marginal fixed point, while stochastic computer simulations of this system yield 
decaying oscillations towards a (quasi-)stable state \cite{provata1999oscillatory,mobilia2007phase, 
washenberger2007influence}. 
This stationary state exhibits large and erratic excursions triggered by fluctuations in the species 
concentrations in zero-dimensional \cite{mckane2005predator} as well as spatially extended systems 
\cite{butler2009predator}. 
Spatially extended stochastic LV model variants also show intriguing spatial patterns and moving activity 
fronts \cite{mobilia2007phase,tauber2011stochastic, dobramysl2018stochastic}. 
Crucially, stochastic variants of the LV model exhibit a large susceptibility to randomness in the 
predator-prey interaction rates \cite{dobramysl2008spatial,dobramysl2013environmental}.

Spatially extended cyclic models such as the RPS or ML systems are influenced by internal reaction noise 
and exhibit differences in species extinction times and resulting spiral pattern wavelengths compared to the 
mean-field approximation \cite{reichenbach2008self,rulands2013global,dobrinevski2012extinction}.
In one dimension, `superdomains' may form in these cyclic models \cite{frachebourg1996spatial}.
Although both models are cyclic in nature, they exhibit different sensitivity to stochastic fluctuations. 
The RPS model, a generalization of the LV model to three cyclically competing species, displays 
comparatively weak fluctuation renormalizations in the quasi-stable coexistence state and minimal 
modifications due to randomized reaction rates \cite{he2010spatial}. 
In contrast, the ML model features a stronger renormalization of the oscillation frequency in the unstable
region where spiral structures form spontaneously, but appears to have an insignificant response to 
randomized reaction rates \cite{he2011coexistence}. 
These observations from Monte Carlo simulations raise the intriguing question: 
Under what conditions will fluctuations significantly alter the system's properties and cause marked 
deviations from simple mean-field predictions?

\begin{figure}[ht]
\subfloat[]{\includegraphics[width=0.45\linewidth]{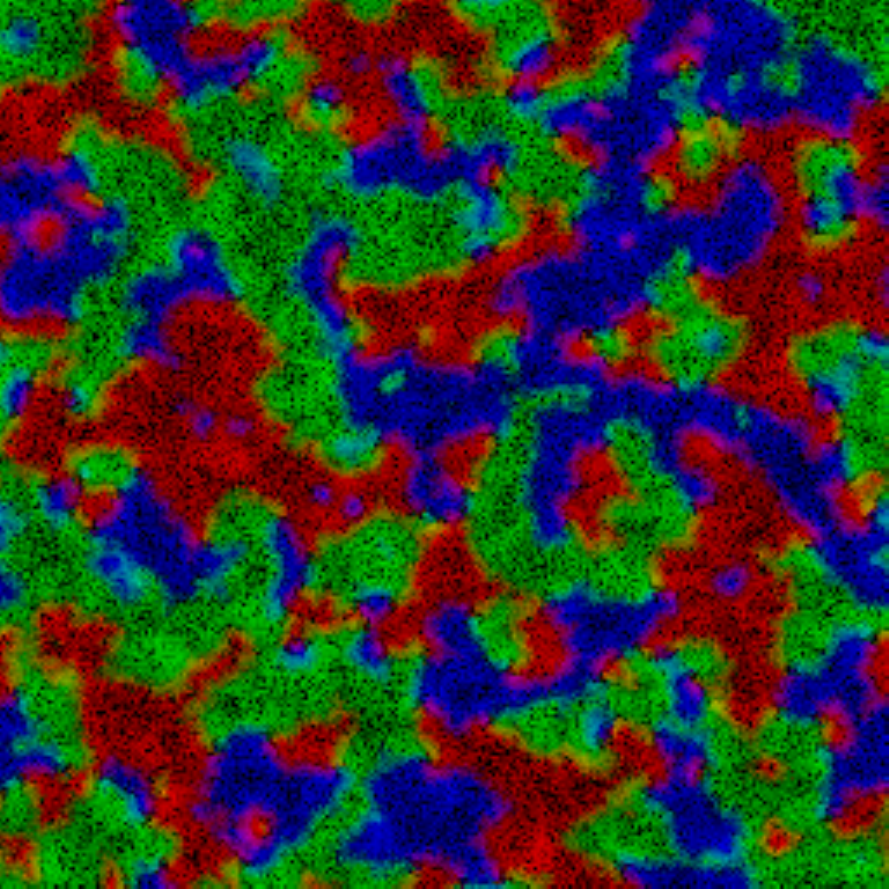}}\qquad
\subfloat[]{\includegraphics[width=0.45\linewidth]{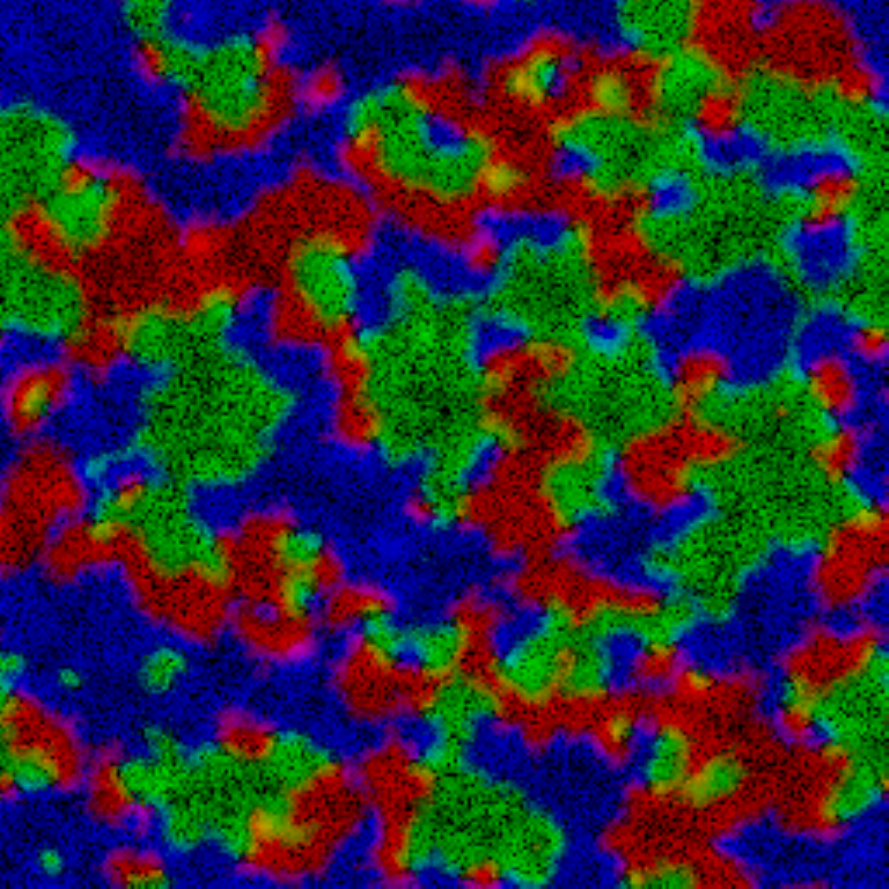}}\\
\subfloat[]{\includegraphics[width=0.45\linewidth]{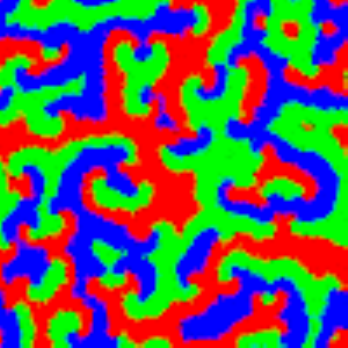}}\qquad
\subfloat[]{\includegraphics[width=0.45\linewidth]{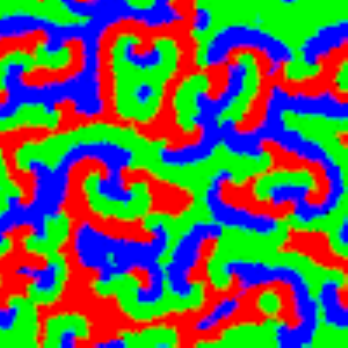}}
\caption{Snapshots of the spatial particle distribution in cyclic three-species RPS and ML models for single 
stochastic simulation runs (system size $100 \times 100$ lattice sites):
Each lattice pixel is assigned an RGB value such that each color value is proportional to the number of 
individuals of a specific species. 
A color value $0$ represents the absence of the species corresponding to that asigned color; therefore, 
black pixels indicate empty sites.  
Top: RPS model with reaction rate parameter $\lambda'=0.5$ at (a) $t=300$ Monte Carlo Steps (MCS) 
and (b) $t=400$ MCS; bottom: ML model with predation rate $\sigma'=0.5$ and reproduction rate 
$\mu=0.5$ at (c) $t=300$ MCS and (d) $t=400$ MCS, respectively. 
The red species predates on the blue species, the blue species on the green species, and the green species 
on the red species in both models.} 
\label{snapshots}
\end{figure}
To at least partially answer this question, a field-theoretical perturbation analysis was applied to the 
stochastic spatially extended LV model in Ref.~\cite{tauber2012population}. 
To one-loop order, this semi-quantitative analysis confirms that i) the fluctuation-induced damping 
renders the system unstable against spatio-temporal structures, and ii) fluctuations significantly 
renormalize the oscillation frequency in the two-species co-existence phases, especially below three 
dimensions. 
Aiming to better understand the fluctuations in spatially extended RPS [Fig.~\ref{snapshots}(a,b)] and 
ML models [Fig.~\ref{snapshots}(c,d)], we utilize a similar Doi--Peliti field theory representation for their 
associated stochastic reaction processes. 
To study the impact of intrinsic fluctuations on system parameters, a one-loop calculation is carried out 
in the perturbative regime, where the reaction rates are small as compared with the diffusivity, and a 
thorough comparison between the RPS, ML, and LV systems is conducted. 
In contrast to earlier observations in numerical simulations, the RPS model exhibits noticeable 
fluctuation-induced corrections in the perturbative regime, similar to the LV model. 
We believe that, as the dissipation becomes non-negligible in the non-perturbative regime, the associated 
infra-red (IR) divergence is regularized,  and thus substantial renormalizations become effectively 
suppressed. 
We note that in all investigated systems, the field-theoretic loop expansion technically only applies to the 
stable regions with spatially homogeneous ground states. 
Our results demonstrate that, at least in the stable region, the dynamical features of the ML model 
conversely do not receive significant modifications from fluctuations.
Based on these explicit calculation results, we also provide pertinent arguments that explain the absence of 
spontaneous spatio-temporal patterns in the RPS model with conserved total population number, as 
opposed to the ML model, which for sufficiently large system sizes develops spiral oscillatory patterns, as 
depicted in Fig.~\ref{snapshots}(c,d).

The paper is organized as follows: 
Detailed perturbative field-theoretical analyses for the cyclic and symmetric RPS and ML models are 
performed in sections II and III, respectively, where we establish the Doi--Peliti functionals for both 
models and state their corresponding generalized Langevin equations. 
Renormalized damping coefficients, oscillation frequencies, as well as diffusivities are calculated up to 
one-loop order in the perturbative fluctuation expansion. 
In section IV, a comprehensive comparison between the LV, RPS, and ML models is provided, and 
pertinent distinctions between these paradigmatic systems are highlighted. 
Specifically, we discuss the influence of fluctuations and the stability of spatio-temporal structures, and also 
briefly address the effect of quenched disorder in the reaction rates.
We conclude with a brief summary and outlook. 
Finally, Appendix~A presents a succinct review of the Doi--Peliti field theory approach and also provides a 
brief analysis of the asymmetric RPS model, demonstrating its effective two-species limit for strong 
asymmetry at the mean-field level. 
The remaining appendices list additional technical and computational details for the symmetric ML model.

\section{Stochastic rock-paper-scissors (RPS) model}

\subsection{RPS model and mean-field rate equations}

The RPS model consists of three particle species, subject to the cyclically coupled stochastic competition reactions
\begin{equation}
\begin{aligned}
    A_1+A_2 &\xrightarrow{\lambda_1'} 2A_1 \ , \\
    A_2+A_3 &\xrightarrow{\lambda_2'} 2A_2 \ , \\
    A_3+A_1 &\xrightarrow{\lambda_3'} 2A_3 \ .
\end{aligned}
\label{rps reaction}
\end{equation}
In this paper, we consider the cyclic-symmetric case, such that $\lambda_1'=\lambda_2'=\lambda_3'=\lambda'$. In this limit, the system displays a discrete $S_3$ symmetry among the three species. A brief analysis of the general asymmetric case is presented in Appendix~A. We note that every species interacts via a standard non-linear Lotka--Volterra predation reaction with the subsequent species in the cycle, consuming a “prey” particle and reproducing at the same instant. The total number of individuals is unchanged by all reactions, hence particle number conservation holds globally and locally (except for hops to neighboring lattice sites, see below).

We consider a model wherein particles from all three species perform random walks on a $d$-dimensional hyper-cubic lattice with $L^d$ sites and lattice constant $c$. We do not restrict the number of particles per lattice site, hence we do not consider finite local carrying capacities here (the total number of particles is fixed). The rate at which particles hop between sites is given by $D/c^2$, where $D$ denotes a macroscopic diffusion constant. The reactions (\ref{rps reaction}) occur on-site, and only if two particles of differing species are present. Reaction products are put on the same lattice point as the reactants.

In the limit of large diffusivities (relative to the reaction rates $\lambda'$) the system can be considered well-mixed. Hence, the RPS rules can be approximated by the three coupled mean-field rate equations for the homogenized species concentrations and with the volume reactivities $\lambda = c^{-d} \lambda'$:
\begin{equation}
\begin{aligned}
    \frac{\mathrm{d}a_1(t)}{\mathrm{d}t} &=\lambda a_1(t) \big[ a_2(t)-a_3(t) \big] \ , \\
    \frac{\mathrm{d}a_2(t)}{\mathrm{d}t} &=\lambda a_2(t) \big[ a_3(t)-a_1(t) \big] \ , \\
    \frac{\mathrm{d}a_3(t)}{\mathrm{d}t} &=\lambda a_3(t) \big[ a_1(t)-a_2(t) \big] \ . 
\end{aligned}
\end{equation}
This system of ordinary differential equations yields non-linear oscillations around a neutral fixed-line which is determined by the initial conditions. The fixed-line steady-state concentrations can be obtained by setting the time derivatives to zero, resulting in
\begin{equation}
    a_i^{\infty}=\frac{\rho}{3} \ , \quad \forall i \in \{1, 2, 3\} \ ,
\end{equation}
with the conserved total population density $\rho = a_1 + a_2 + a_3 = {\rm const}$, parameterizing the fixed-line. Linearization about this three-species coexistence fixed-line yields the stability matrix
\begin{equation}
S_{\mathrm{RPS}} = \frac{\lambda\rho}{3}
\begin{pmatrix}
    0 & +1 & -1 \\
    -1 & 0 & +1 \\
    +1 & -1 & 0
\end{pmatrix} \ ,
\label{rps stability}
\end{equation}
with eigenvalues $\{ 0, -i\omega_0 , i\omega_0 \}$. Since the non-zero eigenvalues are purely imaginary, the mean-field RPS system performs perpetual non-linear oscillations around the coexistence fixed point with frequency (in the linearized approximation) $\omega_0 = \rho\lambda/\sqrt{3}$.

\subsection{Doi--Peliti field theory and generalized Langevin equations}

The bulk part of the Doi--Peliti action for the stochastic spatially extended RPS model follows directly from the reactions (\ref{rps reaction}) and reads \footnote{A brief introduction of the Doi--Peliti field theory representation is presented in Appendix~A. We refer interested readers to Refs.~\cite{tauber2014critical,tauber2005applications} for more details.}
\begin{equation}
    \mathcal{A}^{\mathrm{RPS}} = \sum_{i=1,2,3} \int \mathrm{d}t \, \mathrm{d}^dx \, 
    \hat{a}_i \left( \partial_t-D\nabla^2 \right) a_i + \lambda \sum_{i=1,2,3} \int \mathrm{d}t \, \mathrm{d}^dx \, 
    \hat{a}_i \left( \hat{a}_{i+1}  -\hat{a}_i \right) a_i a_{i+1} \ .
\label{RPS_action}
\end{equation}
For convenience, here we drop all position and time indices $(\vec{x},t)$ on the fields and identify $a_4=a_1$. The first term describes the random nearest-neighbor hopping of the particles in the system, while the second contribution originates from the nonlinear reactions (\ref{rps reaction}). As the auxiliary field $\hat{a}_i(\vec{x}, t)$ corresponds to a projection dual state, with average $\langle \hat{a}_i (\vec{x},t) \rangle = 1$, a Doi shift $\tilde{a}_i(\vec{x}, t) = \hat{a}_i(\vec{x}, t)-1$ is conveniently applied to have the new field averaged to $\langle \tilde{a}_i(\vec{x}, t) \rangle = 0$. After the Doi shift and ignoring boundary terms, the action becomes
\begin{equation}
    \mathcal{A}^{\mathrm{RPS}} = \sum_{i=1,2,3} \int \! \mathrm{d}t \, \mathrm{d}^dx \, \tilde{a}_i 
    \left( \partial_t-D\nabla^2 \right) a_i + \lambda\! \sum_{i=1,2,3} \int \! \mathrm{d}t \mathrm{d}^dx 
    \left( \tilde{a}_i+1 \right) \left( \tilde{a}_{i+1}-\tilde{a}_i \right) a_i a_{i+1} \ .
\label{RPS_action_doi}
\end{equation}

This shifted action may now be viewed as a Janssen--De Dominicis response functional \cite{janssen1976lagrangean, 
dominicis1976technics} that represents the stochastic dynamics in terms of generalized Langevin equations. The $\tilde{a}_i$ 
fields play the role of response fields and their coupling to the particle densities, shown in the terms that are second order in 
these fields, entails the presence of multiplicative noise terms. This comparison leads to the formulation of equivalent Langevin
stochastic differential equations encoded in the action (\ref{RPS_action_doi}),
\begin{equation}
    \partial_t a_i  =D \nabla^2 a_i + \lambda a_i \left( a_{i+1}-a_{i-1} \right) + \zeta_i \ ,
\end{equation}
where $\zeta_i(\vec{x},t)$ are the components of multiplicative noise in the system with vanishing means and correlations
\begin{equation}
    \langle \zeta_i(\vec{x}_1,t_1) \, \zeta_j(\vec{x}_2, t_2) \rangle = 
    2 Z_{ij} \, \delta^{(d)}(\vec{x}_1-\vec{x}_2) \delta(t_1-t_2) \ ,
\end{equation}
with the noise correlation matrix
\begin{equation}
    Z=\lambda
\begin{pmatrix}
    \ a_1a_2 & -\frac12 a_1a_2 & -\frac12 a_1a_3 \\
    -\frac12 a_1a_2 & \ a_2a_3 & -\frac12 a_2a_3 \\
    -\frac12 a_1a_3 & -\frac12 a_2a_3 & \ a_1a_3
\end{pmatrix} \ .
\end{equation}
Note that the noise auto-correlations $Z_{ii}$ are always determined by the concentration of the predator species $A_i$ and its 
respective prey $A_{i+1}$, and the scale is set by the predation rate $\lambda$. Hence the noise directly associated with a 
given species is solely determined by its role as predator.

\subsection{Particle number conservation and Noether's theorem}

Before we proceed with the perturbation theory analysis, we quickly comment on the conserved Noether current associated with the total particle number preservation in the stochastic reaction processes (\ref{rps reaction}). This conservation law corresponds to a global $U(1)$ symmetry in the Doi--Peliti action (\ref{RPS_action}) for the RPS model, namely it remains invariant under the $U(1)$ gauge transformation
\begin{equation}
    \hat{a}_i' = e^{-i\theta} \hat{a}_i \ , \quad  a_i' = e^{i\theta} a_i \ ,
\label{RPS_symmetry}
\end{equation}
where $\theta$ is an arbitrary phase angle. The conservation law follows from the action (\ref{RPS_action}) and the symmetry transformation (\ref{RPS_symmetry}) and assumes the usual form of a continuity equation
\begin{equation}
    \partial_t j_0  + \nabla \cdot \vec{j} = 0 \ ,
\end{equation}
with
\begin{equation}
    j_0 = \sum_i \hat{a}_ia_i \ , \quad \vec{j} = -D \sum_i \left( \hat{a}_i \nabla a_i - a_i \nabla \hat{a}_i \right) .
\label{RPS_conteq}
\end{equation}
When choosing the Doi field $\hat{a}_i = 1$, $a_i$ represents the density of particle species $A_i$ and Eq.~(\ref{RPS_conteq}) turns into the diffusion equation for the conserved total particle number density,
\begin{equation}
    \partial_t  \sum_{i=1,2,3} a_i = D \nabla^2 \sum_{i=1,2,3} a_i \ .
\label{noether}
\end{equation}
We note that the symmetry (\ref{RPS_symmetry}) corresponds to the freedom of choosing the phases of the probability state $a_i$ and its dual projected state $\hat{a}_i$.

\subsection{Diagonalization of the harmonic action}

To start, we transform the fields to describe the fluctuations around the stationary fixed-point species concentrations. To this end we employ the linear transformation
\begin{equation}
    c_i(\vec{x},t) = a_i(\vec{x},t) - \frac{\rho}{3} \ , \quad \tilde{c}_i(\vec{x},t) = \tilde{a}_i(\vec{x},t) \ ,
\end{equation}
which implies $\langle c_i \rangle = 0$. In the symmetric RPS model, there is both total particle number conservation and cyclic permutation symmetry among the three distinct species. These two symmetries combined imply vanishing additive counterterms to the stationary concentrations due to fluctuations. The action for these new fluctuating fields now reads
\begin{equation}
\begin{aligned}
    &\mathcal{A}^{\mathrm{RPS}} = \int \! \mathrm{d}t \, \mathrm{d}^dx \sum_i \bigg[ \tilde{c}_i 
    \left( \partial_t - D\nabla^2 \right) c_i - \frac{\lambda \rho^2}{3} \tilde{c}_i \left( \tilde{c}_i - \tilde{c}_{i+1} \right) 
    - \frac{\lambda\rho}{3} \tilde{c}_i \left( \tilde{c}_{i+1} - \tilde{c}_{i+2} \right) \\
    &\!\! - \frac{\lambda\rho}{3} \tilde{c}_i^2 (c_i+c_{i+1}) + \frac{\lambda\rho}{3} \tilde{c}_i \tilde{c}_{i+1} 
    (c_i+c_{i+1}) - \lambda \tilde{c}_i c_i (c_{i+1}-c_{i+1}) - \lambda \tilde{c}_i^2 c_ic_{i+1} 
    + \lambda \tilde{c}_i \tilde{c}_{i+1} c_i c_{i+1} \bigg] ,
    \end{aligned}
\end{equation}
where we again identify $c_4 = c_1$ and $c_5=c_2$ for convenience. The quadratic part in the above action can be diagonalized by means of the following linear transformation
\begin{equation}
\begin{pmatrix}
    c_1 \\  c_2 \\ c_3
\end{pmatrix} 
= \frac{1}{\sqrt{3}}
\begin{pmatrix}
    1 & -\frac{1+i\sqrt{3}}{2} & -\frac{1-i\sqrt{3}}{2} \\
    1 & -\frac{1-i\sqrt{3}}{2} & -\frac{1+i\sqrt{3}}{2} \\
    1 & 1 & 1
\end{pmatrix}
\begin{pmatrix}
    \phi_o \\ \phi_+ \\ \phi_-
\end{pmatrix} \ ,
\end{equation}
and 
\begin{equation}
\begin{pmatrix}
    \tilde{c}_1 \\  \tilde{c}_2 \\ \tilde{c}_3
\end{pmatrix}
= \frac{1}{\sqrt{3}}
\begin{pmatrix}
    1 & -\frac{1-i\sqrt{3}}{2} & -\frac{1+i\sqrt{3}}{2} \\
    1 & -\frac{1+i\sqrt{3}}{2} & -\frac{1-i\sqrt{3}}{2} \\
    1 & 1 & 1
\end{pmatrix}
\begin{pmatrix}
    \tilde{\phi}_o \\ \tilde{\phi}_+ \\ \tilde{\phi}_-
\end{pmatrix} \ .
\end{equation}

The resulting action becomes $\mathcal{A}^{\mathrm{RPS}} = \mathcal{A}^{\mathrm{RPS}}_0 + 
\mathcal{A}^{\mathrm{RPS}}_{\mathrm{int}}$, with the Gaussian part
\begin{equation}
    \mathcal{A}^{\mathrm{RPS}}_0 = \int\!\mathrm{d}t\,\mathrm{d}^dx \Big[ \tilde{\phi}_+ 
    \left( \partial_t-D\nabla^2+i\omega_0 \right)\! \phi_+ + \tilde{\phi}_o \left( \partial_t-D\nabla^2 \right)\! \phi_o 
    + \tilde{\phi}_- \left( \partial_t-D\nabla^2-i\omega_0 \right)\! \phi_- \Big] ,
\label{rps harmonic}
\end{equation}
and the nonlinear contributions (vertices)
\begin{equation}
\begin{aligned}
    \mathcal{A}^{\mathrm{RPS}}_{\mathrm{int}} &= \int\!\mathrm{d}t \, \mathrm{d}^dx \, \bigg[ - \lambda\rho^2 
    \tilde{\phi}_+ \tilde{\phi}_- + i\frac{\lambda\rho}{3} \tilde{\phi}_o \left( \tilde{\phi}_+\phi_+ - \tilde{\phi}_-\phi_- 
    \right) - \frac{2\lambda\rho}{\sqrt{3}} \, \tilde{\phi}_+ \tilde{\phi}_- \phi_o \\
    &\qquad\qquad\quad\ - i\lambda \left (\tilde{\phi}_+ \phi_-^2 - \tilde{\phi}_- \phi_+^2 - \tilde{\phi}_+ \phi_+ \phi_o
    + \tilde{\phi}_- \phi_- \phi_o \right) \\
    &\qquad\quad - i\frac{\lambda\rho}{6} \left[ (1-i\sqrt{3}) \tilde{\phi}_+^2 \phi_- - (1+i\sqrt{3}) \tilde{\phi}_-^2 
    \phi_+ \right] + \text{four-point vertices} \bigg] \ .
\end{aligned}
\label{rps_int}
\end{equation}
Here, $\omega_0 = \lambda\rho/\sqrt{3}$ denotes the zeroth-order oscillation frequency of the $\phi_{+/-}$ modes. 
$\mathcal{A}^{\mathrm{RPS}}_0$ is the diagonalized harmonic part of the action, while 
$\mathcal{A}^{\mathrm{RPS}}_{\mathrm{int}}$ represents the ``interaction'' contributions for the perturbative expansion. We omit the explicit expressions for the four-point vertices as they will not contribute to the dispersion relations at one-loop order, which shall be clear in the calculation below. It is manifest that $\phi_o=(a_1+a_2+a_3-\rho)/\sqrt{3}$ represents the fluctuation of the total particle number density. Due to the conservation law (\ref{noether}), the $\phi_o$ mode is purely diffusive, and its exact dispersion relation in the harmonic part of the action acquires no fluctuation corrections. The $\phi_+$ and $\phi_-$ modes may be viewed as the left- and right-rotating waves in the system. At tree level, they are purely oscillating modes without dissipation, i.e., the real part of the mass term vanishes. 

\subsection{One-loop fluctuation corrections}

\begin{figure}[t]
    \centering
    \includegraphics[scale=1.2]{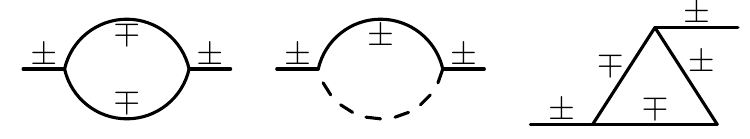}
    \caption{One-loop fluctuation contributions to the two-point vertex function 
    $\Gamma_{\tilde{\phi}_{\pm}\phi_{\pm}}^{(1,1)}$ in RPS model. The solid lines represent the $\phi_+$ and $\phi-$
    modes, whereas the dashed lines represent the purely diffusive $\phi_o$ mode.}
    \label{rps propagator}
\end{figure}
We have applied a field-theoretical perturbation theory to one-loop order and calculated the renormalized diffusion constant $D_r$, damping constant $\gamma_r$, and oscillation frequency 
$\omega_r$\footnote{For more details on the perturbation expansion and Feynman graph representations, we refer to 
Refs.~\cite{tauber2012population,tauber2014critical,zinn2021quantum}.}. To all orders in the fluctuation expansion extending 
beyond the mean-field approximation, there should be no correction to the two-point vertex function 
$\Gamma_{\tilde{\phi}_{o}\phi_{o}}^{(1,1)}$ or propagator self-energy for the $\phi_o$ mode, whence it retains its 
tree-level purely diffusive dispersion relation as dictated by the conservation law. For the $\phi_\pm$ modes, the one-loop 
Feynman diagrams for the two-point vertex functions $\Gamma_{\tilde{\phi}_{\pm}\phi_{\pm}}^{(1,1)}$ are displayed in 
Fig.~\ref{rps propagator}. The solid lines represent the $\phi_+$ and $\phi_-$ propagators, while the dashed lines indicate 
the diffusive mode $\phi_o$. In our convention, time and hence momentum always flow from right to left in the Feynman 
diagrams. The analytic expression for the two-point vertex function $\Gamma_{\tilde{\phi}_{\pm}\phi_{\pm}}^{(1,1)}$ 
reads
\begin{equation}
\begin{aligned}
    \Gamma_{\tilde{\phi}_{\pm}\phi_{\pm}}^{(1,1)}&(p,\omega) = i\omega+Dp^2+u_0 \pm i\omega_0 
    + \frac{\sqrt{3}\lambda\omega_0}{6D}\int_k I\!\left(\frac{u_0\pm i\omega_0}{2D}\right) \\
    &-\frac{\sqrt{3}\lambda\omega_0}{6D}(1\pm i\sqrt{3})\int_k I\!\left(\frac{u_0\mp i\omega_0}{D}\right) 
    - \frac{\lambda\omega_0^2}{D^2}\int_k\frac{1}{k^2+\frac{u_0}{D}} \, I\!\left(\frac{u_0 \mp i\omega_0}{D} \right) ,
\end{aligned}
\label{rps_vertex}
\end{equation}
where $\int_k$ is short-hand for the $d$-dimensional wavevector integral $\int \! \mathrm{d}^dk / (2\pi)^d$, and the
function $I$ is defined as
\begin{equation}
    I(m^2)=\frac{1}{k^2+m^2+\frac{i\omega+Dp^2}{2D}+p\cdot k} \ .
\end{equation}

The damping constant $u_0$ in Eq.~(\ref{rps_vertex}) is introduced to regularize the infrared (IR) singularities that emerge 
in later calculations. A nonzero renormalized $u_r$ will be generated by the fluctuations, but it is of higher order in the perturbation expansion; thus, we need to take $u_0\rightarrow 0$ at the end. This two-point function can also be expressed with the renormalized parameters as
\begin{equation}
    \Gamma_{\tilde{\phi}_{\pm}\phi_{\pm}}^{(1,1)}(p,\omega)=Z_{\phi_\pm} 
    \left( i\omega+D_rp^2+u_r\pm i\omega_r \right) ,
\label{rps renomalized vertex}
\end{equation}
where $Z_{\phi_\pm}$ absorbs all related wave function renormalizations (ultraviolet / UV divergences) in (\ref{rps_vertex}). 
The renormalized diffusivity $D_r$, damping $u_r$, and oscillation frequency $\omega_r$ can be inferred accordingly from 
the explicit one-loop result (\ref{rps_vertex}) and (\ref{rps renomalized vertex}). We obtain the following formal expressions 
for $D_r$, $u_r$, and $\omega_r$,
\begin{equation}
    \begin{aligned}
    D_r=&D + \frac{\sqrt{3}\lambda\omega_0}{6dD}\int_k\frac{k^2}{(k^2 \pm \frac{i\omega_0}{2D})^3} 
    - \frac{\sqrt{3}\lambda\omega_0}{6dD}(1\pm i\sqrt{3})\int_k\frac{k^2}{(k^2 \mp \frac{i\omega_0}{D})^3} \\
    &- \frac{\lambda\omega_0^2}{dD^2}\int_k \frac{1}{(k^2\mp \frac{i\omega_0}{D})^3} \ , \\
    u_r\pm i\omega_r =& \pm i\omega_0 \Bigg[1-\frac{\lambda}{2D}\int_k \frac{1}{k^2 + \frac{u_0}{D}} 
    \pm \frac{i\sqrt{3}\lambda}{6D}\int_k\left(\frac{1}{k^2 \mp \frac{i\omega_0}{D}} 
    - \frac{1}{k^2 \pm \frac{i\omega_0}{2D}}\right) \\
    &+\frac{\sqrt{3}\lambda\omega_0}{12D^2}\int_k\frac{1}{(k^2 \pm \frac{i\omega_0}{2D})^2}  
    - \frac{\sqrt{3}\lambda\omega_0}{12D^2}(1\mp i\sqrt{3})\int_k\frac{1}{(k^2 \pm \frac{i\omega_0}{2D})^2} \Bigg] \ .
    \end{aligned}
\end{equation}
Hence we indeed see that a non-zero damping $u_r$ is generated at one-loop order from the fluctuations, in agreement with 
Monte Carlo simulation data \cite{he2010spatial}. The infra-red (IR) divergence at one-loop order that appears when in the
renormalized oscillation frequency $\omega_r$ in low dimensions $d \leq 2$ is caused by the contribution of the massless 
$\phi_o$ mode as $u_0$ is set to zero. The infra-red (IR) divergence at one-loop order that appears in the renormalized oscillation frequency $\omega_r$ in low dimensions $d \leq 2$ is caused by the superposition of $\phi_+$ and $\phi_-$ 
modes as $u_0$ is set to zero. Our analysis of the one-loop results shows that the $\phi_\pm$ modes are inherently massive, 
as they acquire non-zero damping $u_r$. Thus, the IR divergence can be resolved simply by maintaining a finite value for 
$u_0$. For dimensions $d \geq 2$, there are also UV divergences present. Nevertheless, all systems of interest have a natural
cutoff in the UV limit, which is defined by the lattice constant $c$. The renormalized variables in different physically accessible 
dimensions are presented below.

\subsubsection{$d = 1$:}
\begin{figure}[t]
\subfloat[]{\includegraphics[width=0.45\linewidth]{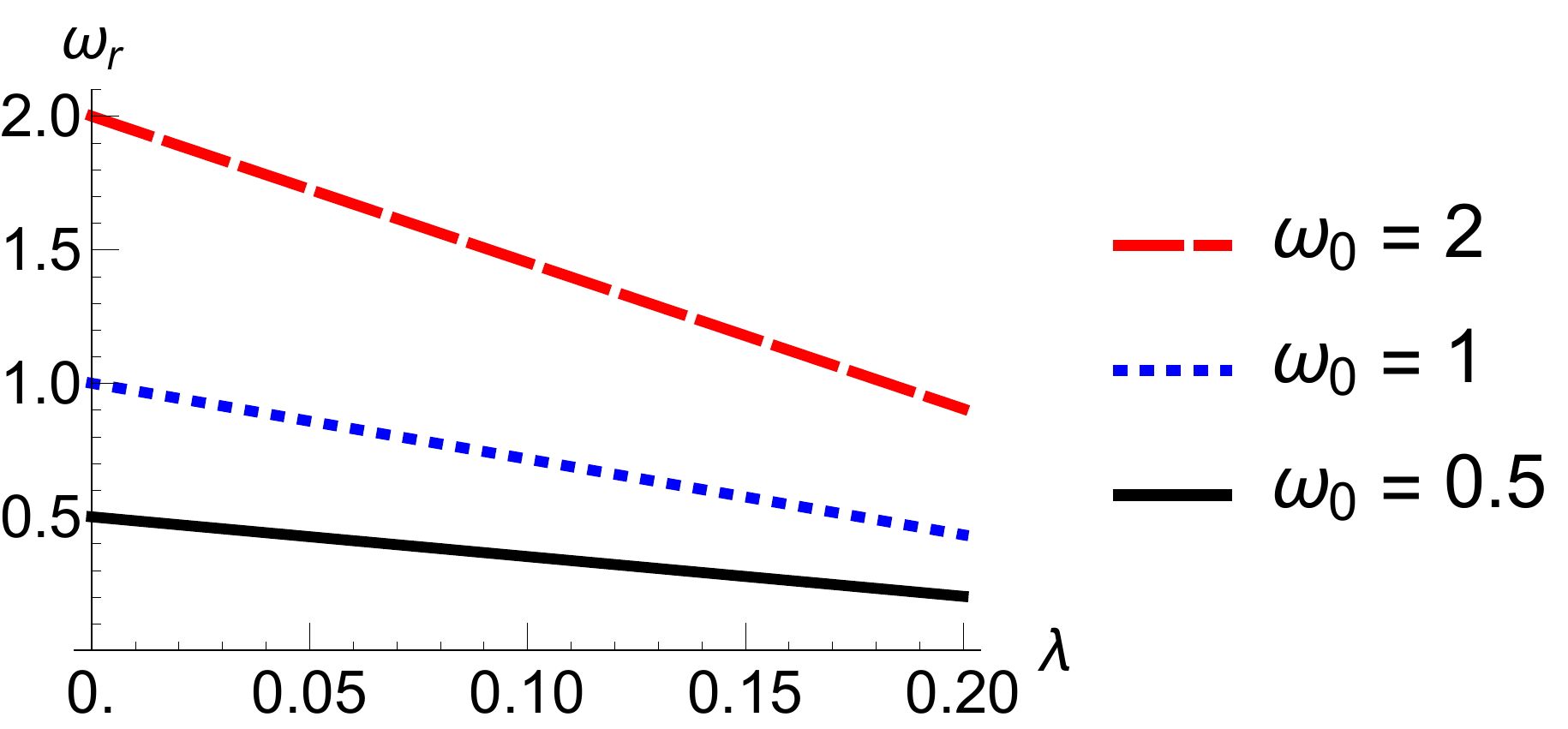}}\qquad
\subfloat[]{\includegraphics[width=0.45\linewidth]{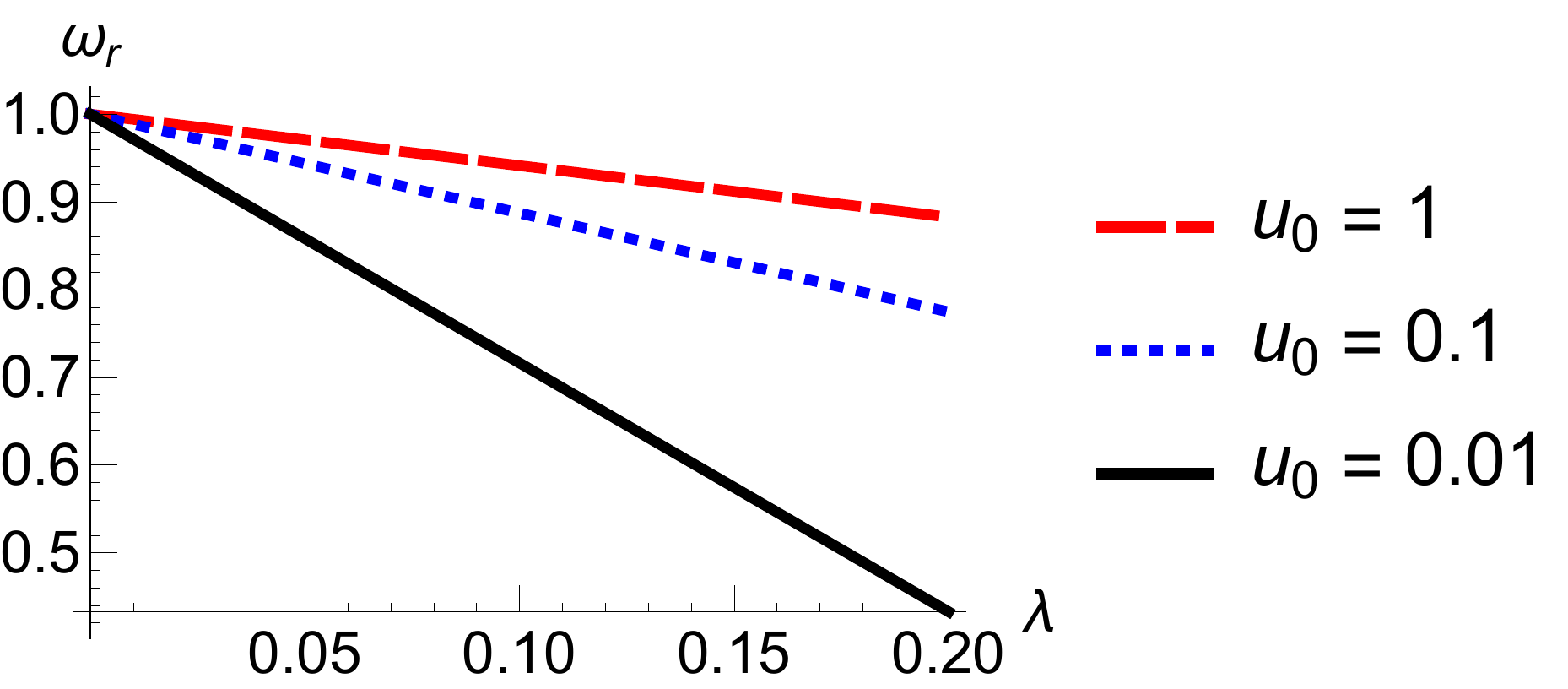}}\\
\subfloat[]{\includegraphics[width=0.45\linewidth]{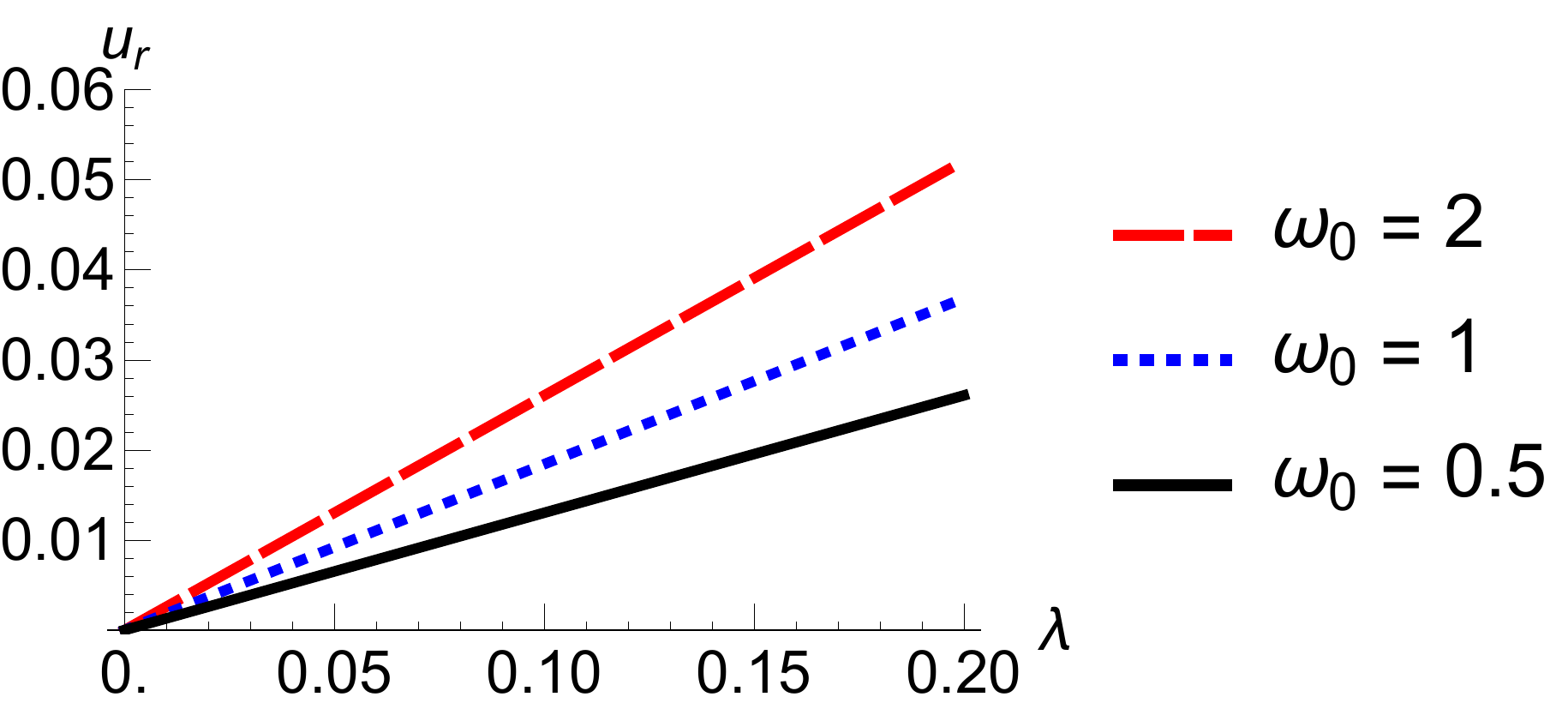}}\qquad
\subfloat[]{\includegraphics[width=0.45\linewidth]{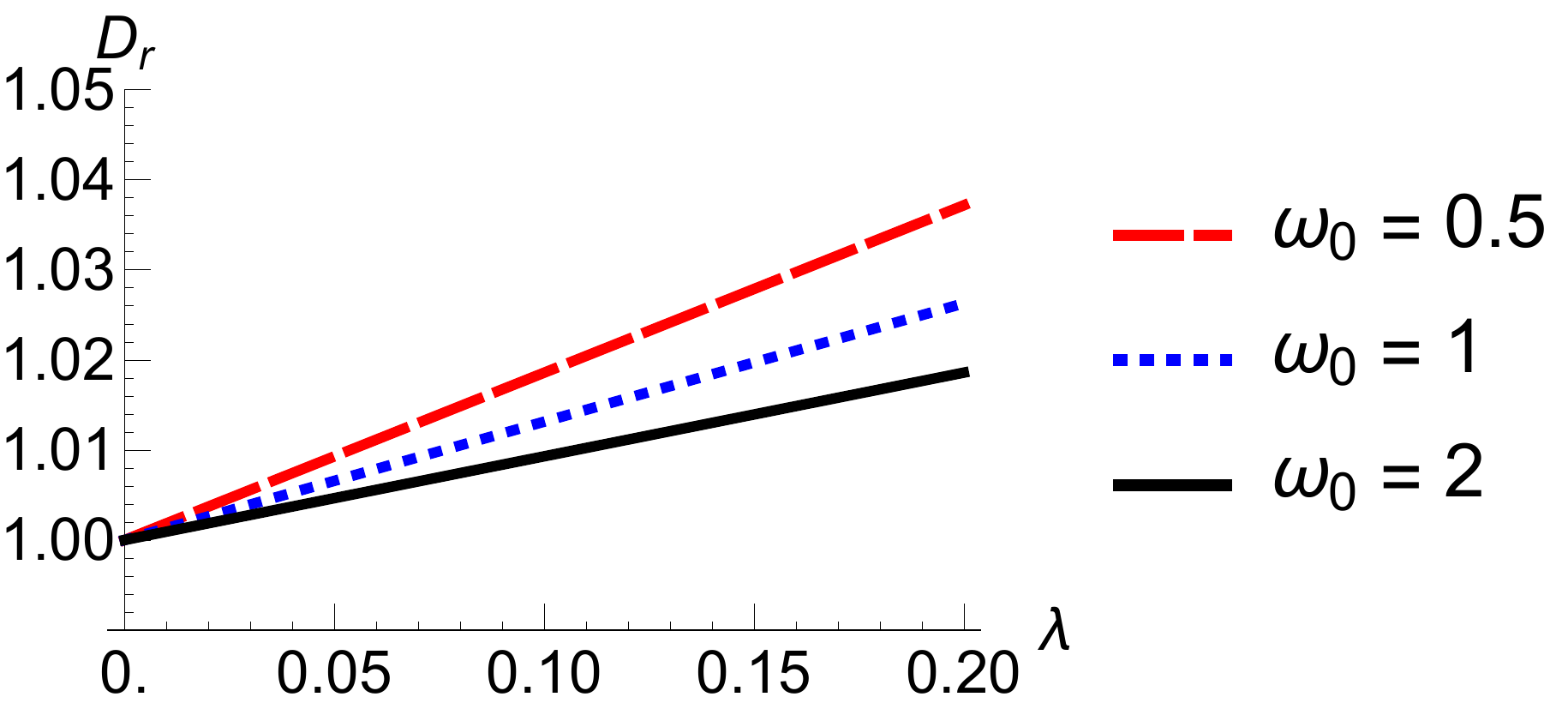}}
\caption{Renormalized parameters in one dimension ($d=1$) in the RPS model: 
	(a) Renormalized frequency $\omega_r$ as a function of the reactivity parameter $\lambda$, with $D=1$, $u_0=0.01$, 
	and for different bare frequencies $\omega_0$.  
	(b) Renormalized frequency $\omega_r$ as a function of the parameter $\lambda$, with $\omega_0 = 1$, $D = 1$, and 
	different bare damping coefficients $u_0$. 
	(c) Renormalized $u_r$ as a function of the parameter $\lambda$ with $D = 1$ and different bare frequencies 
	$\omega_0$.
	(d) Renormalized diffusion constant $D_r$ as a function of the parameter $\lambda$ with $D = 1$ and different bare 
	frequencies $\omega_0$.}
\label{RPS 1d Plot}
\end{figure}

In one dimension, the renormalized parameters are
\begin{equation}
\begin{aligned}
    \mathrm{Re} D_r &= D+\lambda\sqrt{\frac{D}{\omega_0}}\left(\frac{7\sqrt{2}}{64}+\frac{\sqrt{6}}{192}
    -\frac{\sqrt{3}}{48}\right) , \\
    \omega_r &= \omega_0\Bigg[1-\frac{\lambda}{4D}\sqrt{\frac{D}{u_0}}-\frac{\lambda}{D}\sqrt{\frac{D}{\omega_0}}
    \left(\frac{\sqrt{3}}{8} + \frac{\sqrt{6}}{32} + \frac{\sqrt{2}}{32}\right)\Bigg] \ , \\
     u_r &= \frac{\lambda\omega_0}{D}\sqrt{\frac{D}{\omega_0}}\left(\frac{\sqrt{3}}{8} - \frac{\sqrt{6}}{32} + \frac{\sqrt{2}}{32}\right) ;
\end{aligned}   
\end{equation}
some numerical results are depicted in Fig.~\ref{RPS 1d Plot}. We observe that the renormalized frequency $\omega_r$ 
diverges when $u_0 \rightarrow 0$. This IR divergence indicates strong fluctuation corrections to the oscillation frequency in 
the perturbative regime where reaction rates are small, and $u_0 \to u_r \sim \lambda$ is of first order in the reactivity. 
However, numerical simulations are invariably performed outside this regime for the sake of computational efficiency, as large 
reaction rates are used to avoid long relaxation times. Thus, no strong fluctuation-induced renormalization have been
encountered in the simulations. To one-loop order, $u_r > 0$, which indicates the stability of the system's spatially 
homogeneous ground state with respect to fluctuations.

\subsubsection{$d = 2$:}
\begin{figure}[t]
\subfloat[]{\includegraphics[width=0.45\linewidth]{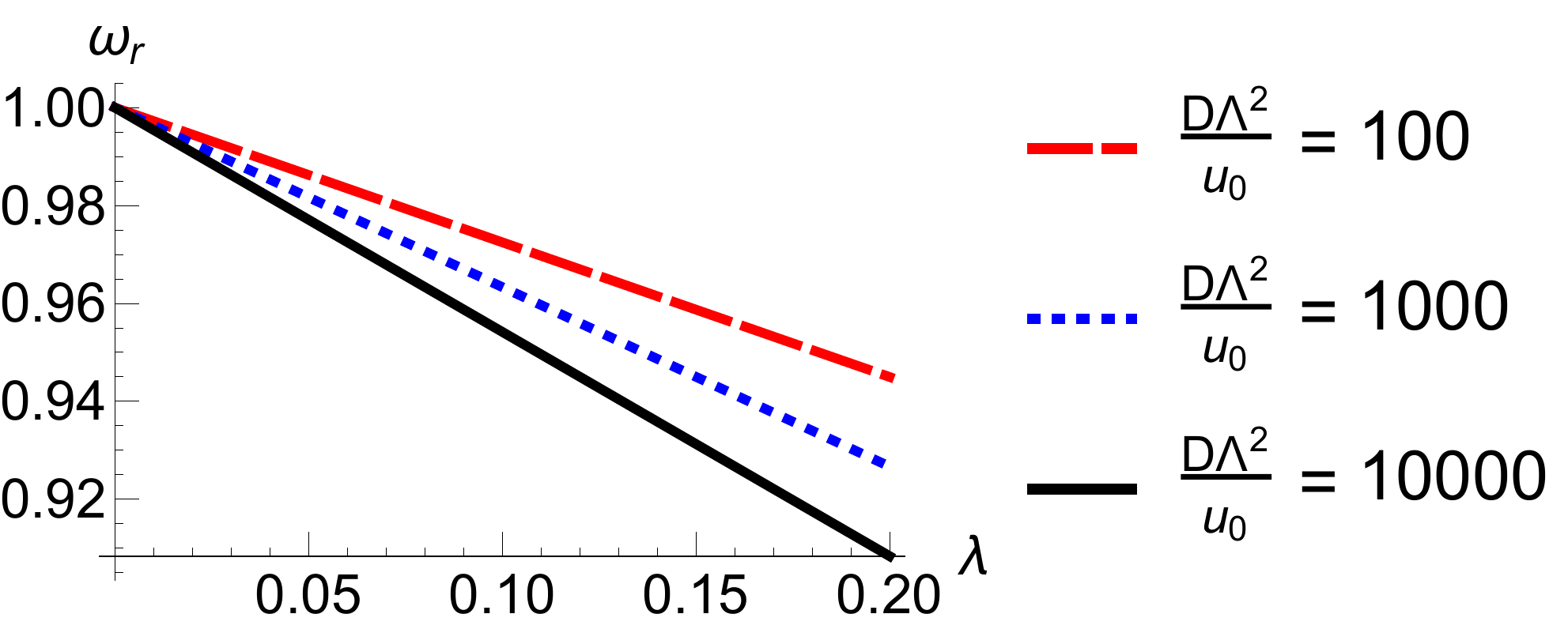}}\qquad
\subfloat[]{\includegraphics[width=0.45\linewidth]{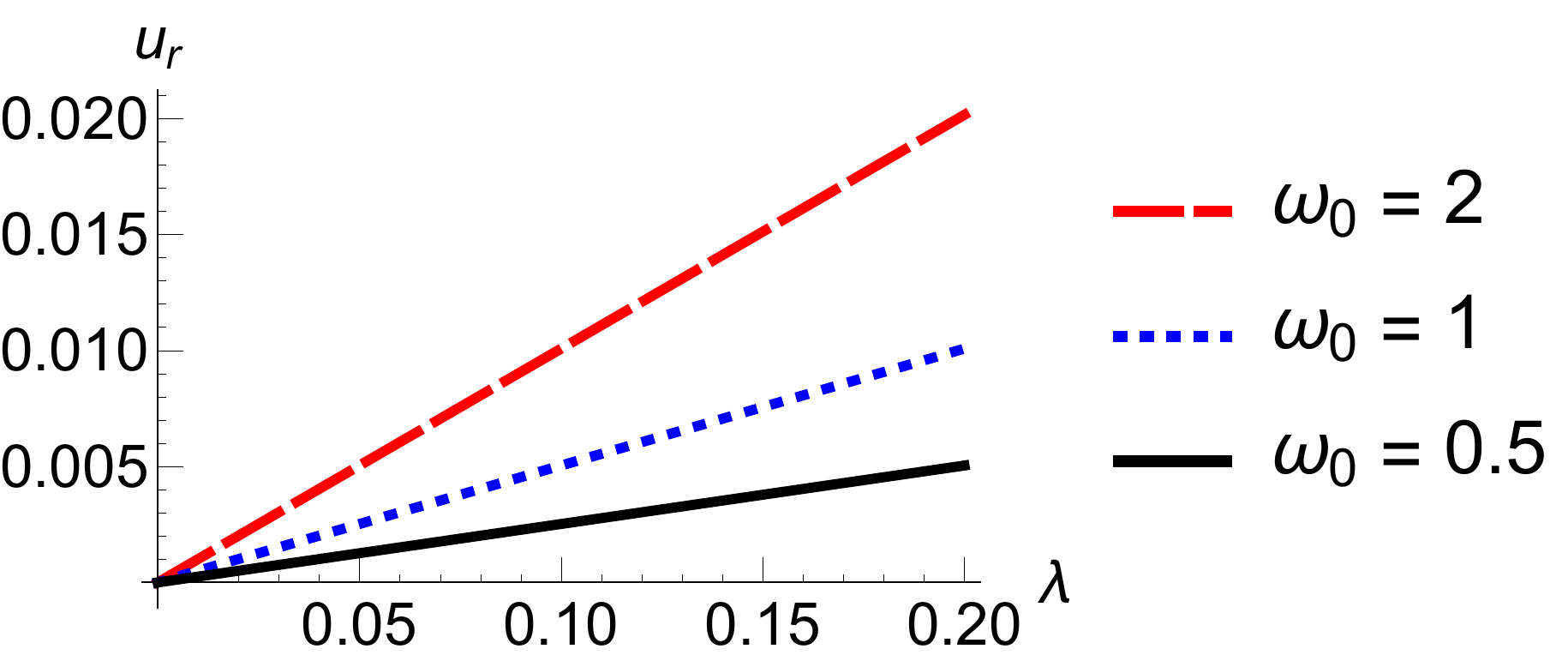}}
\caption{Renormalized parameters in two dimensions ($d=2$) in the RPS model: 
	(a) Renormalized frequency $\omega_r$ as a function of the reactivity parameter $\lambda$ with $D=1$ and different 
	ratios $D\Lambda^2/u_0$.  
	(b) Renormalized damping $u_r$ as a function of the parameter $\lambda$ with $D = 1$ and different bare 
	frequencies $\omega_0$.}
\label{RPS 2d Plot}
\end{figure}

In two dimensions, the renormalized variables read
\begin{equation}
\begin{aligned}
    \mathrm{Re} D_r &= D+ \frac{3\lambda}{32\pi} \ , \\
    \omega_r &= \omega_0\Bigg[1-\frac{\lambda}{8D\pi}\ln \bigg(\frac{D\Lambda^2}{u_0}\bigg)-\frac{\lambda}{D\pi}\left(\frac{1}{16}+\frac{\sqrt{3}\pi}{24}\right)\Bigg] \ , \\
     u_r &= \frac{\sqrt{3}\lambda\omega_0}{48D\pi}\left(3+2\ln 2\right) .
\end{aligned}   
\end{equation}
Note that we have explicitly introduced the cutoff $\Lambda \sim \pi / c$ to regularize the UV divergence for the renormalized 
oscillation frequency. We plot $\omega_r$ and the damping parameter $u_r$ in Fig.~\ref{RPS 2d Plot}. The renormalized diffusion 
constant $D_r$ only linearly depends on the parameter $\lambda$.  As in the one-dimensional case, the oscillation frequency 
$\omega_r$ diverges as $u_0\rightarrow 0$. We note that the damping constant $u_r$ is also positive at $d=2$.

\subsubsection{$d = 3$:}
\begin{figure}[b]
\subfloat[]{\includegraphics[width=0.27\linewidth]{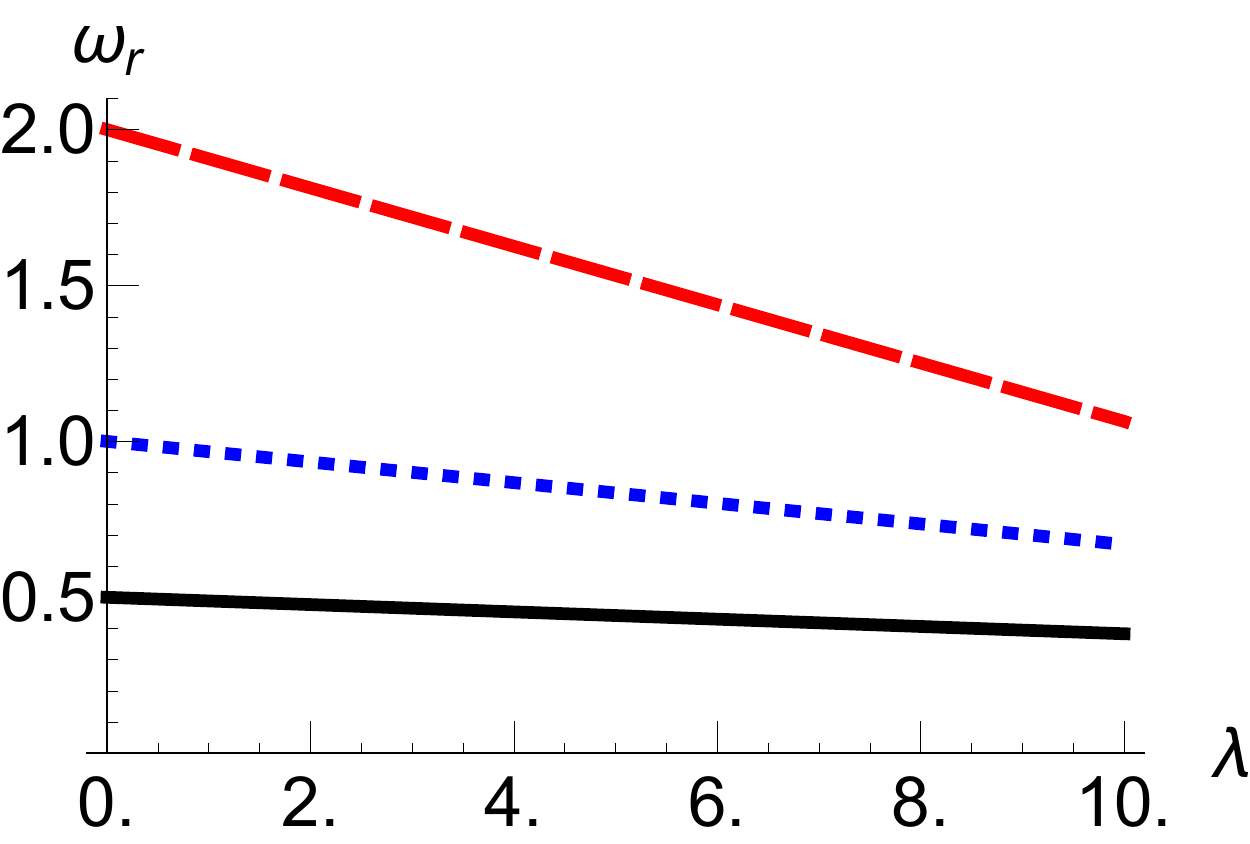}}\
\subfloat[]{\includegraphics[width=0.3\linewidth]{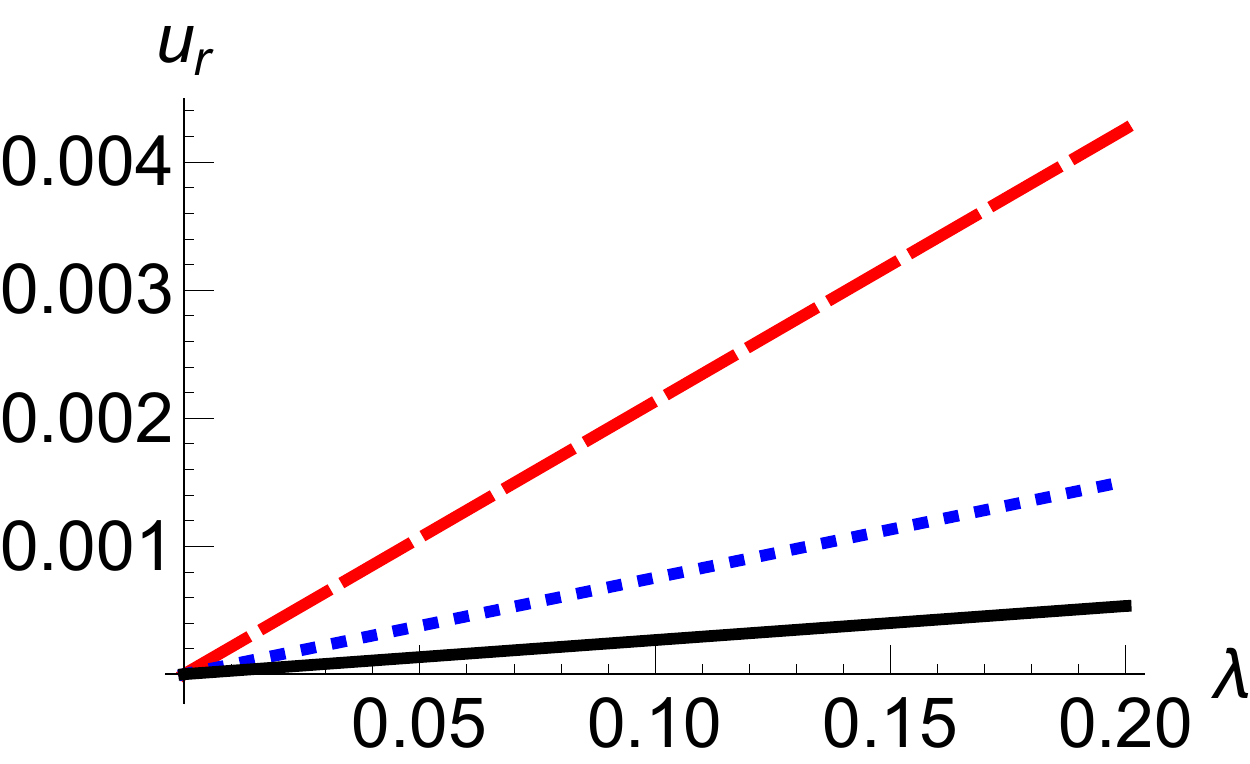}}\
\subfloat[]{\includegraphics[width=0.4\linewidth]{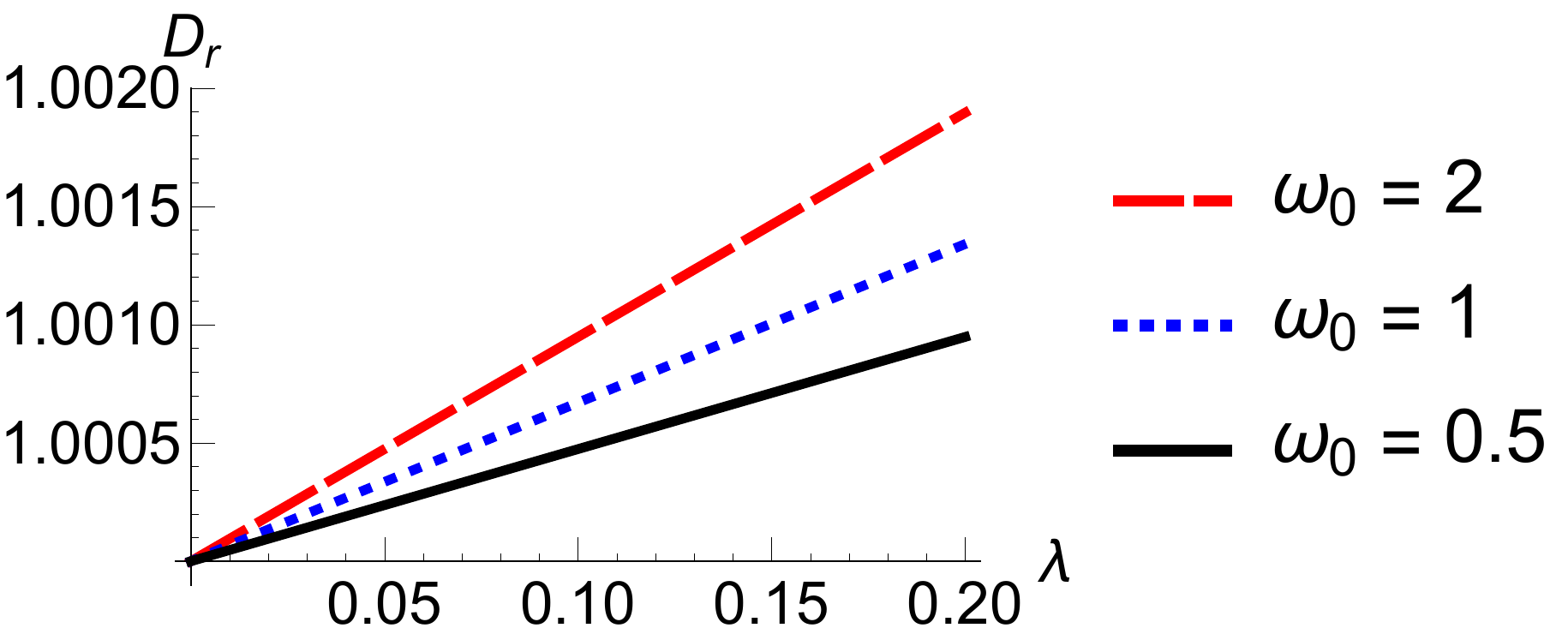}}
\caption{Renormalized parameters in three dimensions ($d=3$) in the RPS model: 
	(a) Renormalized frequency $\omega_r$ as a function of the reactivity parameter $\lambda$ with $D=1$ and different 
	ratios $D\Lambda^2/u_0$.  
	(b) Renormalized damping $u_r$ as a function of the parameter $\lambda$ with $D = 1$ and different bare frequencies 
	$\omega_0$. 
	(c) Renormalized diffusion constant $D_r$ as a function of the parameter $\lambda$ with $D = 1$ and different bare 
	frequencies $\omega_0$.}
\label{RPS 3d Plot}
\end{figure}

In three dimensions, we may safely set $u_0=0$ and the renormalized system parameters follow as
\begin{equation}
\begin{aligned}
    \mathrm{Re} D_r &= D+ \frac{\lambda}{\pi}\sqrt{\frac{\omega_0}{D}}\left(\frac{\sqrt{3}}{192}
    -\frac{\sqrt{6}-5\sqrt{2}}{384}\right) , \\
    \omega_r &= \omega_0\Bigg[1-\frac{\lambda}{D\pi} \sqrt{\frac{\omega_0}{D}}\left(\frac{\sqrt{3}}{96} 
   + \frac{5\sqrt{6}}{192} + \frac{\sqrt{2}}{64}\right)\Bigg] \ , \\
    u_r &= \frac{\lambda\omega_0}{D\pi}\sqrt{\frac{\omega_0}{D}}\left(-\frac{\sqrt{3}}{96} + \frac{5\sqrt{6}}{192} 
    - \frac{\sqrt{2}}{64}\right) .
\end{aligned}   
\end{equation}
We notice that for $d>2$ the IR divergences disappear and fluctuation effects become generally weak. $u_r$ is also positive 
for $d=3$, as displayed in Fig.~\ref{RPS 3d Plot}.

We have found that in dimensions $d=1$, $2$, and $3$, the diffusivity $D$ experiences an upward shift, indicating that 
fluctuations enhance the diffusion. Our results, depicted in Fig. \ref{RPS 1d Plot}, \ref{RPS 2d Plot}, and \ref{RPS 3d Plot}, 
show that as the reaction rate increases, the characteristic frequency $\omega_r$ shifts to smaller values, as the reactions drive 
the system towards a spatially more homogeneous distribution, leading to slower oscillations. The decline in the characteristic 
frequencies is in accordance with the numerical simulation data in Ref.~\cite{he2010spatial}. In contrast with the LV model, 
Monte Carlo simulations of the RPS system do not appear to show strong renormalization effects \cite{he2010spatial}, although both models feature a logarithmic divergence in two dimensions. The IR divergence in the RPS model appears as a 
consequence of the fact that the corrections are built using the Gaussian theory which has zero damping, precisely as in the LV 
model. The positive fluctuation-induced damping $\mu_r$, in contrast to the possibly negative one in the LV model, indicates 
that the system remains stable against the spontaneous emergence of spatio-temporal structures.

\section{Stochastic May--Leonard (ML) model}

\subsection{ML model and mean-field rate equations}

The following discussion of the mean-field theory, Doi-Peliti action, and Langevin representation for the spatially extended 
stochastic ML model was laid out in detail in Ref.~\cite{serrao2017stochastic}. Here we summarize the pertinent points needed 
for our comparison with the RPS model and the computation of the fluctuation corrections to one-loop order. Following the conventions in Ref.~\cite{serrao2017stochastic}, the reactions in the ML model read
\begin{equation}
\begin{aligned}
B_i+B_{i+1} &\xrightarrow{\sigma'} B_i \ , \\
B_ i&\xrightarrow{\mu} 2B_i \ , \\
2B_i &\xrightarrow{\kappa'} B_i \ ,
\end{aligned}
\label{ml reactions}
\end{equation}
where $i=1,2,3$ denotes the three competing species, and we identify $B_4=B_1$ as before. In contrast to the RPS system, 
the reactions in the ML model do not conserve the total particle number. The first two reactions implement predation and reproduction independently, while the third reaction implements ``soft" site occupation constraint to effectively represent a 
finite carrying capacity. As in the RPS model, we consider a model wherein particles from all three species perform random walks on a $d$-dimensional hyper-cubic lattice with $L^d$ sites and lattice constant $c$. In the large diffusivity limit, the 
dynamics is governed by the mean-field rate equations
\begin{equation}
\begin{aligned}
    \frac{\mathrm{d}b_1(t)}{\mathrm{d}t}&=b_1(t)\Big(-\sigma b_3(t)+\mu-\kappa b_1(t) \Big) \ , \\
    \frac{\mathrm{d}b_2(t)}{\mathrm{d}t}&=b_2(t)\Big(-\sigma b_1(t)+\mu-\kappa b_2(t) \Big) \ , \\
    \frac{\mathrm{d}b_3(t)}{\mathrm{d}t}&=b_3(t)\Big(-\sigma b_2(t)+\mu-\kappa b_3(t) \Big) \ ,
\end{aligned}
\end{equation}
where $\sigma=c^d\sigma'$ and $\kappa=c^d\kappa'$ are the volume reaction rates. Instead of a fixed line defined by the
initial condition, the ML system displays a unique fixed point at mean-field level. By setting the time derivatives to be $0$, the 
steady-state concentrations are found to be
\begin{equation}
    b_i^{\infty}=\frac{\mu}{\sigma+\kappa} \ ,\quad \forall i \in \{1, 2, 3\} \ ,
\end{equation}
and the associated stability matrix reads
\begin{equation}
    S_{\mathrm{ML}}=-\frac{\mu}{\sigma+\kappa}
    \begin{pmatrix}
    \kappa & 0 & \sigma\\
    \sigma & \kappa & 0\\
    0 & \sigma & \kappa
    \end{pmatrix} .
\end{equation}
Its eigenvalues at the coexistence fixed point are $\{-\mu, -\mu (2\kappa-\sigma\pm i\sqrt{3}\sigma) / 2(\sigma+\kappa)\}$. The first eigenvalue $-\mu$ is always negative which implies the stability of the corresponding eigenmode, namely the 
exponential decay of the total particle number, see below. The imaginary part of the two complex conjugate eigenvalues, 
$\pm\sqrt{3}\mu\sigma/2(\sigma+\kappa)$, represents the frequency of temporal oscillations for the associate modes, whose
amplitudes are either exponentially damped or growing. When $2\kappa>\sigma$, the real part of the complex eigenvalues is negative and the limit circles are stable. Otherwise, for $2\kappa<\sigma$, the limit circles are unstable and one observes the 
spontaneous formation of spiral structures in the system. In the vicinity of the Hopf bifurcation at $2\kappa=\sigma$, the time 
evolution of the two modes corresponding to the complex conjugate eigenvalues becomes much slower than the fast relaxing 
mode, which introduces a natural time scale separation. As a consequence of the critical slowing down near the Hopf 
bifurcation, the fast relaxing mode can be integrated out and the system is effectively governed by the complex time-dependent 
Ginzburg--Landau equation \cite{serrao2017stochastic}.

\subsection{Doi--Peliti field theory and generalized Langevin equations}

The Doi-Peliti action follows from the reactions of the ML model and reads
\begin{equation}
\begin{aligned}
\mathcal{A}^{\mathrm{ML}}=\sum_{i=1,2,3}\int\!\mathrm{d}t\,\mathrm{d}^dx \bigg[\hat{b}_i \left( \partial_t
-D\nabla^2\right)\! b_i+\mu\,\hat{b}_ib_i \Bigl( 1-\hat{b}_i \Bigr)+\kappa\,\hat{b}_ib_i^2 \Bigl( \hat{b}_i-1 \Bigr) \\
+\sigma\,\hat{b}_ib_ib_{i+1}\Bigl( \hat{b}_{i+1}-1 \Bigr) \bigg] .
\end{aligned}
\label{doi ml}
\end{equation}
This action does not obey the $U(1)$ global symmetry present in the RPS model; indeed, the total particle number is not
conserved under the ML reactions (\ref{ml reactions}). Following the Doi shift to the fluctuating auxiliary fields 
$\tilde{b}_i(\vec{x},t)=\hat{b}_i(\vec{x},t)-1$, the action becomes
\begin{equation}
\begin{aligned}
\mathcal{A}^{\mathrm{ML}}=\sum_{i=1,2,3}\int\!\mathrm{d}t\,\mathrm{d}^dx \bigg[ \tilde{b}_i \left( 
\partial_t-D\nabla^2-\mu \right)\! b_i-\mu\,\tilde{b}_i^2b_i+\kappa\,\tilde{b}_ib_i^2 \Bigl( \tilde{b}_i+1 \Bigr) \\
+\sigma\,\tilde{b}_{i+1}b_ib_{i+1} \Bigl( \tilde{b}_ i+1 \Bigr) \bigg] .
\end{aligned}
\end{equation}
As in the RPS model above, we may now view this shifted action as a Janssen--De Dominicis functional which is equivalent to
the corresponding generalized Langevin equations
\begin{equation}
    \partial_tb_i=D\nabla^2b_i+\mu b_i-\kappa b_i^2-\sigma b_{i}b_{i+2}+\xi_i \ ,
\end{equation}
where $\xi_i(\vec{x},t)$ represent the multiplicative noise components with correlators
\begin{equation}
    \langle\xi_i(\vec{x}_1,t_1)\,\xi_j(\vec{x}_2,t_2)\rangle=2\Xi_{ij}\,\delta^{(d)}(\vec{x}_1-\vec{x}_2)\delta(t_1-t_2) \ ,
\end{equation}
with
\begin{equation}
    \mathrm{\Xi}=\begin{pmatrix}
    \mu b_1-\kappa b_1^2  &  -\frac{\sigma}{2}b_1b_2 &  -\frac{\sigma}{2}b_1b_3 \\
    -\frac{\sigma}{2}b_1b_2  &  \mu b_2-\lambda b_2^2  &  -\frac{\sigma}{2}b_2b_3 \\
    -\frac{\sigma}{2}b_1b_3  &  -\frac{\sigma}{2}b_2b_3  &  \mu b_3-\lambda b_3^2
    \end{pmatrix} .
\end{equation}

\subsection{Diagonalization of the harmonic action}

Before diagonalizing the quadratic action, we first shift to fluctuating fields, $\tilde{d}_i(\vec{x},t) = \tilde{b}_i(\vec{x},t)$
and $d_i(\vec{x},t) = b_i(\vec{x},t)-\frac{\mu}{\sigma+\kappa}-C$. Here $C$ is a counterterm which encodes the fluctuation 
corrections to the average concentrations. Owing to the cyclic symmetry among the three different species, we only need to 
introduce a single counterterm. The harmonic part of the action in terms of the new fields $\tilde{d}_i$ and $d_i$ reads
\begin{equation}
\begin{aligned}
    \mathcal{A}^{\mathrm{ML}}_0=\sum_i\int\!\mathrm{d}t\,\mathrm{d}^dx \bigg[ \tilde{d}_i 
    \left( \partial_t-D\nabla^2+\frac{\kappa\mu}{\kappa+\sigma}+(2\kappa+\sigma)C \right)\!d_i
    +\left( \frac{\sigma\mu}{\kappa+\sigma}+\sigma C \right)\!\tilde{d}_id_{i+1} \bigg] .
\end{aligned}
\end{equation}
Since the counterterm $C$ is of first order in the perturbative expansion parameters, up to zeroth order the harmonic part of 
the action can be diagonalized by the following linear transformation
\begin{equation}
\begin{pmatrix}
\tilde{d}_1 \\
\tilde{d}_2 \\
\tilde{d}_3 \\
\end{pmatrix}=\frac{1}{\sqrt{3}}
\begin{pmatrix}
1 & -\frac{1-i\sqrt{3}}{2} & -\frac{1+i\sqrt{3}}{2} \\
1 & -\frac{1+i\sqrt{3}}{2} & -\frac{1-i\sqrt{3}}{2} \\
1 & 1 & 1
\end{pmatrix}
\begin{pmatrix}
\tilde{\psi}_o\\
\tilde{\psi}_+\\
\tilde{\psi}_-
\end{pmatrix} ,
\label{ML transfromation}
\end{equation}
and 
\begin{equation}
\begin{pmatrix}
d_1 \\
d_2 \\
d_3 \\
\end{pmatrix}=\frac{1}{\sqrt{3}}
\begin{pmatrix}
1 & -\frac{1+i\sqrt{3}}{2} & -\frac{1-i\sqrt{3}}{2} \\
1 & -\frac{1-i\sqrt{3}}{2} & -\frac{1+i\sqrt{3}}{2} \\
1 & 1 & 1
\end{pmatrix} 
\begin{pmatrix}
\psi_o\\
\psi_+\\
\psi_-
\end{pmatrix} .
\label{ML transformation bar}
\end{equation}
After applying this linear transformation, the action can be expressed as $\mathcal{A}^{\mathrm{ML}} =
\mathcal{A}^{\mathrm{ML}}_0+\mathcal{A}^{\mathrm{ML}}_s+\mathcal{A}^{\mathrm{ML}}_{\mathrm{int}}$, representing, respectively, the harmonic, source, and non-linear interaction terms. Again, we omit the four-point vertices,
since they will not contribute to the dispersion relation renormalizations at one-loop order. The other terms are
\begin{equation}
\begin{aligned}
    \mathcal{A}^{\mathrm{ML}}_0=\int\!\mathrm{d}t\,\mathrm{d}^dx  \bigg[ &\tilde{\psi}_o \left( 
    \partial_t-D\nabla^2+\mu+2(\sigma+\kappa)C \right)\!\psi_o +\tilde{\psi}_+
    \left(\partial_t-D\nabla^2+\gamma_0+i\nu_0\right)\!\psi_+ \\
    &+\tilde{\psi}_-\left(\partial_t-D\nabla^2+\gamma_0-i\nu_0\right)\psi_- \\
    &+\frac12 \left( \sigma+4\kappa+i\sqrt{3} \sigma \right)\!C \tilde{\psi}_+\psi_+ 
    +\frac12 \left( \sigma+4\kappa-i\sqrt{3}\sigma \right)\!C \tilde{\psi}_-\psi_- \bigg] ,
\end{aligned}
\label{ml harmonic}
\end{equation}
\begin{equation}
\begin{aligned}
\mathcal{A}^{\mathrm{ML}}_s=\int\!\mathrm{d}t\,\mathrm{d}^dx \bigg[ &\left( \mu C+(\kappa+\sigma)C^2 \right)
\left( \sqrt{3}\tilde{\psi}_o+\tilde{\psi}_o^2 \right) \\
&+ \left( -\frac{3\mu^2\sigma}{(\kappa+\sigma)^2} + 2\mu\frac{\kappa-2\sigma}{\kappa+\sigma} C
+(2\kappa-\sigma)C^2 \right)\! \tilde{\psi}_+\tilde{\psi}_-\bigg] ,
\end{aligned}
\end{equation}
\begin{equation}
\begin{aligned}
    &\mathcal{A}^{\mathrm{ML}}_\mathrm{int} = \int\!\mathrm{d}t\,\mathrm{d}^dx \bigg[ \left( 
    \frac{\mu(\kappa-2\sigma)}{\sqrt{3}(\kappa+\sigma)}+\frac{2\kappa-\sigma}{\sqrt{3}}\,C \right)
    \left( \tilde{\psi}_+^2\psi_-+\tilde{\psi}_-^2\psi_+ \right) + \frac{\kappa+\sigma}{\sqrt{3}}\tilde{\psi}_o\psi_o^2 \\
    &\qquad\ +\left( \frac{\mu}{\sqrt{3}}+\frac{2(\kappa+\sigma)}{\sqrt{3}}\,C \right)\!\tilde{\psi}_o^2\psi_o 
   +\left( \frac{\mu(2\kappa-\sigma)}{\sqrt{3}(\kappa+\sigma)}+\frac{4\kappa+\sigma}{\sqrt{3}}\,C \right)\!
    \left( \tilde{\psi}_o\tilde{\psi}_+\psi_+ + \tilde{\psi}_o\tilde{\psi}_-\psi_- \right) \\
    &\qquad\quad +\frac{\sqrt{3}}{6} \left( 2\kappa-\sigma+\sqrt{3}i\sigma \right)\!\tilde{\psi}_-\psi_+^2 
    + \frac{\sqrt{3}}{6} \left( 2\kappa-\sigma-\sqrt{3}i\sigma \right)\!\tilde{\psi}_+\psi_-^2 
    +\frac{2\kappa-\sigma}{\sqrt{3}} \,\tilde{\psi}_o\psi_+\psi_- \\
    &\!\!+ \frac{\sqrt{3}}{6}\!\left( 4\kappa+\sigma-\sqrt{3}i\sigma \right)\!\tilde{\psi}_-\psi_o\psi_- 
    + \frac{\sqrt{3}}{6}\!\left( 4\kappa+\sigma+\sqrt{3}i\sigma \right)\!\tilde{\psi}_+\psi_o\psi_+ 
    + \text{four-point vertices} \bigg] ,
\end{aligned}
\end{equation}
where $\gamma_0 = \mu(2\kappa-\sigma)/2(\sigma+\kappa)$ and $\nu_0=\sqrt{3}\mu\sigma/2(\sigma+\kappa)$. The 
$\psi_o$ mode corresponds to the fluctuation of the total particle density and decays exponentially at tree level. In contrast to 
the RPS model, the ML $\psi_\pm$ modes display non-vanishing dissipation $\gamma_0$ already on the mean-field level. As
mentioned above, a Hopf bifurcation occurs at $2\kappa=\sigma$; when $2\kappa < \sigma$, the system is unstable and 
spiral structures are spontaneously generated. In the perturbative regime, the assumed tiny fluctuation corrections should not
change the overall stability features, but will only shift the Hopf bifurcation point by a small amount.

\subsection{One-loop fluctuation corrections}

\begin{figure}[t]
    \centering
    \includegraphics[scale=1.2]{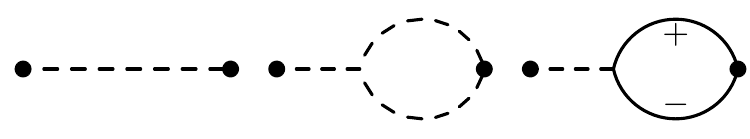}
    \caption{Contributions to $\langle\psi_o\rangle$ in the ML model up to one-loop order. The solid lines represent the
	 $\psi_\pm$ modes as indicated, while the dashed lines denotes the $\psi_o$ mode.}
    \label{ml counter}
\end{figure}
Prior to calculating the renormalized quantities, we need to determine the counterterm $C$ by requiring the average fluctuation 
of the total density to be zero, $\langle\psi_o\rangle=0$. Up to one-loop order, the contributions to $\langle\psi_o\rangle$ are
shown in Fig.~\ref{ml counter}. We note that the second diagram in Fig.~\ref{ml counter} is of second order and thus can be 
dropped. The corresponding analytic expression results in
\begin{equation}
    C=-\frac{\mu\sigma(2\kappa-\sigma)}{2D(\kappa+\sigma)^2}\int_k\frac{1}{k^2+\frac{\gamma_0}{D}} \ .
\end{equation}

\begin{figure}[t]
    \centering
    \includegraphics[scale=1.2]{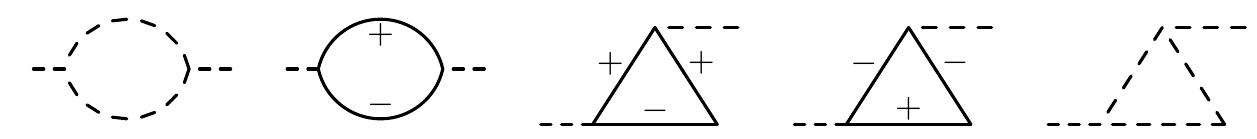}
    \caption{One-loop fluctuation contributions to the two-point vertex function $\Gamma^{(1,1)}_{\tilde{\psi}_o\psi_o}$ in 		
	the ML model.}
    \label{ml0 propagator}
\end{figure}

We may now proceed to the fluctuation renormalization of the two-point vertex functions 
$\Gamma^{(1,1)}_{\tilde{\psi}_o\psi_o}$ and $\Gamma^{(1,1)}_{\tilde{\psi}_{\pm}\psi_{\pm}}$, which encode the
self-energies entering the dispersion relation of the $\psi_o$ and $\psi_\pm$ modes. As the total particle number is not 
conserved in the ML model, the dispersion relation acquires non-trivial corrections from the perturbation expansion. The 
one-loop diagrams that contribute to the vertex function $\Gamma^{(1,1)}_{\tilde{\psi}_o\psi_o}$ are pictured in 
Fig.~\ref{ml0 propagator}. The last diagram is of higher order and thus can be omitted; this results in
\begin{equation}
\begin{aligned}
    &\Gamma^{(1,1)}_{\tilde{\psi}_o\psi_o}(p,\omega) = i\omega+Dp^2+\mu+2(\kappa+\sigma)C
    -\frac{\mu(\kappa+\sigma)}{3D}\int_k I\!\left(\frac{\mu}{D}\right) \\
    &\, -\frac{2(\sigma+\kappa)\gamma_0}{3D\mu} \left(\gamma_0-\sqrt{3}\nu_0\right) \int_k I\!\left(\frac{\gamma_0}{D}
    \right) - \frac{2\sqrt{3}(\sigma+\kappa)\gamma_0\nu_0}{3D^2\mu}\,(\gamma_0+\mu)
    \int_k\frac{1}{k^2+\frac{\gamma_0}{D}} \, I\!\left(\frac{\gamma_0}{D}\right) .
\end{aligned}
\end{equation}
Provided $\gamma_0 > 0$, all corrections at one-loop level are real, and no imaginary part appears in the mass term of the 
$\psi_o$ mode, hence there are no total particle number oscillations. However, for $\gamma_0 < 0$ the system exhibits emergent oscillations of the total particle number, indicating the spontaneous formation of spatio-temporal structures. As the 
renormalized two-point vertex function can also be written as
\begin{equation}
    \Gamma^{(1,1)}_{\tilde{\psi}_o\psi_o}(p,\omega)=Z_{\phi_o} \left( i\omega + D_{r}^op^2 +\mu_r \right) ,
\end{equation}
the renormalized diffusivity $D_{r}^o$ and mass parameter $\mu_r$ can be calculated accordingly. Here, $Z_{\phi_o}$ 
absorbs all wave function renormalizations. Since the rotating wave modes $\psi_\pm$ acquires a different diffusivity renormalization from the $o$ mode, we carefully distinguish the renormalized quantities $D_r^{\pm}$ and $D_r^{o}$. 

The explicit expressions for $D_{r}^o$ and $\mu_r$ read
\begin{equation}
\begin{aligned}
&\mu_r = \mu \bigg[ 1-\frac{2(\sigma+\kappa)}{3D}\frac{\gamma_0}{\mu} \int_k 
\frac{1}{k^2+\frac{\gamma_0}{D}}-\frac{\sigma+\kappa}{3D} \int_k \frac{1}{k^2+\frac{\mu}{D}}
- \frac{\mu(\sigma+\kappa)}{6D^2} \int_k \frac{1}{(k^2+\frac{\mu}{D})^2} \\
&\ -\frac{(\sigma+\kappa)\gamma_0}{3D^2} \left( 1+\frac{2\sqrt{3}\nu_0\gamma_0}{\mu^2} \right)
\int_k\frac{1}{(k^2+\frac{\gamma_0}{D})^2} - \frac{\sqrt{3}(\sigma+\kappa)}{3D^3}
\frac{\nu_0\gamma_0(\gamma_0+\mu)}{\mu}\int_k \frac{1}{(k^2+\frac{\gamma_0}{D})^3} \bigg] , \\
&D_{r}^o = D - \frac{\mu(\sigma+\kappa)}{3dD} \int_k \frac{k^2}{(k^2+\frac{\mu}{D})^3}
-\frac{2(\sigma+\kappa)}{3dD}\frac{\gamma_0}{\mu}\left(\gamma_0-\sqrt{3}\nu_0\right) 
\int_k \frac{k^2}{(k^2+\frac{\gamma_0}{D})^3} \\
&\qquad - \frac{2\sqrt{3}(\sigma+\kappa)}{3dD^2}\frac{\nu_0\gamma_0}{\mu} \left( \gamma_0+\mu \right)
\int_k \frac{k^2}{(k^2+\frac{\gamma_0}{D})^4} \ .
\end{aligned}
\end{equation}
After evaluating the integrals one arrives at
\begin{equation}
\begin{aligned}
&\mu_r = 1 - \frac{\Gamma(1-d/2)}{2^d\pi^{d/2}} \left( \frac{\sigma+\kappa}{3D}\Big(\frac{\mu}{D}\Big)^{d/2-1}
+\frac{2(\sigma+\kappa)\gamma_0}{3D\mu} \Big(\frac{\gamma_0}{D}\Big)^{d/2-1} \right) \\
&\qquad - \frac{\Gamma(2-d/2)}{2^d\pi^{d/2}} \biggl[\frac{\mu(\sigma+\kappa)}{6D^2}\Big(\frac{\mu}{D}\Big)^{d/2-2}
+\frac{(\sigma+\kappa)\gamma_0}{3D^2} \biggl( 1+\frac{2\sqrt{3}\nu_0\gamma_0}{\mu^2} \biggr) 
\Big(\frac{\gamma_0}{D}\Big)^{d/2-2}\biggr] \\
&\qquad - \frac{\Gamma(3-d/2)}{2^d\pi^{d/2}}\frac{\sqrt{3}(\sigma+\kappa)\nu_0\gamma_0}{3D^3\mu}\left(\gamma_0+\mu\right)\Big(\frac{\gamma_0}{D}\Big)^{d/2-3} \, , \\
&D_r^o = D - \frac{\Gamma(2-d/2)}{2^{d+2}\pi^{d/2}} \left[ \frac{\mu(\sigma+\kappa)}{3D} 
\Big(\frac{\mu}{D}\Big)^{d/2-2} + \frac{2(\sigma+\kappa)\gamma_0}{3D\mu}\left(\gamma_0-\sqrt{3}\nu_0\right)
\Big(\frac{\gamma_0}{D}\Big)^{d/2-2} \right] \\
&\qquad - \frac{\Gamma(3-d/2)}{3 \cdot 2^{d+2}\pi^{d/2}}\frac{2\sqrt{3}(\sigma+\kappa)}{3D^2\mu}\nu_0\gamma_0 (\gamma_0+\mu) \Big(\frac{\gamma_0}{D}\Big)^{d/2-3} \, .
\end{aligned}
\label{ml renormalized}
\end{equation}
For $d\geq 2$, UV divergences in $\mu_r$ are manifest; but since the lattice constant $c$ serves as a natural UV cutoff in 
lattice models, we will not discuss these UV divergences further. 

\begin{figure}[t]
    \centering
    \includegraphics[scale=1.2]{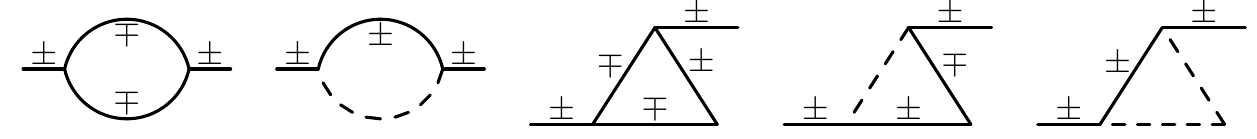}
    \caption{One-loop fluctuation contributions to the two-point vertex functions 
    	$\Gamma^{(1,1)}_{\tilde{\psi}_{\pm}\psi_{\pm}}$ in the ML model.}
    \label{ml propagator}
\end{figure}
The renormalized vertex function $\Gamma^{(1,1)}_{\tilde{\psi}_{\pm}\psi_{\pm}}$ can also be calculated according to the one-loop diagrams in Fig.~\ref{ml propagator}, resulting in 
\begin{equation}
\begin{aligned}
    \Gamma^{(1,1)}_{\tilde{\psi}_{\pm}\psi_{\pm}}(p,\omega) &= i\omega+Dp^2+\gamma_0 \pm i \nu_0
    +\frac{\sigma+\kappa}{\mu} \left(\gamma_0+\mu\pm i\nu_0\right) C \\
    &\quad - \frac{(\kappa+\sigma)(\gamma_0-\sqrt{3}\nu_0)}{3D\mu}\,(\gamma_0\mp i\nu_0) 
    \int_k I\!\left(\frac{\gamma_0\mp i\nu_0}{D}\right) \\
    &\quad -\frac{(\sigma+\kappa)\gamma_0}{3D\mu}\,(\gamma_0+\mu\pm i\nu_0)
    \int_k I\!\left(\frac{\gamma_0+\mu\pm i\nu_0}{2D}\right) \\
    &\quad -\frac{2\sqrt{3}(\sigma+\kappa)\nu_0}{3D^2\mu}\,(\gamma_0^2+\nu_0^2)
    \int_k\frac{1}{k^2+\frac{\gamma_0}{D}}\, I\!\left(\frac{\gamma_0\mp i\nu_0}{D}\right) \\
    &\quad -\frac{\sqrt{3}(\sigma+\kappa)\gamma_0\nu_0}{3D^2\mu}\,(\gamma_0+\mu\pm i\nu_0)
    \int_k\frac{1}{k^2+\frac{\gamma_0}{D}}\,I\!\left(\frac{\gamma_0+\mu\pm i\nu_0}{2D}\right) .
\end{aligned}
\end{equation}
Upon invoking the relation with renormalized quantities 
\begin{equation}
    \Gamma^{(1,1)}_{\tilde{\psi}_{\pm}\psi_{\pm}}(p,\omega) = Z_{\phi_\pm}
    \left( i\omega + D_{r}^\pm p^2 +\gamma_r \pm i\nu_r \right) ,
\end{equation}
the expressions for the renormalized parameters $\gamma_r$ and $\nu_r$ are readily inferred,
\begin{equation}
\begin{aligned}
\gamma_r \pm i\nu_r = &\, \gamma_0 \pm i\nu_0 +(\sigma+\kappa) \bigg[ 
M_1^{(\pm)}\frac{\Gamma(1-d/2)}{2^d\pi^{d/2}}\left(\frac{\gamma_0}{D}\right)^{d/2-1}\\
&+M_2^{(\pm)}\frac{\Gamma(1-d/2)}{2^d\pi^{d/2}}\left(\frac{\gamma_0^2+\nu_0^2}{D^2}\right)^{\!(d-2)/4}
\exp\!\left(\mp i \frac{d-2}{2} \theta\right) \\
&+M_3^{(\pm)}\frac{\Gamma(1-d/2)}{2^d\pi^{d/2}}\left(\frac{(\mu+\gamma_0)^2+\nu_0^2}{4D^2}\right)^{\!(d-2)/4}
\exp\!\left(\pm i \frac{d-2}{2} \eta\right) \\
&+M_4^{(\pm)}\frac{\Gamma(2-d/2)}{2^d\pi^{d/2}}\left(\frac{\gamma_0^2+\nu_0^2}{D^2}\right)^{\!(d-4)/4}
\exp\!\left(\mp i \frac{d-4}{2} \theta\right) \\
&+M_5^{(\pm)}\frac{\Gamma(2-d/2)}{2^d\pi^{d/2}}\left(\frac{(\mu+\gamma_0)^2+\nu_0^2}{4D^2}\right)^{\!(d-4)/4}
\exp\!\left(\pm i \frac{d-4}{2} \eta\right) \bigg] ,
\end{aligned}
\label{ml_mass_r}
\end{equation}
where the coefficients $M_i^{(\pm)}$ are defined in Eq.~(\ref{coefficients_ML}) in the appendix, and the angles $\theta$ and $\eta$ are given by $\tan \theta = \nu_0 / \gamma_0$ and $\tan \eta = \nu_0 / (\gamma_0+\mu)$. We note that at odd 
dimensions $d$, the first term in the bracket in Eq.~(\ref{ml_mass_r}) switches from real to imaginary as $\gamma_0$ 
changes its sign; however, at even dimensions, this term is always real. 
Finally, the renormalized diffusivity reads
\begin{equation}
\begin{aligned}
D_r^{\pm} = &\, D -\frac{\kappa+\sigma}{d\,2^d\pi^{d/2}} \bigg[ \Gamma(1-d/2)P_1^{(\pm)}
\left(\frac{\gamma_0^2+\nu_0^2}{D^2}\right)^{\!(d-2)/4} \exp\!\left(\mp i\frac{d-2}{2}\theta\right) \\
&\quad +\Gamma(2-d/2)P_2^{(\pm)}\left(\frac{\gamma_0^2+\nu_0^2}{D^2}\right)^{(d-4)/4}
\exp\!\left(\mp i\frac{d-4}{2}\theta\right) \\
&\quad +\frac{\Gamma(3-d/2)}{2}P_3^{(\pm)}\left(\frac{\gamma_0^2+\nu_0^2}{D^2}\right)^{\!(d-6)/4}
\exp\!\left(\mp i\frac{d-6}{2}\theta\right) \\
&\quad + \Gamma(1-d/2)Q_1^{(\pm)}\left(\frac{(\mu+\gamma_0)^2+\nu_0^2}{4D^2}\right)^{\!(d-2)/4}
\exp\!\left(\pm i\frac{d-2}{2}\eta\right) \\
&\quad + \Gamma(2-d/2)Q_2^{(\pm)}\left(\frac{(\mu+\gamma_0)^2+\nu_0^2}{4D^2}\right)^{\!(d-4)/4}
\exp\!\left(\pm i\frac{d-4}{2}\eta\right) \\
&\quad + \frac{\Gamma(3-d/2)}{2}Q_3^{(\pm)}\left(\frac{(\mu+\gamma_0)^2+\nu_0^2}{4D^2}\right)^{\!(d-6)/4}
\exp\!\left(\pm i\frac{d-6}{2}\eta\right)\\
&\quad -\Gamma(1-d/2)\left(P_1^{(\pm)}+Q_1^{(\pm)}\right)\left(\frac{\gamma_0}{D}\right)^{\!(d-2)/2} \bigg] .
\end{aligned}
\end{equation}
In the appendices, we provide additional details and the definitions of the various coefficients, as well as explicit evaluations for $d=1,2,3$. Here, we focus on the behavior of the damping parameter $\mu_r$ across different dimensions. It is important to 
note that $\mu_r$ is generally a complex number: Its imaginary part conveys information about spatial oscillations, while its 
real part represents either exponential decay or growth of the average particle density.

\subsubsection{$d=1$:} 
\begin{figure}[t]
\subfloat[]{\includegraphics[width=0.45\linewidth]{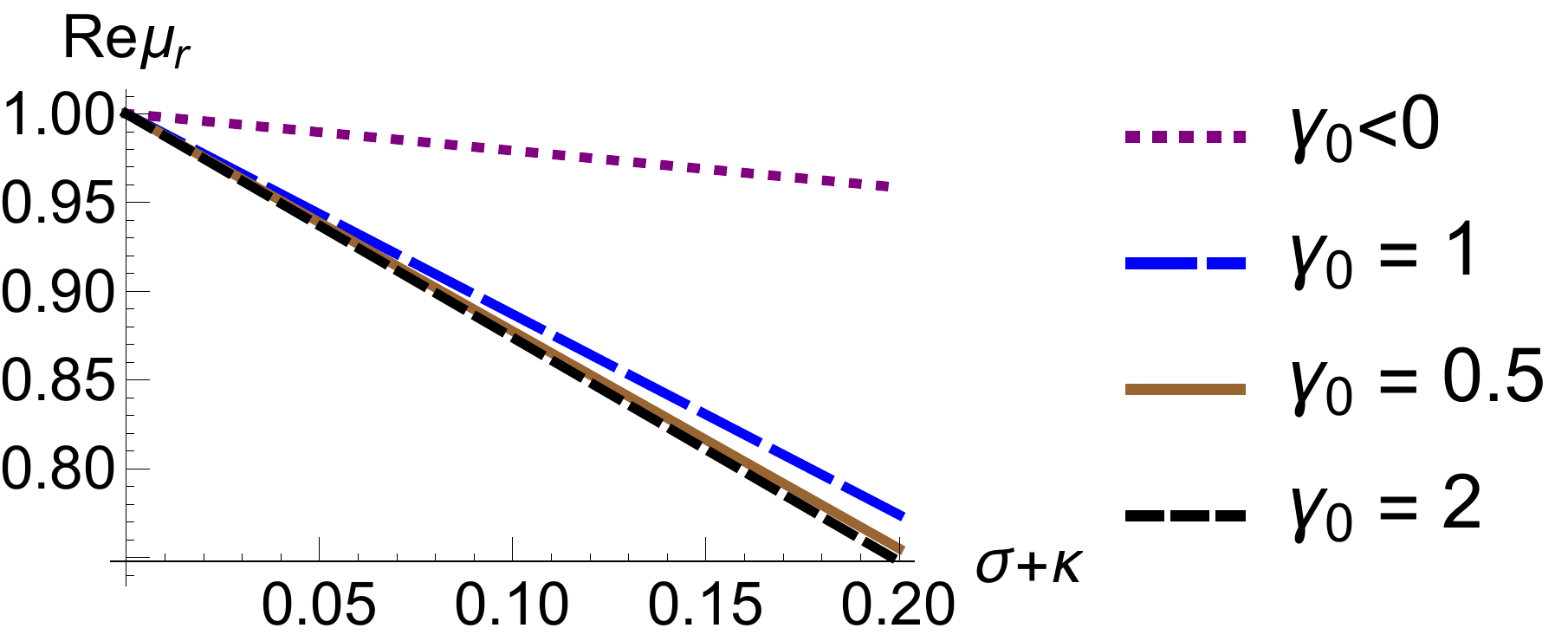}}\qquad
\subfloat[]{\includegraphics[width=0.45\linewidth]{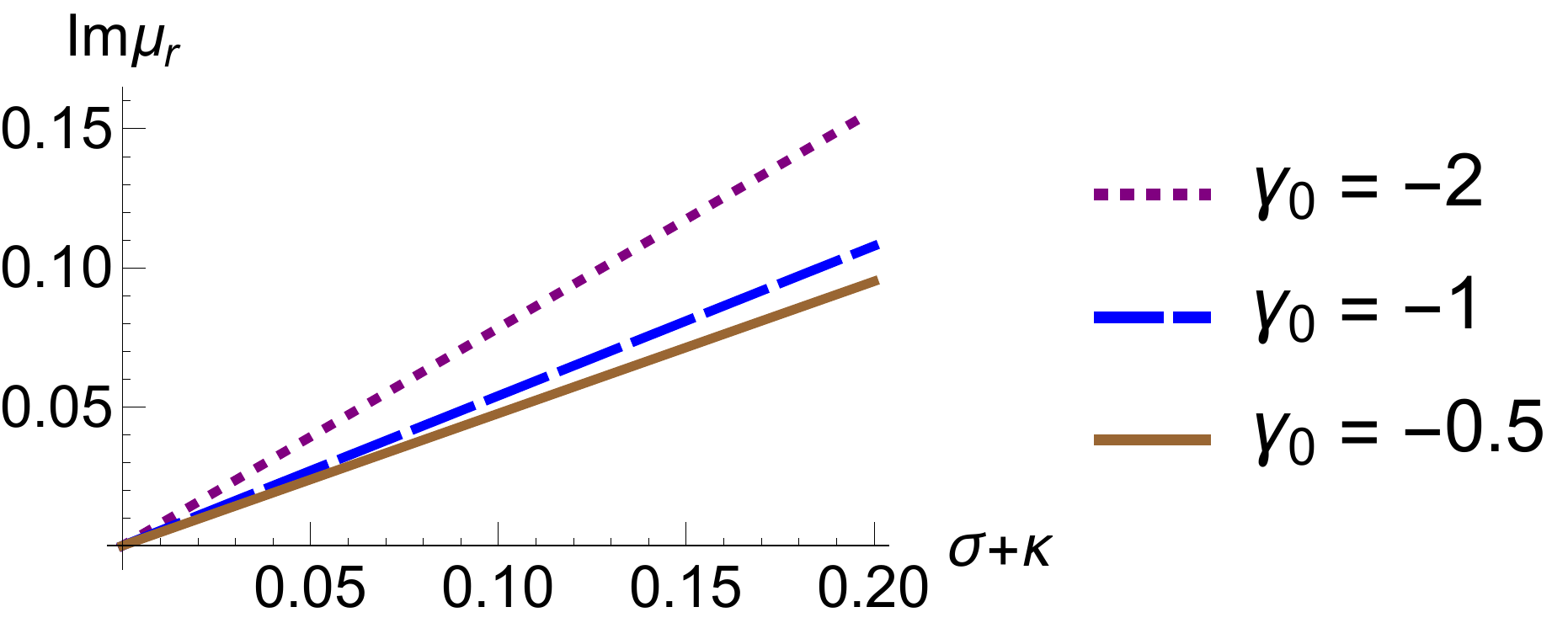}}
\caption{Renormalized $\mu_r$ in one dimension ($d=1$): (a) Real part of  $\mu_r$ as a function of $\sigma+\kappa$ with 
$D=1$, $\mu=1$, $\nu_0=1$, and different values of $\gamma_0$. (b) Imaginary part of $u_r$ as a function of $\lambda$ 
with $D=1$, $\mu=1$, $\nu_0=1$, and different $\gamma_0 <0$.}
\label{ML 1d Plot}
\end{figure}

In one dimension, the renormalized damping parameter $\mu_r$ reads
\begin{equation}
    \mu_r = \mu \left[ 1-\frac{5(\sigma+\kappa)}{24D}\sqrt{\frac{D}{\mu}}-\frac{\sigma+\kappa}{D}
    \sqrt{\frac{D}{\gamma_0}} \left( \frac{1}{12}+\frac{\gamma_0}{3\mu}+\frac{\sqrt{3}\nu_0}{16\mu}
    +\frac{\sqrt{3}\nu_0}{16\gamma_0}+\frac{\sqrt{3}\nu_0\gamma_0}{6\mu^2} \right) \right] .
\end{equation}
For $\gamma_0 < 0$, $\mu_r$ acquires an imaginary part, indicating oscillatory behavior. However, if $\gamma_0 > 0$, 
there is only damping. We display different scenarios in Fig.~\ref{ML 1d Plot}.

\subsubsection{$d=2$:} 
\begin{figure}[b]
\subfloat[]{\includegraphics[width=0.45\linewidth]{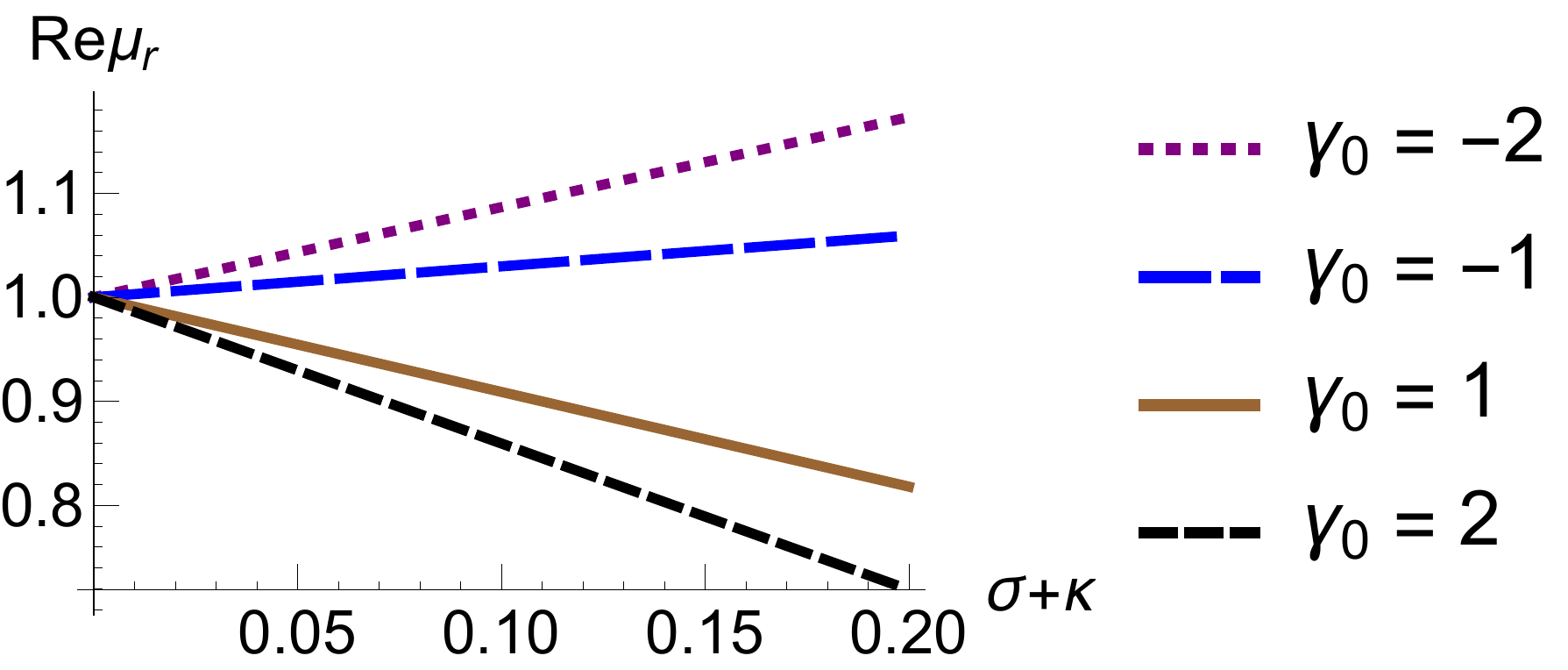}}\qquad
\subfloat[]{\includegraphics[width=0.45\linewidth]{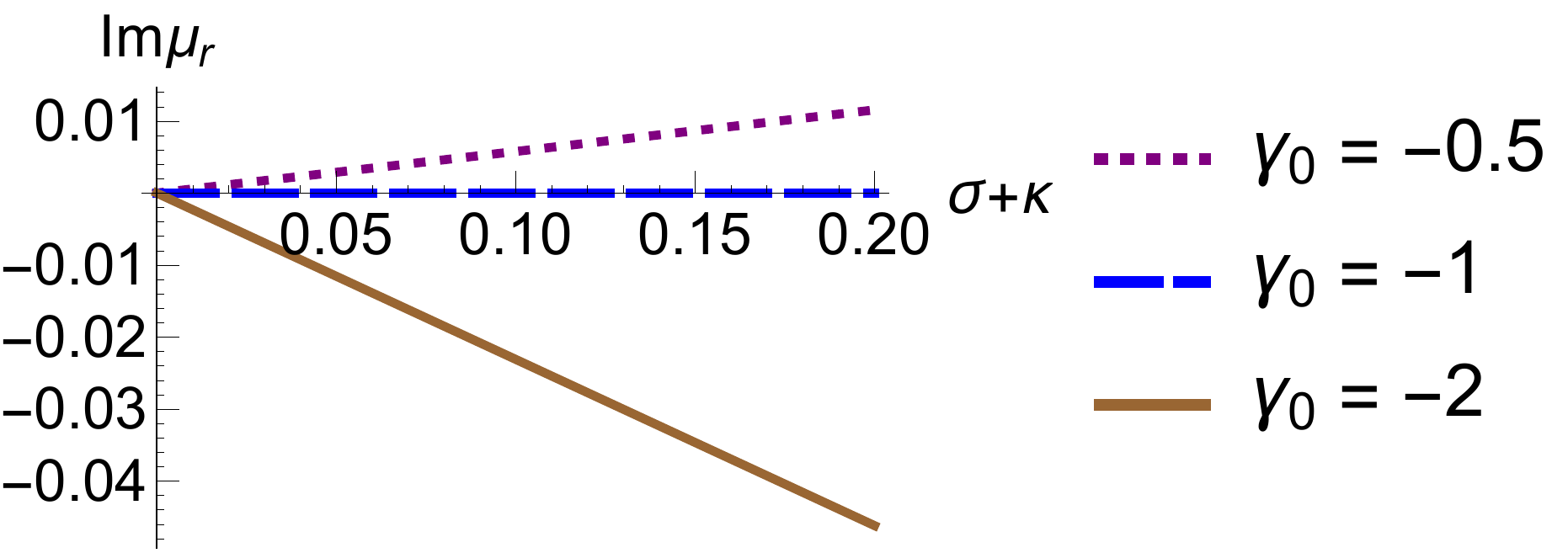}}
\caption{Renormalized $\mu_r$ at two dimensions ($d=2$) with cutoff $\Lambda^2 = 10000$: (a) Real part of $\mu_r$ as 
a function of $\sigma+\kappa$ with $D=1$, $\mu=1$, $\nu_0=1$, and different values of $\gamma_0$.  (b) Imaginary part 
of $u_r$ as a function of $\lambda$ with $D=1$, $\mu=1$, $\nu_0=1$, and different $\gamma_0 <0$.}
\label{ML 2d Plot}
\end{figure}

At two dimensions, we need to introduce the UV cutoff $\Lambda \sim \pi/c$; the damping parameter becomes
\begin{equation}
\begin{aligned}
\mu_r = \mu \Bigg[ 1-\frac{\sigma+\kappa}{6D\pi}\frac{\gamma_0}{\mu}\ln\frac{D\Lambda^2}{\gamma_0}
-\frac{\sigma+\kappa}{12D\pi}\ln\frac{D\Lambda^2}{\mu}-\frac{\sigma+\kappa}{24D\pi} \\
- \frac{\sigma+\kappa}{12D\pi} \left( 1+\frac{\sqrt{3}\nu_0}{2\mu}+\frac{\sqrt{3}\nu_0}{2\gamma_0}
+\frac{2\sqrt{3}\nu_0\gamma_0}{\mu^2} \right) \Bigg] .
\end{aligned}
\end{equation}
As in the one-dimensional case, when $\gamma_0 > 0$, we have pure damping, whereas the system displays population oscillations if $\gamma_0 <0$. The damping coefficients $\mu_r$ at $d=2$ for different bare parameters are plotted in 
Fig.~\ref{ML 2d Plot}.

\subsubsection{$d=3$:} 
\begin{figure}[t]
\subfloat[]{\includegraphics[width=0.45\linewidth]{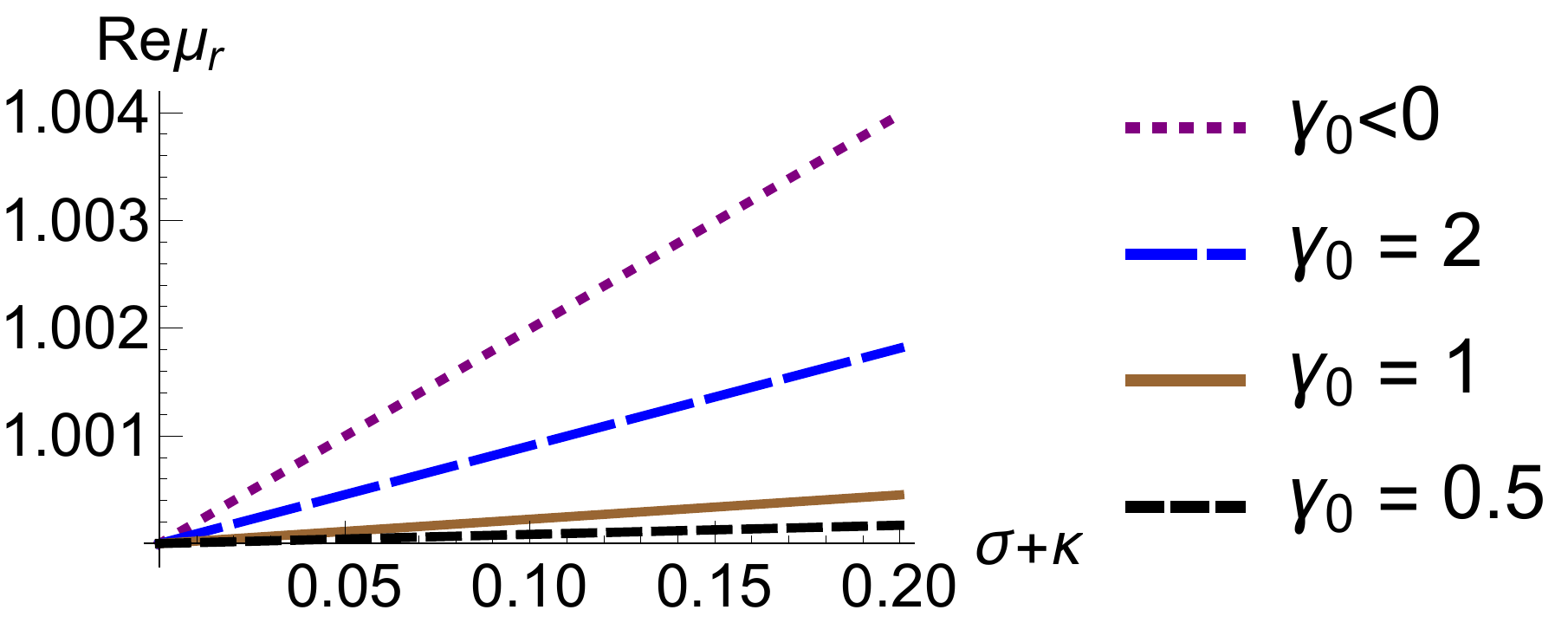}}\qquad
\subfloat[]{\includegraphics[width=0.45\linewidth]{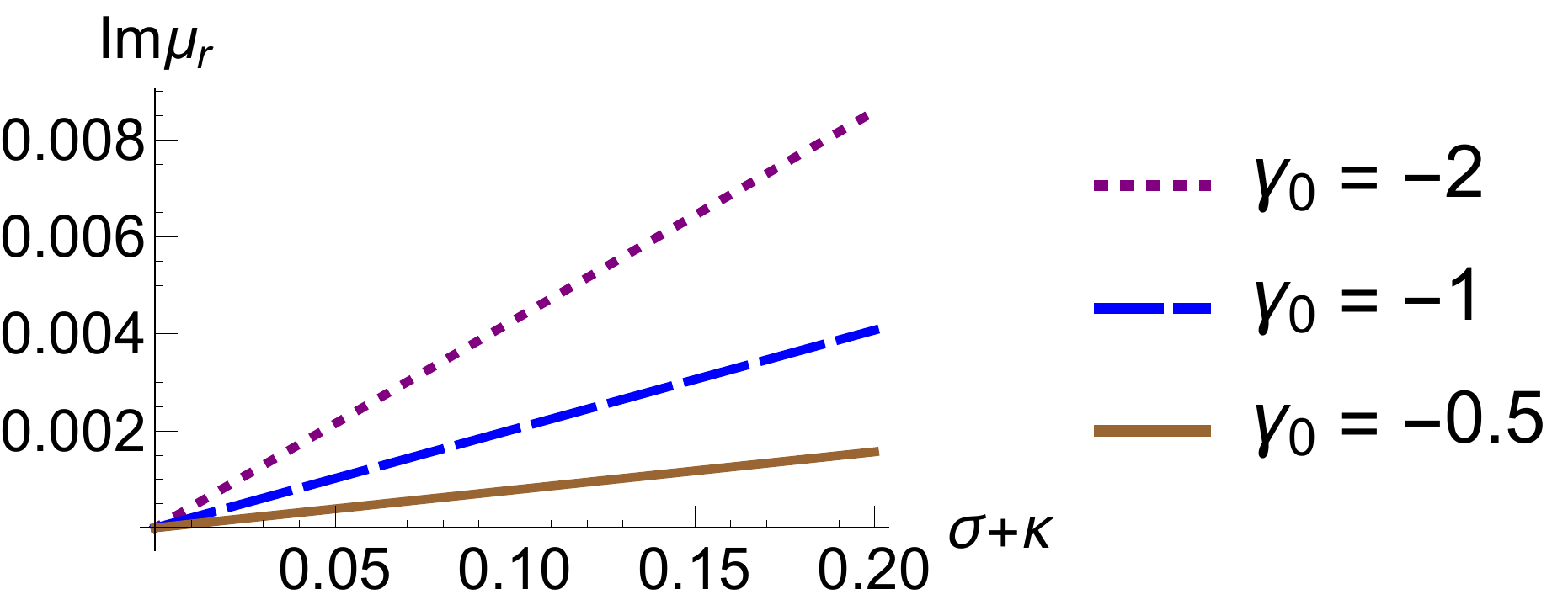}}
\caption{Renormalized $\mu_r$ at three dimensions ($d=3$): (a) Real part of $\mu_r$ as a function of $\sigma+\kappa$
with $D=1$, $\mu=1$, $\nu_0=1$, and different values of $\gamma_0$. (b) Imaginary part of $u_r$ as a function of 
$\lambda$ with $D=1$, $\mu=1$, $\nu_0=1$, and different $\gamma_0 <0$.}
\label{ML 3d Plot}
\end{figure}

In three dimensions, the damping parameter reads
\begin{equation}
    \mu_r = \mu \left[1 + \frac{\sigma+\kappa}{16D\pi}\sqrt{\frac{\mu}{D}}+\frac{\sigma+\kappa}{6D\pi}
    \sqrt{\frac{\gamma_0}{D}} \left( -\frac{1}{4}+\frac{\gamma_0}{\mu}-\frac{\sqrt{3}\nu_0}{16\mu}
    -\frac{\sqrt{3}\nu_0}{16\gamma_0}-\frac{\sqrt{3}\nu_0\gamma_0}{2\mu^2} \right) \right] .
\end{equation}
Again, oscillations emerge in the region where $\gamma_0 <0$. The real and imaginary parts of the damping parameter 
$\mu$ are depicted in Fig.~\ref{ML 3d Plot}.

For $\nu_0 = 1$, the damping $\mu_r$ decreases in one and two dimensions, but increases in three dimensions in the stable 
region where $\gamma_0 > 0$, as shown in Figs.~\ref{ML 1d Plot}, \ref{ML 2d Plot}, and \ref{ML 3d Plot}. This indicates 
that the reactions effectively slow down the relaxation processes in one and two dimensions, but speed them up in three 
dimensions. However, due to the highly nonlinear dependence on the parameters $\mu_0$, $\gamma_0$, and $\nu_0$, a 
more general conclusion cannot be made. It is worth noting that in the unstable region, an imaginary part of $\mu_r$ is 
generated, formally resulting from the subtraction of an inadequate homogeneous steady state. Yet these emergent oscillations 
manifestly indicate the instability with respect to spontaneous formation of spatio-temporal patterns in this regime. We
emphasize again that the one-loop fluctuation corrections should merely induce small quantitative corrections, and cannot
induce qualitative changes in the region where perturbation theory is applicable. The ML system thus maintains a bifurcation 
point, below which the homogeneous ground state is rendered unstable and spiral structures emerge.

\section{Comparison between the RPS, ML, and LV models}

In this section, we present a thorough comparison between the spatically extended stochastic LV, RPS, and ML models. We 
specifically discuss the spontaneous formation of spatio-temporal structures and the stability of the homogeneous state up to 
perturbative one-loop order in the fluctuation corrections. We also briefly address the influence of quenched spatial disorder in
the reaction rates in the perturbative regime.

\subsection{Spiral formation from a single lattice site point of view}

It is well-established that in sufficiently large spatial systems, the ML model displays spontaneously emerging dynamic spiral 
structures in individual simulation runs (these are of course averaged out in ensemble averages). Here, we propose that one crucial necessary condition for the formation of such persistent spatio-temporal patterns is the existence of a stable and
uniform oscillation frequency at the local lattice site level, which then allows spatially extended coherent oscillatory features.

Each site in a lattice subject to stochastic reactions and spreading processes can be regarded as a separate system that is 
coupled to a particle reservoir (in the thermodynamic limit). In the ML model, at the (linearized) mean-field level, the local
oscillation frequency is uniquely determined by the reaction rates, as is also apparent in the diagonalized ML Doi--Peliti action 
(\ref{ml harmonic}) at tree level. A similar definition of the oscillation frequency at linearized mean field level can also be 
obtained in the LV model \cite{tauber2012population}. However, in the RPS model at tree level, the oscillation frequency is
set by the global conserved particle number $\rho$. As the particle number at each site is changing all the time owing to 
its coupling to the environment that serves as a nonlocal reservoir, there does not exist a unique characteristic oscillation 
frequency for each site during any single run stochastic realization. 
Ultimately, these nonlocal effects originate from the long-range correlations introduced by the global particle number 
conservation law, whose relevance in the context of pattern formation was demonstrated in Ref.~\cite{tainaka1994vortices}.
The average of the oscillation frequency with a fixed total particle number is given by the expression in the diagonalized RPS
Doi--Peliti action (\ref{rps harmonic}).

These straightforward tree-level arguments are readily generalized to all (loop) orders in the perturbation expansion. In the ML 
model, no nonlinear terms that depend on the total particle density (or any other global quantity) are present in the action,  
and thus the renormalized frequency will also be independent of the particle number density. This is not the case for the RPS 
model, where $\rho$ manifestly enters the vertices. A perhaps more straightforward way to understand this distinction invokes
the fixed points (stationary species densities) of the systems. In the RPS system, there exists a fixed line that is parameterized 
by the conserved total particle number. For each lattice site in the RPS model, the population oscillations wander along this 
fixed line with a mean that corresponds to the averaged local total particle number at this specific site. In contrast, the ML and 
LV models have unique fixed points, independent of any global constraints that originate from conservation laws, and 
consequently all lattice sites are governed by identical oscillation frequencies determined by these fixed points.

\begin{figure}[t]
\subfloat[]{\includegraphics[width = 0.4\linewidth]{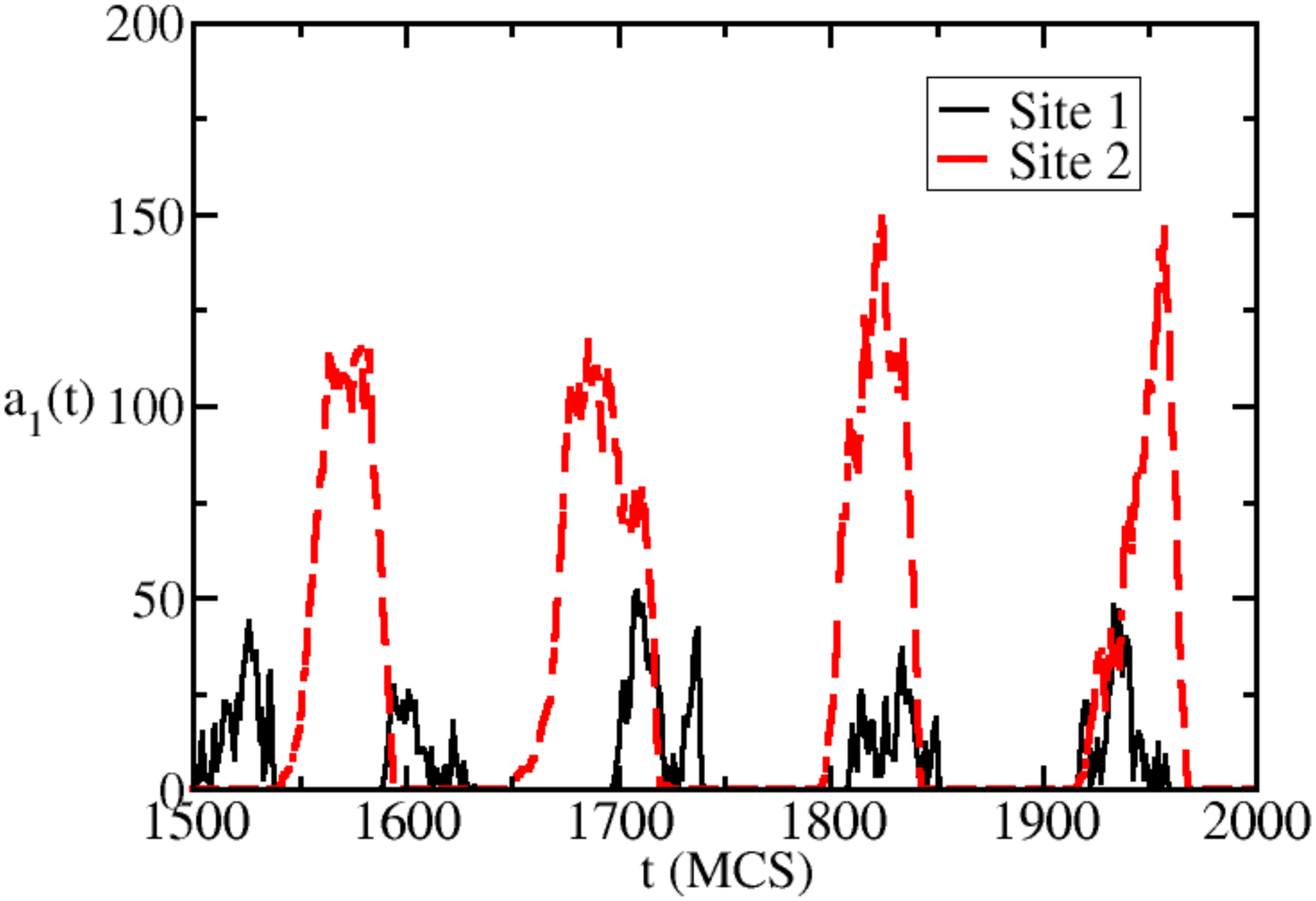}}\qquad\qquad
\subfloat[]{\includegraphics[width = 0.4\linewidth]{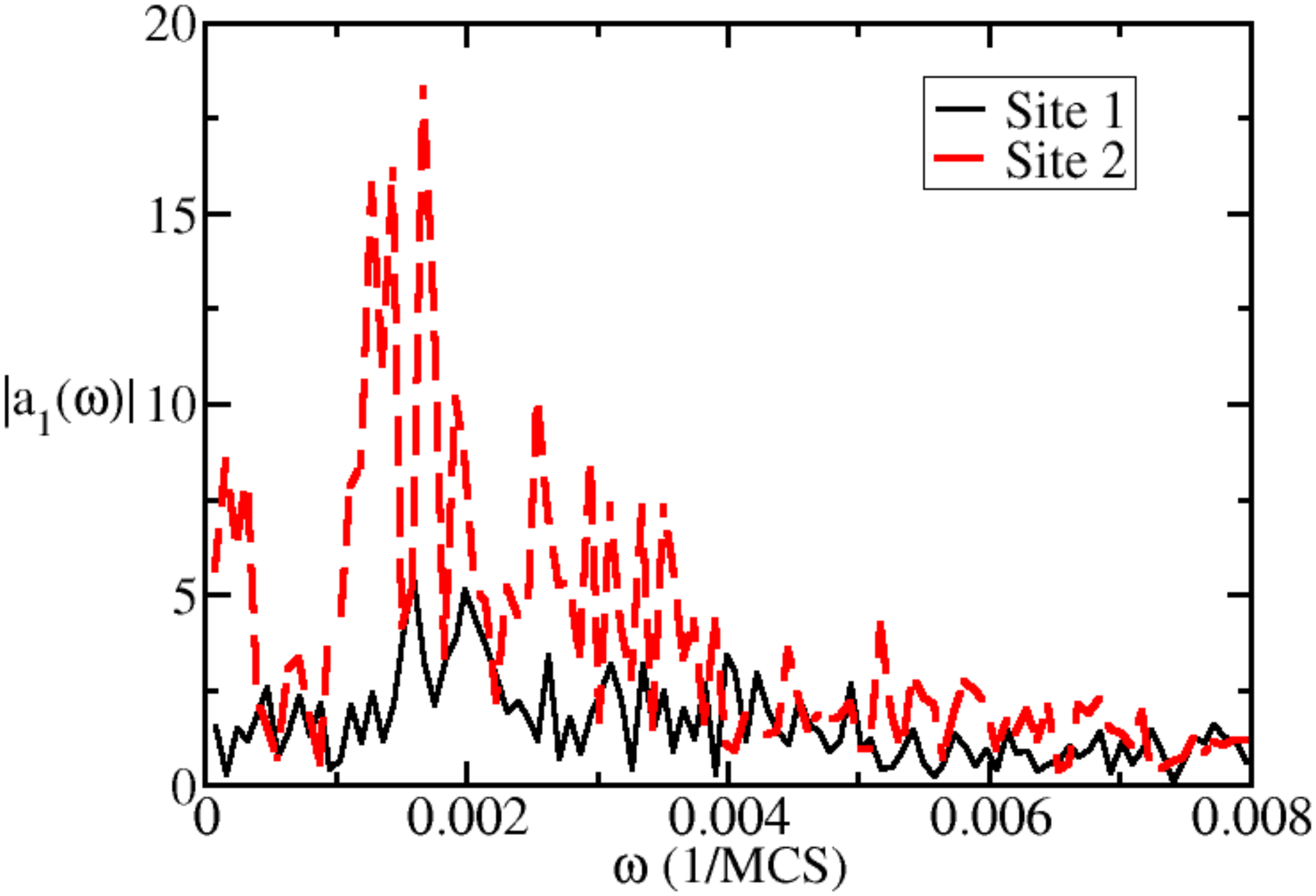}}
\caption{(a) Time evolution of the density of species $A_1$ in the RPS model on two randomly chosen sites; (b) frequency spectrum of the time series in (a) obtained through a Fourier transform. The simulation was run with a reaction rate $0.5$ and 
on a lattice with $100\times100$ sites. The reactions in the simulation take place off-site and no additional hopping processes 
are incorporated.}
\label{Single run results RPS}
\end{figure}
\begin{figure}[t]
\subfloat[]{\includegraphics[width = 0.4\linewidth]{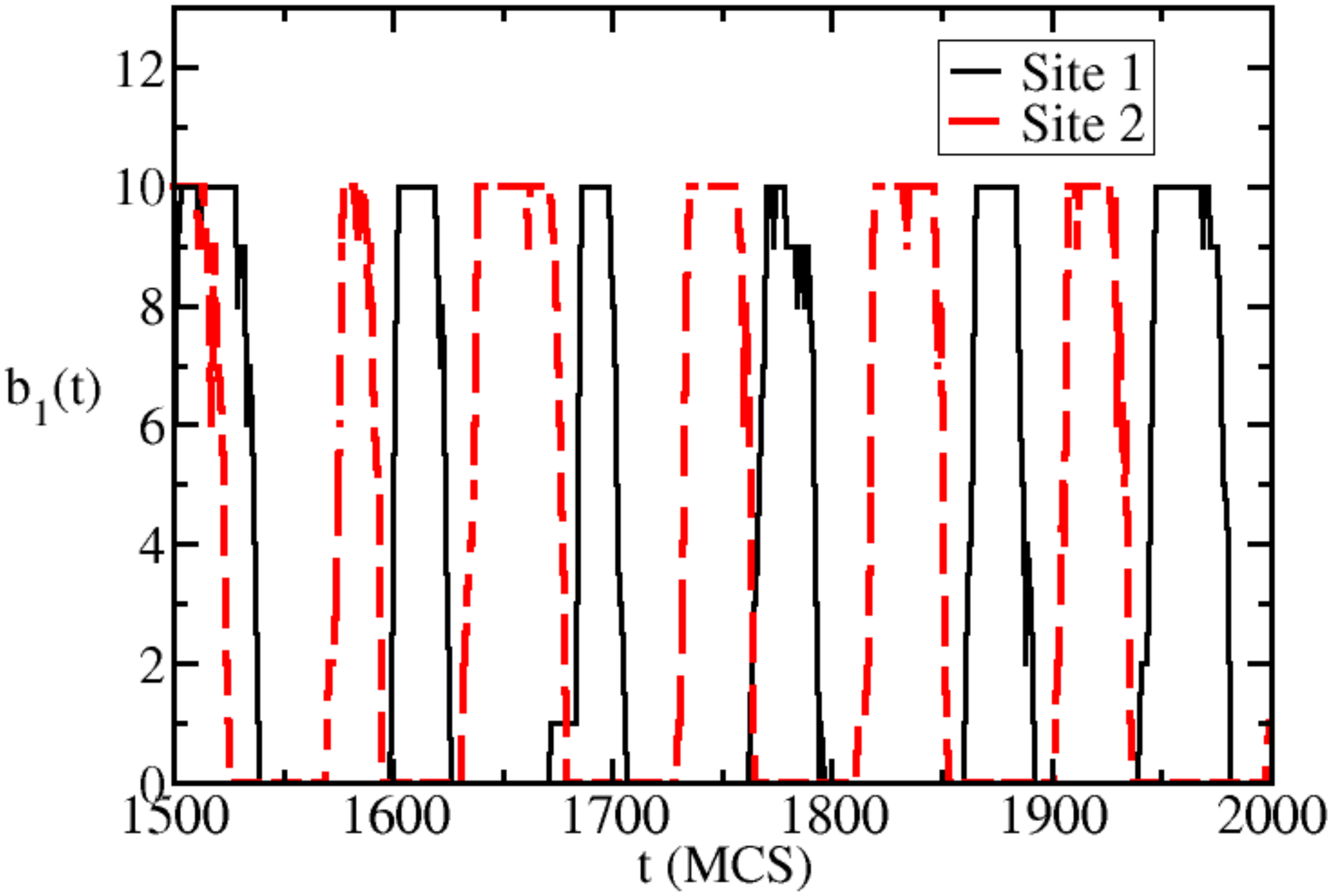}}\qquad\qquad
\subfloat[]{\includegraphics[width = 0.4\linewidth]{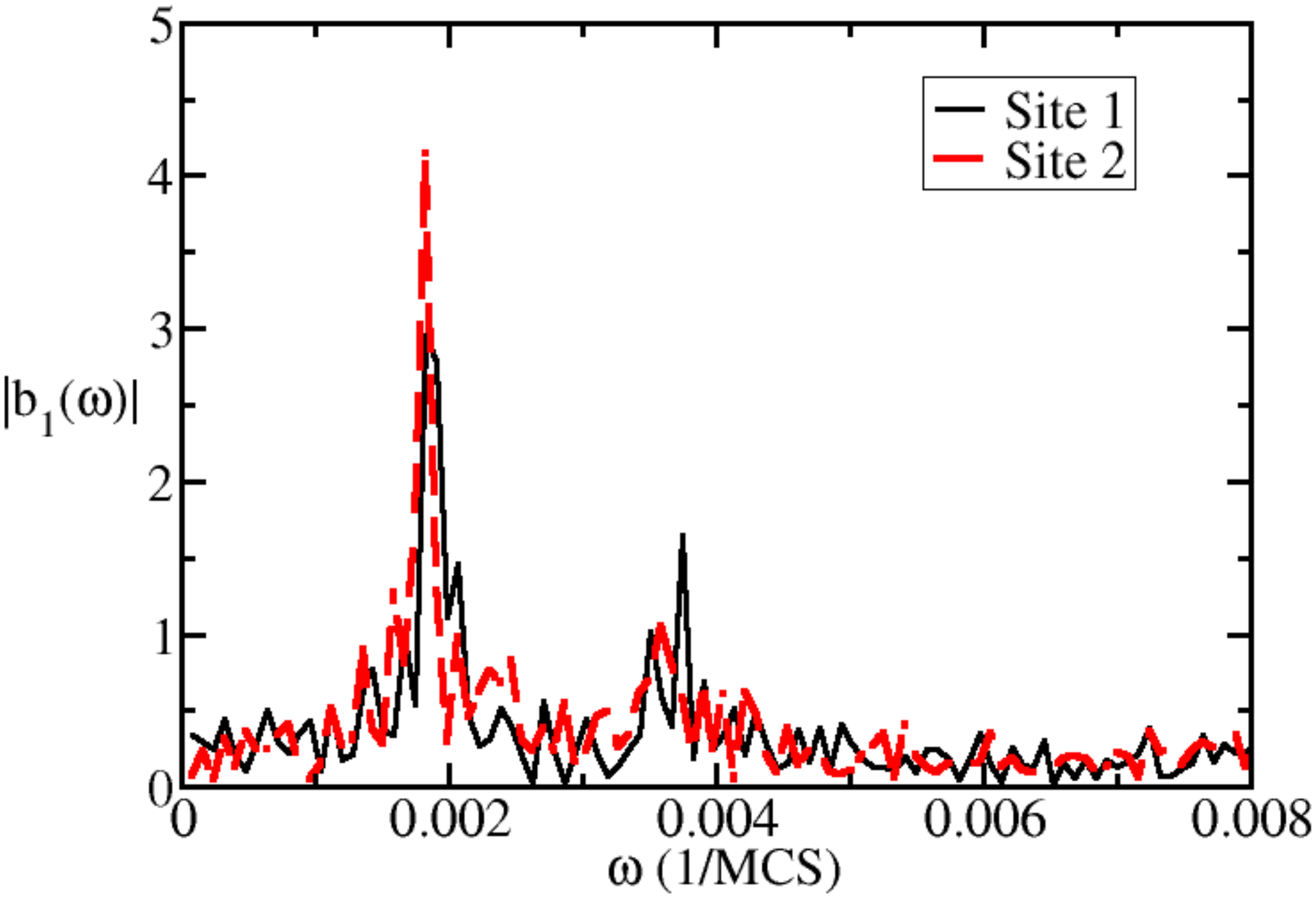}}
\caption{(a) Time evolution of the density of species $B_1$ in the ML model on two randomly chosen sites; (b) frequency 
spectrum of the time series in (a) obtained through a Fourier transform. In the simulation, the predation and reproduction 
rates are both set to $0.5$ and the system size is $100\times100$ lattice sites. The reactions in the simulation take place 
off-site, and no additional hopping processes are incorporated.}
\label{Single run results ML}
\end{figure}
To illustrate these points, we plot the time evolution of the population density of a single species at two randomly chosen 
lattice sites in a single simulation run for the RPS  and ML models in Figs.~\ref{Single run results RPS} and
\ref{Single run results ML}, respectively. For the RPS species density, the time intervals separating different peaks are not 
regularly spaced for the two lattice sites. Correspondingly, its Fourier transform displays multiple peaks with almost equal
intensities, and no well-defined characteristic frequency is discernible. In contrast, the peaks in the ML species density time 
evolution are evenly separated, and hence a dominant Fourier peak emerges. These numerical observations support our
analysis that the frequency in a single run in the ML model is well-defined, while that is clearly not the case for the RPS 
system. Yet a well-defined oscillation frequency clearly constitutes a necessary condition to form spatio-temporal patterns.

Although the arguments in this section are formulated in the framework of individual sites in a regular lattice, they are
readily extended to an effective unit cell that is similar in size to the characteristic diffusion length scale, or to models
defined on a continuum which require a finite reaction range. 
As long as the system size is much larger than any of these small length scales (that provide suitable ultraviolet cutoffs for
the continuum field theory), the specific microscopic details implemented in numerical simulations are not expected to have a 
significant impact on our conclusions.

\subsection{Stability of the homogeneous ground states against fluctuations (to one-loop order)}

While we posit above that a stable oscillation frequency at each lattice site is a necessary requirement to form oscillatory
spatio-temporal patterns such as spirals, this is not a sufficient condition. Another crucial ingredient is of course that the
spatially homogeneous stationary state is rendered unstable; fluctuations of a certain wavevector range will then 
spontaneously generate spatial patterns \cite{cross1993pattern,hohenberg2005introduction}.

In the stochastic spatially extended LV model \cite{tauber2012population}, in the absence of site restrictions, the population 
oscillation damping vanishes in the Gaussian approximation. To one-loop order, a negative damping frequency is generated which indicates an instability of the spatially uniform system, inducing nontrivial particle transport that in turn drives the 
formation of expanding evasion-pursuit activity waves. In the ML model, the damping coefficient is already nonzero in the 
Gaussian approximation. As stated above, when $2\kappa < \sigma$, the limit cycles of the ML model become unstable and 
spiral structures emerge. Yet within the realm of a perturbative approach, any fluctuation corrections will only shift the Hopf 
bifurcation point, but cannot qualitatively change the system's overall features. In contrast, the RPS system behaves similarly
to the LV system, as we encounter a vanishing damping coefficient in the Gaussian approximation. However, at one-loop level,
the generated mass term is positive indicating merely a generated finite damping constant, as opposed to its negative 
counterpart in the LV model. Hence the RPS model remains inert against spontaneous pattern formation. Higher-order terms
in the fluctuation expansion should not overturn the sign of the one-loop results within the perturbative regime, which
renders this argument perturbatively robust.

\subsection{Influence of quenched spatially disordered reaction rates}

In this part, we briefly investigate the effect of spatially disordered, uniformly distributed quenched reaction probabilities
on the fluctuation corrections.
We start with the Doi--Peliti action 
\begin{equation}
\mathcal{A} = \int\!\mathrm{d}t\, \mathrm{d}^dx\left[\mathcal{L}_0({\hat{a}_i,a_i})+\mathcal{L}'({\hat{a}_i,a_i})\right] ,
\label{eq: original action}
\end{equation}
where $\mathcal{L}_0$ and $\mathcal{L}'$ are (usually polynomial) functions of the fields $\{\hat{a}_i\}$ and $\{a_i\}$. 
We introduce spatial disorder to $\mathcal{L}'$, which we take to represent stochastic reactions while $\mathcal{L}_0$ 
encodes particle diffusions. The action with quenched spatial disorder is then given by
\begin{equation}
\mathcal{A}_D = \int\!\mathrm{d}t\, \mathrm{d}^dx \, \mathcal{L}_0 + \int \mathrm{d}^dx \, \eta(\vec{x})
\int\!\mathrm{d}\mathcal{L}' ;
\end{equation}
here, we assume $\eta(\vec{x})$ to be uniformly distributed in the interval $[0,2]$ with mean $\overline{\eta}=1$. Note
that the reaction rates are only spatially disordered and remain fixed in time. The average of any observables follows from 
Eq.~(\ref{eq:avg-coh-path-int}) through
\begin{equation}
\overline{\langle O(t) \rangle}\propto \prod_{\vec{x}}\int_0^2 \frac{\mathrm{d}\eta(\vec{x})}{2}\ \int \prod_{i}
\mathcal{D}[\hat{a}_i]\mathcal{D}[a_i]\,O(\{\hat{a}_i,a_i\}) \, e^{-\mathcal{A}_D} \ ,
\end{equation}
where the overline denotes the quenched disorder average. We can readily integrate out the disorder $\eta(\vec{x})$ first,
since the observable $O$ is independent of the random variables $\eta$, and arrive at
\begin{equation}
\overline{\langle O(t)\rangle}\propto \int\prod_{i}\mathcal{D}[\hat{a}_i]\mathcal{D}[a_i]\,O(\{\hat{a}_i,a_i\})\,
e^{-\mathcal{A}_D'} \ ,
\end{equation}
with
\begin{equation}
\mathcal{A}_D' = \mathcal{A} + \int\mathrm{d}^dx
\ln\left[ \frac{\sinh(\int\!\mathrm{d}t\ \mathcal{L}')}{\int\!\mathrm{d}t\ \mathcal{L}'}\right] .
\end{equation}
Incorporating disorder in the original Doi--Peliti action (\ref{eq: original action}) thus effectively leads to an additional term 
that is nonlocal in the time domain, owing to the temporally fixed reaction rates. However, in the perturbative regime where 
$\mathcal{L}'$ is small, this extra term can be expanded near $\mathcal{L}'=0$:
\begin{equation}
    \ln\left[\frac{\sinh(\int\mathrm{d}t\ \mathcal{L}')}{\int\mathrm{d}t\ \mathcal{L}'}\right] = \frac{1}{3}\int\!\mathrm{d}t 
    \,\mathrm{d}t' \,\mathcal{L}'_t\,\mathcal{L}_{t'} + \mathcal{O}(\mathcal{L}'^3) \ ,
\end{equation}
where the labels $t$ and $t'$ distinguish the different time dependences. It is evident from this expansion that the extra 
temporally nonlocal term ``entangles" different replicas of the system. However, as the first term is already second-order in 
$\mathcal{L}'$, it is of higher order than the original action $\mathcal{A}$, and should not markedly affect the system's 
fluctuation corrections in the naive perturbation regime, at least to low loop orders. This observation may account for the insensitivity of the RPS and ML models to quenched randomness in the reaction rates, as reported in 
Refs.~\cite{he2010spatial,he2011coexistence}. We remark that the stronger effects of varying predation rates $\lambda$ in 
the LV system can be largely traced to the sensitivity $\sim 1/\lambda$ of the stationary species densities already on the 
mean-field approximation level \cite{dobramysl2008spatial}.

\section{Summary and conclusions}

In this paper, we have investigated the dynamics of the RPS and ML models up to one-loop order in the fluctuation 
corrections by means of a perturbative field-theoretical analysis. We utilized the Doi--Peliti formalism to obtain the dynamical
probability functional for the stochastic Markovian dynamics, and also extracted the equivalent generalized Langevin equations. 
In the Gaussian theory, as expected, the RPS model displays only purely oscillatory modes in addition to the strictly diffusive 
conserved total particle density. In the ML model, a Hopf bifurcation point appears at $2\kappa - \sigma = 0$ that separates 
the parameter space into stable and unstable regions; in the latter regime, spiral structures are spontaneously generated.

The one-loop fluctuation corrections in the RPS model,
 which are of first order in the effective nonlinear coupling $\lambda/D \ll 1$, generate dissipation. 
We have found that in the physically accessible dimensions $d=1$, $2$, $3$, the damping coefficient is always positive. 
This indicates the stability of the spatially uniform stationary state in the RPS model, at least in the perturbative regime. 
Thus, our analysis sheds light on the absence of spatio-temporal structures in the RPS model. 
In addition, the one-loop correction to the oscillation frequency is IR divergent due to the dissipation-free nature of the 
mean-field modes, very similar to the LV model. 
However, outside the range of validity of the perturbation expansion, the damping terms become sizeable; hence we argue 
that this IR divergence becomes naturally regularized. 
This explains why no significant fluctuation corrections to the oscillation frequency have as yet been numerically 
observed in computer simulations, which have invariably been situated far away from the perturbative regime. 

In the ML model, as both the dissipation and oscillation frequencies are already finite at mean-field level, the one-loop 
fluctuation corrections should not qualitatively modify the mean-field conclusions. Since both propagating modes are massive 
due to the finite damping coefficients, the ML system does not display any IR singularities. Hence the ML model is insensitive 
to fluctuations at least perturbatively, and in the region where the homogeneous steady state is stable. Moreover, we have
argued that uniformly distributed quenched random disorder in the reaction rates only weakly influences fluctuation 
corrections in either system, which is in agreement with earlier Monte Carlo simulation data.

Finally, we provided two decisive criteria that determine the possibility of spatio-temporal structures in the LV, RPS, and ML models. The first argument considers a single lattice site point of view, while the second is based on studying the global 
stability of the spatially homogeneous stationary state of the system. From a single lattice site perspective, a necessary (but 
not sufficient) condition for the emergence of spatially extended coherent oscillatory behavior is that the oscillation frequency 
is constant over space and time in each run. Different oscillation frequencies on distinct sites would not allow the formation of  
stable coherent patterns. From a global point of view, only if the spatially uniform quasi-steady state is unstable against
finite-wavelength fluctuations can non-trivial spatio-temporal structures be generated, as is evident  in the ML model. Both 
these criteria explain the absence of spatio-temporal patterns in the RPS system, as a consequence of the relevant conservation
law for the total particle number, in contrast with the otherwise apparently similar LV and ML models.
We remark that adding some external noise to the RPS model that explicitly invalidates total particle number conservation 
might induce the formation of spiral patterns at sufficiently large length and long time scales.

\ack
The authors gratefully acknowledge inspiring discussions with Erwin Frey, Nigel Goldenfeld, Qian He, Mauro Mobilia, 
Michel Pleimling, Alastair Rucklidge, and Royce~K.~P.~Zia. This research was partially supported by the U.S National Science 
Foundation, Division of Mathematical Sciences under Award No. NSF DMS-2128587.

\appendix
\section{Doi--Peliti formalism and the asymmetric RPS model}

In this appendix, we provide a brief overview of the construction of the time evolution operator and the resulting 
field-theoretic action via the Doi--Peliti mapping of stochastic reactions to a non-Hermitean many-body quantum 
action \cite{doi1976second, doi1976stochastic, grassberger1980fock, Peliti1985path} (for recent reviews and 
additional details, see Refs.~\cite{tauber2014critical, tauber2005applications, mattis1998uses, cardy2008non}). 
We then proceed to construct coupled Langevin equations describing the dynamics of the system similar to previous 
work in the context of the LV system~\cite{butler2009predator, tauber2012population, tauber2014critical}, 
plankton-based predator-prey models~\cite{butler2009robust}, and Turing patterns~\cite{butler2011fluctuation}.
An interesting corner limit is analyzed via the generalized Langevin equations; we will show that the strongly 
asymmetric RPS system reduces to an effective LV model. Finally, we construct the diagonalized action for the RPS 
model and briefly compare the general situation with the symmetric version discussed in the bulk of this paper. 
This appendix is written in a self-consistent way and we hope it will be of use to readers who would like to delve into
the Doi--Peliti formalism.

\subsection{Stochastic time evolution operator}

The RPS rules~(\ref{rps reaction}) mandate that particle numbers are discrete, hence the occupation numbers of
lattice sites can be written as positive integers $n_{i \alpha}$, where the index $i$ accounts for different particle 
species, while $\alpha$ enumerates the sites on a $d$-dimensional hyper-cubic lattice. The master equation for the 
local, on-site RPS reactions then reads:
\begin{equation}
\label{eq:master-eq-reactions}
  \begin{split}
  \frac{\partial P(n_{i \alpha};t)}{\partial t}=\sum_{i=1,2,3} \lambda'_i \bigl[
  (n_{i \alpha}-1)(n_{i+1 \alpha}+1)P(n_{i \alpha}-1,n_{i+1 \alpha}+1;t) \\
  -n_{i \alpha}n_{i+1 \alpha} P(n_{i \alpha},n_{i+1 \alpha};t) \bigr] \ ,
  \end{split}
\end{equation}
where the index $i$ wraps around (i.e., $i=4$ is to be identified with $i=1$). Note that for brevity we have not 
included hopping to adjacent lattice sites here. As an initial state, we assume a uniform distribution of particles with
an average initial number of particles $N_i$ per lattice site of species $i$. This corresponds to a Poisson distribution 
for the occupation number of each species $i$ at all lattice sites $\alpha$, 
$P(n_{i \alpha};t) = \prod_{i=1,2,3} N_i^{n_{i \alpha}} e^{-N_i}/(n_{i \alpha}!)$. The discrete nature of the 
possible states of the RPS systems suggests the introduction of a product Fock space state vector
\begin{equation}
  \label{eq:fock-space-vector}
  |\Phi(t)\rangle = \sum_{\{n_{i \alpha}\}}P(n_{i \alpha};t) \prod_{i=1,2,3}
  \prod_{\alpha=1}^{L^d}|n_{i \alpha}\rangle \ ,
\end{equation}
where the $|n_{i \alpha}\rangle$ represent the occupation states of species $i$ on lattice site $i$. In analogy with 
the quantum-mechanical harmonic oscillator, the single-site states (and thereby the full state vector) can be acted 
upon by bosonic ladder operators obeying the commutation relations $[a_{i \alpha},a_{j \beta}]=0$ and 
$[a_{i \alpha},a_{j \beta}^\dagger]=\delta_{\alpha\beta}\delta_{ij}$. The occupation number eigenstates are 
constructed via $a_{i \alpha}|n_{i \alpha}\rangle=n_{i \alpha}|n_{i \alpha}-1\rangle$, 
$a_{i \alpha}^\dagger|n_{i \alpha}\rangle=|n_{i \alpha}+1\rangle$, and the empty state $|0\rangle$ is defined by 
$a_{i \alpha}|0\rangle=0$. 

The time evolution of the state vector~(\ref{eq:fock-space-vector}) follows directly from the master 
equation~(\ref{eq:master-eq-reactions}) and can be written in the form
\begin{equation}
  \label{eq:fock-time-evol}
  \frac{\partial}{\partial t}|\Phi(t)\rangle = -H|\Phi(t)\rangle \Longleftrightarrow 
 |\Phi(t)\rangle=e^{-Ht}|\Phi(0)\rangle \ ,
\end{equation}
where $H$ denotes the (time-independent) Liouville operator which can be split into a diffusion and a reaction term, $H=H_{\rm diff}+H_{\rm reac}$, where the on-site reaction contribution is a sum of local terms 
$H_{\rm reac}=\sum_{i=1}^{L^d}H_\alpha$, and specifically for the RPS model
\begin{equation}
  \label{eq:react-liouville}
  H_\alpha=\sum_{i=1,2,3}\lambda_i'(a_{i+1 \alpha}^\dagger - a_{i \alpha}^\dagger) a_{i \alpha}^\dagger 
  a_{i \alpha}a_{i+1 \alpha} \ .
\end{equation}
Similarly, since on-lattice diffusion is implemented by particles performing simple jumps between nearest-neighbor 
lattice sites, the diffusion part of $H$ reads
\begin{equation}
  \label{eq:diff-liouville}
  H_{\rm diff}=\sum_{i=1,2,3}\frac{D_i}{c^2}\sum_{\langle \alpha\beta \rangle}
  (a_{i \alpha}^\dagger-a_{i \beta}^\dagger)(a_{i \alpha}-a_{i \beta}) \ ,
\end{equation}
where $\langle \alpha\beta \rangle$ indicates a sum over all possible nearest-neighbor lattice site pairs in the system.

\subsection{Coherent-state path integral and equivalent Langevin partial differential equations}
\label{sec:janssen-dedominicis-functional}

Following the steps of Refs.~\cite{tauber2012population, tauber2014critical, tauber2005applications}, we write 
averages for observables $O = O(\{n_{i \alpha}\})$ as a multi-dimensional integral over coherent states
\begin{equation}
  \label{eq:average-path-int}
  \langle O(t)\rangle\propto\int\prod_{i=1,2,3}\prod_{\alpha=1}^{L^d} d\psi_{i \alpha}^*d\psi_{i \alpha} \,
  O(\{\psi_{i \alpha}\}) \, e^{-\mathcal{A}(\psi_{i \alpha}^*,\psi_{i \alpha};t)} \ ,
\end{equation}
where the $\psi_{i \alpha}$ and $\psi_{i \alpha}^*$ are complex eigenvalues describing the coherent right and left 
eigenstates of the ladder operators $a_{i \alpha}$ and $a_{i \alpha}^\dagger$, respectively. The coherent-state 
``action'' is given by 
\begin{equation}
  \label{eq:coh-state-action}
  \begin{split}
  \mathcal{A}(\psi_{i \alpha}^*,\psi_{i \alpha};t') = & \sum_{i=1,2,3} \sum_{\alpha=1}^{L^d} \biggl[ \int_0^{t'}\mathrm{d}t\
  \psi_{i \alpha}^*(t) \frac{\partial\psi_{i \alpha}(t)}{\partial t}dt-\psi_{i \alpha}(t') - N_i \psi_{i \alpha}^*(0) 
  \biggr] \\ 
  &+ \int_0^{t'} \mathrm{d}t\ H\bigl( a_{i \alpha}^\dagger\to\psi_{i \alpha}^*(t),a_{i \alpha} \to \psi_{i \alpha}(t)\bigr) \ ,
  \end{split}
\end{equation}
where we have to replace the ladder operators by their eigenvalues in the Liouville operator $H$. 

In the spatial continuum limit (lattice constant $c \to 0$) we may replace the sum over lattice sites with a 
$d$-dimensional volume integral $\sum_{\alpha=1}^{L^d} \to c^{-d}\int d^dx$, and the discretely spaced 
coherent-state values with continuous fields $\psi_{i \alpha}(t)\to c^d a_i(\vec{x},t)$ and 
$\psi_{i \alpha}^*(t)\to 1+\tilde{a}_i(\vec{x},t)$. Hence, the ``bulk'' part of action (not considering the terms from
the initial conditions at $t=0$ and the projection states at $t=t'$) of the RPS system is given by 
\begin{equation}
  \begin{split}
  \mathcal{A} = \int \mathrm{d}t\ \mathrm{d}^dx \Biggl [\sum_{i=1,2,3}\tilde{a}_i 
  \biggl( \partial_t-D_\alpha\nabla^2\biggr) a_i + \lambda_1a_1a_2(\tilde{a}_1+1)(\tilde{a}_2-\tilde{a}_1) \\
  + \lambda_2a_2a_3(\tilde{a}_2+1)(\tilde{a}_3-\tilde{a}_2)  
  + \lambda_3a_3a_1(\tilde{a}_3+1)(\tilde{a}_1-\tilde{a}_3) \Biggr] \ .
  \end{split}
\end{equation}
In the continuum limit we can thus write averages in the following coherent-state path integral form
\begin{equation}
  \label{eq:avg-coh-path-int}
  \langle O(t)\rangle\propto\int\prod_{i=1,2,3} \mathcal{D}[\tilde{a}_i]\mathcal{D}[a_i]\,O(\{a_i\})\,
  e^{-\mathcal{A}(\tilde{a}_i,a_i;t)} \ .
\end{equation}

Our aim here is to derive stochastic partial differential (Langevin) equations for the species concentrations that 
accurately capture the intrinsic reaction noise. To this end, we note that the Janssen--de~Dominicis response 
functional~\cite{tauber2014critical, janssen1976lagrangean, dominicis1976technics, bausch1976renormalized} 
\begin{equation}
\label{eq:janssen-dedominicis-functional}
    \mathcal{A} = \int \mathrm{d}t\ \mathrm{d}^dx \sum_i\tilde{a}_i\biggl(\partial_t a_i-D_i\nabla^2a_i-F_i[\{a_i\}]
     -\sum_j L_{ij}[\{a_i\}]\tilde{a}_j \biggr)
\end{equation}
is equivalent to the set of SPDEs
\begin{equation}
  \label{eq:janssen-dedominicis-spde}
  \partial_t a_i = D_i\nabla^2 a_i(\vec{x},t)+F_i[\{a_i(\vec{x},t)\}]+\zeta_i(\vec{x},t) \ ,
\end{equation}
with the associated noise (cross-)correlations
\begin{equation}
  \label{eq:janssen-dedominicis-noise}
  \langle\zeta_i(\vec{x},t)\zeta_j(\vec{x}',t')\rangle=2L_{ij}[\{a_i(\vec{x},t)\}] \delta(t-t')\delta(\vec{x}-\vec{x}') \ .
\end{equation}
This correspondence allows the immediate derivation of a coupled Langevin equation formulation of any system 
that exhibits an action functional of the form~\eqref{eq:janssen-dedominicis-functional}. Hence, via a direct 
comparison with the action of the RPS system~\eqref{eq:coh-state-action}, we can extract the deterministic part 
of the SPDEs describing the RPS system
\begin{equation}
  F_1=(\lambda_1a_2-\lambda_3a_3)a_1 \ , \quad
  F_2=(\lambda_2a_3-\lambda_1a_1)a_2 \ , \quad
  F_3=(\lambda_3a_1-\lambda_2a_2)a_3 \ , 
\end{equation}
which equal the right-hand side of the mean-field equations, as they should. Furthermore, the effective noise 
correlations are given by the matrix $L_{ij}$:
\begin{align}
  &L_{11}=\lambda_1a_1a_2 \ , \quad L_{12}=-\frac{\lambda_1}{2}a_1a_2 \ , \quad
  L_{13}=-\frac{\lambda_3}{2}a_1a_3 \ , \nonumber\\
  &L_{22}=\lambda_2a_2a_3 \ , \quad L_{23}=-\frac{\lambda_2}{2}a_2a_3 \ , \quad
  L_{33}=\lambda_3a_1a_3 \ .
\end{align}
Hence, the SPDEs~(\ref{eq:janssen-dedominicis-spde}) can be constructed from the mean-field equations by
including a term that accounts for diffusion and multiplicative noise terms obeying the given (cross-)correlations. 
Note that the noise auto-correlations $L_{ii}$ are always determined by the concentration of the predator species 
$A_i$ and its respective prey $A_{i+1}$, and the scale is set by the associated predation rate $\lambda_i$. Thus, 
the noise directly associated with a given species is solely determined by its role as predator.

\subsection{Strongly asymmetric RPS model: mapping to the LV system}

In order to investigate the asymmetric ``corner'' limit of the RPS system, we re-define the interaction rates as 
$\lambda_1=\lambda/x$, $\lambda_2=\lambda$ and $\lambda_3=\kappa\lambda$. The dimensionless variable 
$x$ varies in the interval $(0,1]$ and describes the asymmetry of the rates, while the equally dimensionless 
parameter $\kappa$ is of order unity and describes the difference between the predation rates of species $A_2$ 
and $A_3$. We are interested in the limit $x \to 0$ in which the predation reactions between species $A_1$ and 
$A_2$ dominate. The concentrations at the coexistence fixed point become
\begin{equation}
(\Omega_1,\Omega_2,\Omega_3)=\rho(x,\kappa x,1-[1+\kappa]x)+\mathcal{O}(x^2) \ . 
\end{equation}
Hence, the densities of species $A_1$ and $A_2$ become small as $x \to 0$, while species $A_3$ makes up 
most of the overall species abundance. This is the ``corner'' limit in which RPS can be approximated by a 
two-species Lotka--Volterra system, with the third, most abundant species serving as a mean-field like reservoir 
to feed the first species, and to provide the effective spontaneous death reaction for the second species, as 
explained above \cite{he2012relationship}. 
Species $A_1$ thus effectively turns into prey, while $A_2$ becomes the sole predator species. The noise 
correlation matrix $L$ in this corner case reads
\begin{align*}
  &L_{11}=\frac{\lambda}{x}a_1a_2 \ , \quad L_{12}=-\frac{\lambda}{2x}a_1a_2 \ , \quad
  L_{13}\approx-\frac{\kappa\lambda}{2}a_1\rho \ , \\
  &L_{22}\approx \lambda a_2\rho \ , \quad\,\ L_{23}\approx-\frac{\lambda}{2}a_2\rho \ , \qquad\!
  L_{33}=\kappa\lambda a_1\rho \ .
\end{align*}
The noise strength of species $A_1$, as well as the noise cross-correlations between species $A_1$ and $A_2$, 
are inversely proportional to the large rate scaling factor $x$, indicating that fluctuations of species $A_1$ and 
$A_2$ (the LV predator and prey, respectively) become strong in the limit $x \to 0$. Indeed, writing the resulting 
effective SPDEs in the limit of large $\lambda_1$ and assuming a homogeneous and stationary distribution of 
species $A_3$ yields
\begin{align}
  \partial_t a_1 &\approx D_1\nabla^2a_1 + \lambda_1a_1a_2 - \rho 
  \left( 1 - \frac{\lambda_2+\lambda_3}{\lambda_1} \right) \lambda_3a_1 + \zeta_1 \ , \\
  \partial_t a_2 &\approx D_2\nabla^2a_2 + \rho 
  \left( 1 - \frac{\lambda_2+\lambda_3}{\lambda_1} \right) \lambda_2a_2 - \lambda_1a_1a_2+ \zeta_2 \ ,
\end{align}
with the noise correlations
\begin{align}
  \langle\zeta_1(\vec{x},t)\zeta_1(\vec{x}',t')\rangle &= 2\lambda_1a_1a_2 \delta(\vec{x}-\vec{x}')\delta(t-t')\ , \\
  \langle\zeta_1(\vec{x},t)\zeta_2(\vec{x}',t')\rangle&=-\lambda_1a_1a_2 \delta(\vec{x}-\vec{x}')\delta(t-t')\ ,\\
  \langle\zeta_2(\vec{x},t)\zeta_2(\vec{x}',t')\rangle&=2\lambda_2\rho 
  \left( 1 - \frac{\lambda_2+\lambda_3}{\lambda_1} \right) a_2 \delta(\vec{x}-\vec{x}')\delta(t-t') \ .
\end{align}
This set of Langevin equations precisely matches those derived directly for the LV 
model~\cite{tauber2012population}.

\subsection{Fluctuation corrections}

In order to gain more insight into the role of fluctuations in the RPS system, we study the non-linear vertices 
arising from the Doi--Peliti action~(\ref{eq:coh-state-action}). To this end, we first need to diagonalize the action 
by transforming to appropriate field combinations. We then list the resulting vertices that capture fluctuation
corrections beyond the Gaussian mean-field approximation.

To start, we transform the fields to describe the fluctuations around the fixed-point species concentrations. To this 
end we employ the linear transformation
\begin{equation}
  \label{eq:fluct-transform}
  a_i(\vec{x},t)=\Omega_i+c_i(\vec{x},t) \ , \quad \tilde{a}_i(\vec{x},t)=\tilde{c}_i(\vec{x},t) \ ,
\end{equation}
here ignoring higher-order shifts of the steady-state coexistence concentrations induced by stochastic fluctuations 
(i.e., the counter-terms or additive renormalizations in Ref.~\cite{tauber2012population}). The action for these 
new fluctuating fields becomes 
\begin{equation}
  \mathcal{A}=\int \mathrm{d}t\ \mathrm{d}^dx \Bigl[ \sum_{i=1,2,3} \tilde{c}_i 
  \left( \partial_t-D_i\nabla^2 \right) c_i+\bar{\mathcal{A}} \Bigr] \ ,
\end{equation}
with the reduced part
\begin{equation}
  \label{eq:action-fluctuating-fields}
  \begin{split}
  \bar{\mathcal{A}}=\frac{-1}{\bar{\lambda}^2}\Bigl[ &\lambda_1(\tilde{c}_1+1)(\tilde{c}_1-\tilde{c}_2)
  (\bar{\lambda}c_1+\lambda_2'\rho)(\bar{\lambda}c_2+\lambda_3\rho)\\
  +&\lambda_2'(\tilde{c}_2+1)(\tilde{c}_2-\tilde{c}_3) 
  (\bar{\lambda}c_2+\lambda_3\rho)(\bar{\lambda}c_3+\lambda_1\rho)\\
  +&\lambda_3(\tilde{c}_3+1)(\tilde{c}_3-\tilde{c}_1)
  (\bar{\lambda}c_3+\lambda_1\rho)(\bar{\lambda}c_1+\lambda_2\rho)\Bigr] \ ,
  \end{split}
\end{equation}
and $\bar{\lambda}=\lambda_1+\lambda_2+\lambda_3$. The harmonic part of this action can be cast in a 
bilinear matrix form $\bar{s}_h=\sum_{ij} \tilde{c}_j A_{ij}c_i$ with the mass matrix
\begin{equation}
  \label{eq:massmatrix-flucfields}
  A=\frac{\rho}{\bar{\lambda}} \begin{pmatrix}
    0 & -\lambda_1\lambda_2 & \lambda_2\lambda_3\\
    \lambda_1\lambda_3 & 0 & -\lambda_2\lambda_3\\
    -\lambda_1\lambda_3 & \lambda_1\lambda_2 & 0 \end{pmatrix}=-A_s \ ,
\end{equation}
where $A_s$ is the stability matrix of the system at mean-field level. We note that it reduces to the stability 
matrix~(\ref{rps stability}) in the symmetric limit $\lambda_1=\lambda_2=\lambda_3=\lambda$.

\begin{table*}[tb]
  \centering
  \begin{tabular}{|p{0pt}p{36pt}|p{0.83\textwidth}|}
    \hline
    \multicolumn{3}{|l|}{Two-point (noise) sources \tikz[scale=0.3,baseline=-0.1cm]
    {\draw[->] (1,0) -- (0,0.5); \draw[->] (1,0) -- (0,-0.5);}}\\
    \hline
    & $\tilde{\phi}_\pm^2$ & $+\frac{\rho\lambda_2}{(\lambda_1+\lambda_3)^2}
    \bigl(\rho\frac{\lambda_1^2(\lambda_2-\lambda_3)-\lambda_3^2(\lambda_1-\lambda_2)}
    {\lambda_1+\lambda_2+\lambda_3}\pm i\omega_0[\lambda_1-\lambda_3]\bigr)$ \\
    & $\tilde{\phi}_+\tilde{\phi}_-$ & $-\frac{2\rho^2\lambda_2}{\lambda_1+\lambda_3}
    \frac{\lambda_1\lambda_2+\lambda_1\lambda_3+\lambda_2\lambda_3}
    {\lambda_1+\lambda_2+\lambda_3}$ \\
    \hline 
    \multicolumn{3}{|l|}{Merging three-point vertices \tikz[scale=0.3,baseline=-0.1cm]
    {\draw[->] (2,0.5) -- (1,0); \draw[->] (2,-0.5) -- (1,0); \draw[->] (1,0) -- (0,0);}} \\
    \hline
    & $\tilde{\phi}_\pm\phi_\pm\psi$ & $\pm 2i\frac{\omega_0}{\rho\lambda_2}
    \frac{(\lambda_1+\lambda_2+\lambda_3)^2}{\lambda_1+\lambda_3}$ \\
    & $\tilde{\phi}_\pm\phi_\pm^2$ & $+\bigl(\frac{\lambda_1+\lambda_2+\lambda_3}
    {\lambda_1+\lambda_3}\bigr)^2\bigl([\lambda_1-\lambda_3]\pm i\frac{\omega_0}{\rho\lambda_2}
    [\lambda_1-2\lambda_2+\lambda_3]\bigr)$ \\
    & $\tilde{\phi}_\pm\phi_\mp^2$ & $-\bigl(\frac{\lambda_1+\lambda_2+\lambda_3}
    {\lambda_1+\lambda_3}\bigr)^2\bigl([\lambda_1-\lambda_3]\pm i\frac{\omega_0}{\rho\lambda_2}
    [\lambda_1+2\lambda_2+\lambda_3]\bigr)$ \\
    \hline
    \multicolumn{3}{|l|}{Splitting three-point vertices \tikz[scale=0.3,baseline=-0.1cm]
    {\draw[->] (2,0) -- (1,0); \draw[->] (1,0) -- (0,0.5); \draw[->] (1,0) -- (0,-0.5);}} \\
    \hline
    & $\psi\tilde{\phi}_\pm^2$ & $\frac{2}{(\lambda_1+\lambda_3)^2}\bigl(\rho[\lambda_1^2
    (\lambda_2-\lambda_3)+\lambda_3^2(\lambda_2-\lambda_1)]\pm i\omega_0[\lambda_1-\lambda_3]
    [\lambda_1+\lambda_2+\lambda_3]\bigr)$ \\
    & $\psi\tilde{\phi}_+\tilde{\phi}_-$ & $-\frac{4\rho}{\lambda_1+\lambda_3}
    \bigl(\lambda_1\lambda_2+\lambda_1\lambda_3+\lambda_2\lambda_3\bigr)$ \\
    & $\phi_\pm\tilde{\psi}\tilde{\phi}_\pm$ & $\pm2i\omega_0\frac{\lambda_1+\lambda_2+\lambda_3}
    {\lambda_1+\lambda_3}$ \\
    & $\phi_\pm\tilde{\phi}_\pm^2$ & $\frac{\rho}{(\lambda_1+\lambda_3)^2}\bigl(\lambda_1\lambda_2
    [\lambda_1-\lambda_2]-\lambda_1\lambda_3[\lambda_1+\lambda_3]+2\lambda_2\lambda_3^2\pm
    \frac{\lambda_2\rho}{i\omega_0}[\lambda_2(\lambda_1^2-\lambda_3^2)+\lambda_1\lambda_3
    (\lambda_2+\lambda_3-2\lambda_1)]\bigr)$ \\
    & $\phi_\pm\tilde{\phi}_\mp^2$ & $-\frac{\rho}{(\lambda_1+\lambda_3)^3}
    \bigl(\lambda_1^2\lambda_2[\lambda_1+3\lambda_2-\lambda_3]+\lambda_1\lambda_3
    [\lambda_1 \lambda_2]^2+2(\lambda_1^2+\lambda_2^2)\lambda_3^2+\lambda_1\lambda_3^3\pm
    \frac{\lambda_2\rho}{i\omega_0}[-\lambda_1^3\lambda_2+\lambda_1\lambda_3^3
    +\lambda_2\lambda_3^3+\lambda_1^2\lambda_3(2\lambda_2+\lambda_3)]\bigr)$ \\
    & $\phi_\pm\tilde{\phi}_+\tilde{\phi}_-$ & $\frac{2\rho}{(\lambda_1+\lambda_3)^2}
    \bigl(2\lambda_1\lambda_2^2+[\lambda_2^2-\lambda_1^2]\lambda_3-[\lambda_1+\lambda_2]
    \lambda_3^2\mp\frac{\lambda_2\rho}{i\omega_0}[\lambda_1^2(\lambda_2+\lambda_3)
    -\lambda_2\lambda_3^2-\lambda_1\lambda_2\lambda_3]\bigr)$ \\
    \hline
  \end{tabular}
  \caption{Coefficients of the vertices in the action after the transformation to fluctuating fields.}
  \label{tab:field_coefficients}
\end{table*}

Our goal is to find a transformation that diagonalizes the mass matrix $A$, and thus the harmonic part of the action~(\ref{eq:coh-state-action}), if we set all diffusivities equal, $D_i =D$. The matrix $A$ is asymmetric, 
hence we make use of its orthogonal left and right eigenvectors $\vec{u}_i A=e_i\vec{u}_i$ and 
$A\vec{v}_i=e_i\vec{v}_i$, respectively. The resulting eigenvector matrices 
$Q=(\vec{u}_1,\vec{u}_2,\vec{u}_3)$ and $P=(\vec{v}_1,\vec{v}_2,\vec{v}_3)$ read
\begin{equation}
  \label{eq:right-eigenv}
  Q=
  \begin{pmatrix}
    1 & 1 & 1 \\
    1 & -\frac{\lambda_2(\lambda_1+\lambda_2)}{\lambda_2\lambda_3+i\Lambda} 
    & -\frac{\lambda_2(\lambda_1+\lambda_2)}{\lambda_2\lambda_3-i\Lambda} \\
    1 & -\frac{\lambda_2(\lambda_2+\lambda_3)}{\lambda_1\lambda_2-i\Lambda} 
    & -\frac{\lambda_2(\lambda_2+\lambda_3)}{\lambda_1\lambda_2+i\Lambda}
  \end{pmatrix} \ ,
\end{equation}
where 
$\Lambda=\sqrt{\lambda_1\lambda_2\lambda_3(\lambda_1+\lambda_2+\lambda_3)}$, and
\begin{equation}
  \label{eq:left-eigenv}
  P=
  \begin{pmatrix}
    1 & 1 & 1 \\
    \frac{\lambda_3}{\lambda_2} & -\frac{\lambda_3(\lambda_1+\lambda_2)}{\lambda_2\lambda_3-i\Lambda} 
    & -\frac{\lambda_3(\lambda_1+\lambda_2)}{\lambda_2\lambda_3+i\Lambda} \\
    \frac{\lambda_1}{\lambda_2} & -\frac{\lambda_1(\lambda_2+\lambda_3)}{\lambda_1\lambda_2+i\Lambda} 
    & -\frac{\lambda_1(\lambda_2+\lambda_3)}{\lambda_1\lambda_2-i\Lambda}
  \end{pmatrix} \ .
\end{equation}
The right and left eigenvector matrices then transform the mass matrix to the diagonal form 
$Q^TAP(Q^TP)^{-1}=\text{diag}(e_i)$. Defining new fields $\tilde{\phi}_i$ and $\phi_i$ according to the 
transformation $\tilde{c}_i=\sum_j Q^T_{ji}\tilde{\phi}_j$ and $c_i=\sum_j P_{ij}\phi_j$, we arrive 
at\footnote{Note that we have employed a different diagonalization convention in this appendix as compared to the 
main text. This difference is reflected in the constant factors in the propagators.}
\begin{align}
  \tilde{c}_1=&\tilde{\psi}+\tilde{\phi}_++\tilde{\phi}_- \ , \nonumber \\ 
  \label{eq:tilde-transform-23}
  \tilde{c}_{2/3}=&\tilde{\psi}-\frac{\lambda_2}{\lambda_1+\lambda_3}(\tilde{\phi}_++\tilde{\phi}_-)\pm
  \frac{i\Lambda/\lambda_{^3/_1}}{\lambda_1+\lambda_3}(\tilde{\phi}_+-\tilde{\phi}_-) \ , \\
  c_1=&\psi+\phi_++\phi_- \ , \nonumber \\
  \label{eq:transform-23}
  c_{2/3}=&\frac{\lambda_{^3/_1}}{\lambda_2}\psi-\frac{\lambda_{^3/_1}}{\lambda_1+\lambda_3}
  (\phi_++\phi_-)\mp\frac{i\Lambda/\lambda_{2}}{\lambda_1+\lambda_3}(\phi_+-\phi_-) \ .
\end{align}
It is already obvious from this structure  that the fields $\tilde{\psi}$ and $\psi$ describe the fluctuation of the 
total population, while the other fields are oscillatory in nature. Employing these transformations, one arrives at 
the diagonalized harmonic action
\begin{equation}
  \label{eq:trans-harm-action}
  \begin{split}
    \mathcal{A}_0=&\int dt\int d^dx(\lambda_1+\lambda_2+\lambda_3)
    \Bigl[\frac{1}{\lambda_2}\tilde{\psi} \left( \partial_t-D\nabla^2 \right) \psi \\
    &+\frac{1}{\lambda_1+\lambda_3}\tilde{\phi}_+ \left( \partial_t-D\nabla^2+i\omega_0 \right) \phi_+
    +\frac{1}{\lambda_1+\lambda_3}\tilde{\phi}_- \left( \partial_t	-D\nabla^2-i\omega_0 \right) \phi_- \Bigr] \ .
  \end{split}
\end{equation}
The field $\psi=\lambda_2(c_1+c_2+c_3)/(\lambda_1+\lambda_2+\lambda_3)$ is massless, encodes no 
reactions, and is purely diffusive, as it represents the total local concentration of all three species. The 
corresponding harmonic propagator is given in Fourier space by
\begin{equation}
  \langle\tilde{\psi}(\vec{q},\omega)\,\psi(\vec{q}',\omega')=\frac{\lambda_2}
  {\lambda_1+\lambda_2+\lambda_3}\frac{(2\pi)^{d+1}\delta(\vec{q}+\vec{q}')\delta(\omega+\omega')}
  {-i\omega+Dq^2} \ ,
\end{equation}
while the oscillating field propagators display poles at finite eigenfrequencies $\mp\omega_0$, similar to the LV 
case~\cite{tauber2012population}. The action transformed to the new fields becomes quite cumbersome to write 
out in full, hence we merely provide the coefficients of the possible field combinations in the vertices in 
Table~\ref{tab:field_coefficients}. Note that we omit the coefficients of any four-point vertices as these do not 
contribute to one-loop order corrections.

\section{Detailed calculations for the ML model}

Here we provide intermediate steps for the one-loop calculation of the ML model. The renormalized frequencies are
\begin{equation}
\begin{aligned}
&\gamma_r \pm i\nu_r = \gamma_0 \pm i\nu_0 +(\sigma+\kappa) \bigg[ M_1^{(\pm)}\!
\int_k\frac{1}{k^2+\frac{\gamma_0}{D}}+ M_2^{(\pm)}\! \int_k\frac{1}{k^2+\frac{\gamma_0\mp i\nu_0}{D}} \\
&\ +M_3^{(\pm)}\! \int_k\frac{1}{k^2+\frac{\mu+\gamma_0\pm i\nu_0}{2D}} + M_4^{(\pm)}\!
\int_k\frac{1}{(k^2+\frac{\gamma_0\mp i\nu_0}{D})^2} +M_5^{(\pm)}\!
\int_k\frac{1}{(k^2+\frac{\mu+\gamma_0\pm i\nu_0}{2D})^2} \bigg] ,
\end{aligned}
\end{equation}
with the coefficients $M_i^{(\pm)} = \mathrm{Re}M_i \pm \mathrm{Im}M_i$, where
\begin{equation}
\begin{aligned}
\mathrm{Re}M_1 =& \frac{\gamma_0}{\sqrt{3} D \mu^2 \nu_0  \left[ (\gamma_0-\mu)^2+\nu_0^2 \right]^2} \Big[ 
\gamma_0^2 \mu  (\gamma_0-\mu)^4-\nu_0^4 (4 \gamma_0^3-19 \gamma_0^2 \mu+2 \gamma_0 \mu^2+8 \mu^3) \\
&-\nu_0^2 (\gamma_0 -\mu )^2 (2 \gamma_0^3-5 \gamma_0^2 \mu +2 \gamma_0 \mu^2+3 \mu^3)
-\nu_0^6 (2 \gamma_0 +5 \mu) \Big] , \\
\mathrm{Im}M_1 =& \frac{1}{\sqrt{3} D \mu^2 \left[ (\gamma_0-\mu)^2+\nu_0 ^2 \right]^2} \Big[ -\gamma_0^2 \mu  
(\gamma_0-\mu)^4-\nu_0^4 (4 \gamma_0 ^3-17 \gamma_0^2 \mu+2 \gamma_0 \mu^2+2 \mu^3) \\
&-\nu_0^2 (2 \gamma_0^5-\gamma_0^4 \mu +6 \gamma_0^3 \mu^2-8 \gamma_0^2 \mu^3+\mu^5) 
- \nu_0^6 (2 \gamma_0 +\mu) \Big] , \\
\mathrm{Re}M_2 =& -\frac{\gamma_0^2 (\sqrt{3} \gamma_0 +\nu_0)}{3 D \mu  \nu_0} \ , \qquad
\mathrm{Im}M_2 = \frac{\gamma_0 (\sqrt{3} \gamma_0 +\nu_0)}{3 D \mu} \ , \\
\mathrm{Re}M_3 =& -\frac{\gamma_0}{3 D \mu \left[ (\gamma_0 -\mu )^2+\nu_0^2 \right]^2} 
\Big[ 2 \sqrt{3} \nu_0^3 (6 \gamma_0^2-\gamma_0 \mu -3 \mu^2)+\nu_0^4 (\gamma_0 +\mu) \\
&\!\!\!\!\!\!\!\!\!\!\!\!\!\!\!\!\!\! +2 \nu_0^2 (\gamma_0 -\mu)^2 (\gamma_0 +\mu)-2 \sqrt{3} \mu \nu_0 (\gamma_0 -\mu)^2 
(\gamma_0 +\mu)+(\gamma_0 -\mu)^4 (\gamma_0 +\mu)-4 \sqrt{3} \nu_0 ^5 \Big] , \\
\mathrm{Im}M_3 =&  \frac{\gamma_0 \nu_0}{3 D \mu  \left[ (\gamma_0 -\mu)^2+\nu_0^2\right]^2} \Big[ 2 \sqrt{3} 
\nu_0 (2 \gamma_0^3+\gamma_0^2 \mu -4 \gamma_0 \mu^2+\mu^3) +2 \sqrt{3} \nu_0^3 (\mu -6 \gamma_0) \\
&-2 \nu_0^2 (\gamma_0 -\mu)^2-(\gamma_0 -\mu)^4-\nu_0^4 \Big] , \\
\mathrm{Re}M_4 =& -\frac{(\gamma_0 +\sqrt{3} \nu_0) (\gamma_0 ^2+\nu_0 ^2)}{6 D^2 \mu} \ , \qquad 
\mathrm{Im}M_4 = \frac{\gamma_0 (\gamma_0^2+\nu_0^2)}{\sqrt{3} D^2 \mu} \ , \\
\mathrm{Re}M_5 =& -\frac{\gamma_0}{6 D^2 \mu \left[ (\gamma_0 -\mu)^2+\nu_0^2 \right]} 
\Big[ \gamma_0^4-\gamma_0^3 (\mu -2 \sqrt{3} \nu_0) -\gamma_0^2 \mu^2 \\
&+\gamma_0 (\mu -2 \sqrt{3} \nu_0) (\mu^2+3 \nu_0^2)-\nu_0 ^2 (\mu^2+\nu_0^2) \Big] , \\
\mathrm{Im}M_5 =&-\frac{\gamma_0 \nu_0} {6 D^2 \mu \left[ (\gamma_0 -\mu)^2+\nu_0^2 \right]}
\Big[ 2 \gamma_0^3-3 \gamma_0^2 (\mu -2 \sqrt{3} \nu_0)+2 \gamma_0 \nu_0^2 \\
&+(\mu -2 \sqrt{3} \nu_0) (\mu^2+\nu_0^2) \Big] . 
\end{aligned}
\label{coefficients_ML}
\end{equation}
The renormalized diffusivity is
\begin{equation}
\begin{aligned}
D_r^{\pm} = &D -(\kappa+\sigma) \bigg[ \frac{(\gamma_0-\sqrt{3}\nu_0)}{3dD\mu}(\gamma_0\mp i\nu_0)
\int_k \frac{k^2}{\left(k^2+\frac{\gamma_0\mp i\nu_0}{D}\right)^3} \\
&-\frac{\gamma_0}{3dD\mu}(\gamma_0+\mu\pm i\nu_0)
\int_k \frac{k^2}{\left(k^2+\frac{\mu+\gamma_0\pm i\nu_0}{2D}\right)^3} \\
&\mp i \frac{2\sqrt{3}\gamma_0}{3dD^2\mu}(\gamma_0^2+\nu_0^2)(\gamma_0 \mp i\nu_0) 
\int_k \frac{1}{\left(k^2+\frac{\gamma_0\mp i\nu_0}{D}\right)^3}\\
&+ \frac{2\sqrt{3}\gamma_0}{3dD\mu\nu_0}(\gamma_0^2+\nu_0^2)
\int_k \frac{1}{\left(k^2+\frac{\gamma_0\mp i\nu_0}{D}\right)^2}\pm i \frac{2 \sqrt{3} \gamma_0}{3d\mu\nu_0^2}(\gamma_0^2+\nu_0^2) \int_k \frac{1}{k^2+\frac{\gamma_0\mp i\nu_0}{D}} \\
&-\frac{\sqrt{3}\gamma_0\nu_0(\gamma_0+\mu\pm i\nu_0)^2}{3dD^2\mu(-\gamma_0 + \mu \pm i\nu_0)}
\int_k \frac{1}{\left(k^2+\frac{\mu+\gamma_0\pm i\nu_0}{2D}\right)^3} \\
&-\frac{4\sqrt{3}\gamma_0^2\nu_0(\gamma_0+\mu\pm i\nu_0)}{3dD\mu(-\gamma_0+\mu\pm i\nu_0)^2}
\int_k \frac{1}{\left(k^2+\frac{\mu+\gamma_0\pm i\nu_0}{2D}\right)^2} \\
&-\frac{8\sqrt{3}\gamma_0^2\nu_0(\gamma_0+\mu\pm i\nu_0)}{3d\mu(-\gamma_0+\mu\pm i\nu_0)^3}
\int_k \frac{1}{k^2+\frac{\mu+\gamma_0\pm i\nu_0}{2D}} \\
&+\frac{2 \gamma}{\sqrt{3} d \mu \nu_0^2} \left[ \frac{4 \gamma \nu_0^3 (\gamma +\mu +i \nu_0)}
{(-\gamma +\mu +i \nu_0 )^3}-i \left(\gamma ^2+\nu_0 ^2\right) \right] 
\int_k \frac{1}{k^2+\frac{\gamma_0}{D}} \bigg] ,
\end{aligned}
\end{equation}
and can be further simplified to
\begin{equation}
\begin{aligned}
D_r^\pm = &D +\frac{\kappa+\sigma}{d} \bigg[ P_3^{(\pm)}\!
\int_k \frac{1}{\left(k^2+\frac{\gamma_0\mp i\nu_0}{D}\right)^3} +P_2^{(\pm)}\!
\int_k \frac{1}{\left(k^2+\frac{\gamma_0\mp i\nu_0}{D}\right)^2} +P_1^{(\pm)}\!
\int_k \frac{1}{k^2+\frac{\gamma_0\mp i\nu_0}{D}} \\
&+ Q_3^{(\pm)}\! \int_k \frac{1}{\left(k^2+\frac{\mu+\gamma_0\pm i\nu_0}{2D}\right)^3} 
+Q_2^{(\pm)}\!\int_k \frac{1}{\left(k^2+\frac{\mu+\gamma_0\pm i\nu_0}{2D}\right)^2} + Q_1^{(\pm)}\!
\int_k \frac{1}{k^2+\frac{\mu+\gamma_0\pm i\nu_0}{2D}} \\ 
&- \left( P_1^{(\pm)}+Q_1^{(\pm)} \right) \int_k \frac{1}{k^2+\frac{\gamma_0}{D}} \ \bigg] ,
\end{aligned}
\end{equation}
where $P_i^{(\pm)} = \mathrm{Re}P_i \pm i \mathrm{Im}P_i$ and 
$Q_i^{(\pm)} = \mathrm{Re}Q_i \pm i \mathrm{Im}Q_i$, with
\begin{equation}
\begin{aligned}
\mathrm{Re}P_1 =& 0 \ , \qquad 
\mathrm{Im}P_1 = \frac{2 \gamma_0 \left(\gamma_0 ^2+\nu_0^2 \right) }{\sqrt{3} \mu \nu_0^2} \ , \\
\mathrm{Re}P_2 =& \frac{\gamma_0 \left(2 \sqrt{3} \gamma_0^2-\gamma_0 \nu_0 +3 \sqrt{3} \nu_0^2\right)}
{3 D \mu  \nu_0} \ , \qquad
\mathrm{Im}P_2 =  \frac{\nu_0 \left(\gamma_0 -\sqrt{3} \nu_0 \right)}{3 D \mu} \ , \\
\mathrm{Re}P_3 = & \frac{\left(\gamma_0^3-3 \sqrt{3} \gamma_0^2 \nu_0 -\gamma_0 \nu_0^2-\sqrt{3} \nu_0^3\right)}
{3 D^2 \mu} \ , \qquad
\mathrm{Im}P_3 = -\frac{2 \gamma_0^2 \left(\sqrt{3} \gamma_0 +\nu_0 \right)}{3 D^2 \mu } \ , \\
\mathrm{Re}Q_1 =& \frac{8 \gamma_0^2 \nu_0 \left[ 6 \gamma_0  \nu_0^2 (\mu -\gamma_0)+(\gamma_0 -\mu)^3 
(\gamma_0 +\mu)+\nu_0 ^4 \right]}{\sqrt{3} \mu  \left[ (\gamma_0 -\mu )^2+\nu_0 ^2 \right]^3} \ , \\ 
\mathrm{Im}Q_1 =& \frac{16 \gamma_0^2 \nu_0^2 \left(2 \gamma_0^3-3 \gamma_0^2 \mu -2 \gamma_0 \nu_0^2
+\mu^3+\mu \nu_0^2\right)}{\sqrt{3} \mu \left[ (\gamma_0 -\mu)^2+\nu_0^2 \right]^3} \ , \\
\mathrm{Re}Q_2 =& -\frac{\gamma_0}{3 D \mu \left[ (\gamma_0 -\mu)^2+\nu_0^2 \right]^2} \Big[ 
\gamma_0^5+\gamma_0^4 \left(4 \sqrt{3} \nu_0 -3 \mu \right) \\
&+2 \gamma_0^3 \left( \mu^2-2 \sqrt{3} \mu \nu_0 +\nu_0^2\right) +2 \gamma_0^2 \left( \mu^3
-2 \sqrt{3} \mu^2 \nu_0 -\mu \nu_0^2-6 \sqrt{3} \nu_0^3 \right) \\
&-\gamma_0 \left( 3 \mu^2-4 \sqrt{3} \mu \nu_0 -\nu_0^2\right) \left(\mu^2+\nu_0^2\right) 
+\mu \left( \mu^2+\nu_0 ^2\right)^2 \Big] \ , \\
\mathrm{Im}Q_2 =& \frac{\gamma_0 \nu_0 \left[ -2 \nu_0 ^2 (\gamma_0 -\mu )^2-4 \sqrt{3} \gamma \nu_0  
(3 \gamma_0 +\mu) (\gamma_0 -\mu)-(\gamma_0 -\mu )^4+4 \sqrt{3} \gamma_0 \nu_0^3-\nu_0^4 \right]}
{3 D \mu \left[ (\gamma_0 -\mu)^2+\nu_0 ^2\right]^2} , \\
\mathrm{Re}Q_3 =& \frac{\gamma_0}{6 D^2 \mu \left[ (\gamma_0 -\mu)^2+\nu_0^2\right)} \Big[ 
\gamma_0^4+2 \sqrt{3} \gamma_0^3 \nu_0 -2 \gamma_0^2 \mu \left(\mu -\sqrt{3} \nu_0 \right) \\
&-2 \gamma_0 \nu_0 \left(\sqrt{3} \mu^2-2 \mu \nu_0+3 \sqrt{3} \nu_0^2\right)
+\left(\mu^2-2 \sqrt{3} \mu \nu_0 -\nu_0 ^2\right) \left(\mu^2+\nu_0^2\right) \Big] \ , \\
\mathrm{Im}Q_3 =& \frac{\gamma_0 \nu_0 \left[ \nu_0^2 (\gamma_0 +\mu )+\sqrt{3} \nu_0 (3 \gamma_0 -\mu) 
(\gamma_0 +\mu )+(\gamma_0 -\mu )^2 (\gamma_0 +\mu )-\sqrt{3} \nu_0^3 \right]}
{3 D^2 \mu  \left[ (\gamma_0 -\mu )^2+\nu_0 ^2 \right]} \ .
\end{aligned}
\end{equation}

\section{Renormalized variables in the ML model at physically accessible dimensions}

Finally, we list the expressions of the renormalized variables at physically accessible dimensions $d=1$, $2$, and $3$.

\subsection{$d=1$:} For the $\psi_o$ mode, the renormalized parameters read 
\begin{equation}
\begin{aligned}
\mu_r =& \mu \left[ 1-\frac{5(\sigma+\kappa)}{24D}\sqrt{\frac{D}{\mu}}-\frac{\sigma+\kappa}{D}
\sqrt{\frac{D}{\gamma_0}} \left( \frac{1}{12}+\frac{\gamma_0}{3\mu}+\frac{\sqrt{3}\nu_0}{16\mu}
+\frac{\sqrt{3}\nu_0}{16\gamma_0}+\frac{\sqrt{3}\nu_0\gamma_0}{6\mu^2} \right)  \right] , \\
D^{0}_r =& D-\frac{\sigma+\kappa}{48}\sqrt{\frac{D}{\mu}}-\frac{\sigma+\kappa}{24}\sqrt{\frac{D}{\gamma_0}}
\left( \frac{\gamma_0}{\mu}-\frac{\sqrt{3}\nu_0}{2\mu}+\frac{\sqrt{3}\nu_0}{2\gamma_0} \right) .
\end{aligned}
\end{equation}
For the $\psi_\pm$ modes, when $\gamma_0 \geq 0$, the renormalized parameters are
\begin{equation}
\begin{aligned}
\gamma_r = &\gamma_0 + (\sigma+\kappa)\sqrt{\frac{D}{\nu_0}} \ \bigg[ \frac12 \mathrm{Re}M_1 
\sqrt{\frac{\nu_0}{\gamma_0}}+\frac{1}{2} \left( 1+\frac{\gamma_0^2}{\nu_0^2} \right)^{\!-1/4}
\left( \mathrm{Re}M_2\cos\frac{\theta}{2}-\mathrm{Im}M_2\sin\frac{\theta}{2} \right) \\
&+\frac{\sqrt{2}}{2} \left( 1+\frac{(\gamma_0+\mu)^2}{\nu_0^2} \right)^{\!-1/4} 
\left( \mathrm{Re}M_3\cos\frac{\eta}{2}+\mathrm{Im}M_3\sin\frac{\eta}{2} \right) \\
&+ \frac{D}{4\nu_0}\left(1+\frac{\gamma_0^2}{\nu_0^2}\right)^{\!-3/4}
\left( \mathrm{Re}M_4\cos\frac{3\theta}{2}-\mathrm{Im}M_4\sin\frac{3\theta}{2} \right) \\
&+ \frac{\sqrt{2}D}{2\nu_0} \left (1+\frac{(\mu+\gamma_0)^2}{\nu_0^2} \right)^{\!-3/4}
\left( \mathrm{Re}M_5\cos\frac{3\eta}{2}+\mathrm{Im}M_5\sin\frac{3\eta}{2} \right) \bigg] \ , \\
\nu_r =& \nu_0 +(\sigma+\kappa)\sqrt{\frac{D}{\nu_0}} \ \bigg[ \frac12 \mathrm{Im}M_1 
\sqrt{\frac{\nu_0}{\gamma_0}}+\frac{1}{2} \left( 1+\frac{\gamma_0^2}{\nu_0^2}\right)^{\!-1/4}
\left( \mathrm{Re}M_2\sin\frac{\theta}{2}+\mathrm{Im}M_2\cos\frac{\theta}{2} \right) \\
&+\frac{\sqrt{2}}{2} \left( 1+\frac{(\gamma_0+\mu)^2}{\nu_0^2} \right)^{\!-1/4}
\left( -\mathrm{Re}M_3\sin\frac{\eta}{2}+\mathrm{Im}M_3\cos\frac{\eta}{2} \right) \\
&+ \frac{D}{4\nu_0} \left( 1+\frac{\gamma_0^2}{\nu_0^2} \right)^{\!-3/4} 
\left( \mathrm{Re}M_4\sin\frac{3\theta}{2}+\mathrm{Im}M_4\cos\frac{3\theta}{2} \right) \\
&+ \frac{\sqrt{2}D}{2\nu_0} \left( 1+\frac{(\mu+\gamma_0)^2}{\nu_0^2}\right)^{\!-3/4}
\left( -\mathrm{Re}M_5\sin\frac{3\eta}{2}+\mathrm{Im}M_5\cos\frac{3\eta}{2} \right) \bigg] \ ;
\end{aligned}
\end{equation}
yet if $\gamma_0 < 0$, the first term in both parameters changes:
\begin{equation}
   \begin{aligned}
\gamma_r = &\gamma_0 + (\sigma+\kappa)\sqrt{\frac{D}{\nu_0}} \, \bigg[ \!-\frac12 \mathrm{Im}M_1 
\sqrt{\frac{\nu_0}{|\gamma_0|}}+\frac12 \left( 1+\frac{\gamma_0^2}{\nu_0^2} \right)^{\!-1/4}\!
\left( \mathrm{Re}M_2\cos\frac{\theta}{2}-\mathrm{Im}M_2\sin\frac{\theta}{2} \right) \\
&+\frac{\sqrt{2}}{2} \left( 1+\frac{(\gamma_0+\mu)^2}{\nu_0^2} \right)^{\!-1/4}
\left( \mathrm{Re}M_3\cos\frac{\eta}{2}+\mathrm{Im}M_3\sin\frac{\eta}{2} \right) \\
&+ \frac{D}{4\nu_0} \left( 1+\frac{\gamma_0^2}{\nu_0^2} \right)^{\!-3/4}
\left( \mathrm{Re}M_4\cos\frac{3\theta}{2}-\mathrm{Im}M_4\sin\frac{3\theta}{2} \right) \\
&+ \frac{\sqrt{2}D}{2\nu_0} \left( 1+\frac{(\mu+\gamma_0)^2}{\nu_0^2} \right)^{\!-3/4}
\left( \mathrm{Re}M_5\cos\frac{3\eta}{2}+\mathrm{Im}M_5\sin\frac{3\eta}{2} \right) \bigg] \ , \\
\nu_r =& \nu_0 +(\sigma+\kappa)\sqrt{\frac{D}{\nu_0}} \ \bigg[ \frac12 \mathrm{Re}M_1 
\sqrt{\frac{\nu_0}{|\gamma_0|}} +\frac12 \left( 1+\frac{\gamma_0^2}{\nu_0^2} \right)^{\!-1/4}
\left( \mathrm{Re}M_2\sin\frac{\theta}{2}+\mathrm{Im}M_2\cos\frac{\theta}{2} \right) \\
&+\frac{\sqrt{2}}{2} \left(1 +\frac{(\gamma_0+\mu)^2}{\nu_0^2} \right)^{\!-1/4}
\left( -\mathrm{Re}M_3\sin\frac{\eta}{2}+\mathrm{Im}M_3\cos\frac{\eta}{2}\right) \\
&+ \frac{D}{4\nu_0} \left( 1+\frac{\gamma_0^2}{\nu_0^2} \right)^{\!-3/4}
\left( \mathrm{Re}M_4\sin\frac{3\theta}{2}+\mathrm{Im}M_4\cos\frac{3\theta}{2} \right) \\
&+ \frac{\sqrt{2}D}{2\nu_0} \left( 1+\frac{(\mu+\gamma_0)^2}{\nu_0^2} \right)^{\!-3/4}
\left( -\mathrm{Re}M_5\sin\frac{3\eta}{2}+\mathrm{Im}M_5\cos\frac{3\eta}{2} \right) \bigg] \ .
\end{aligned} 
\end{equation}
The renormalized diffusivity is not affected by the sign of $\gamma_0$,
\begin{equation}
\begin{aligned}
\mathrm{Re}D_r = D &+(\kappa+\sigma)\sqrt{\frac{D}{\nu_0}} \ \bigg[ -\frac12 
\left( 1+\frac{\gamma_0^2}{\nu_0^2} \right)^{\!-1/4} \mathrm{Im}P_1\sin\frac{\theta}{2} 
- \frac12 \mathrm{Re}Q_1 
\sqrt{\frac{\nu_0}{\gamma_0}} \\
&+\frac{D}{4\nu_0} \left( 1+\frac{\gamma_0^2}{\nu_0^2} \right)^{\!-3/4}
\left(\mathrm{Re}P_2\cos\frac{3\theta}{2}-\mathrm{Im}P_2\sin\frac{3\theta}{2} \right) \\
&+\frac{3D^2}{16\nu_0^2} \left( 1+\frac{\gamma_0^2}{\nu_0^2}\right)^{\!-5/4}
\left( \mathrm{Re}P_3\cos\frac{5\theta}{2}-\mathrm{Im}P_3\sin\frac{5\theta}{2} \right) \\
&+\frac{\sqrt{2}}{2} \left(1 +\frac{(\mu+\gamma_0)^2}{\nu_0^2} \right)^{\!-1/4}
\left( \mathrm{Re}Q_1\cos\frac{\eta}{2}+\mathrm{Im}Q_1\sin\frac{\eta}{2} \right) \\
&+\frac{\sqrt{2}D}{2\nu_0} \left( 1+\frac{(\mu+\gamma_0)^2}{\nu_0^2} \right)^{\!-3/4}
\left( \mathrm{Re}Q_2\cos\frac{3\eta}{2}+\mathrm{Im}Q_2\sin\frac{3\eta}{2} \right) \\
&+\frac{3\sqrt{2}D^2}{4\nu_0^2} \left( 1+\frac{(\mu+\gamma_0)^2}{\nu_0^2} \right)^{\!-5/4}
\left( \mathrm{Re}Q_3\cos\frac{5\eta}{2}+\mathrm{Im}Q_3\sin\frac{5\eta}{2} \right) \bigg] \ .
\end{aligned}
\end{equation}

\subsection{$d=2$:} For the $\psi_o$ mode, the renormalized parameters read
\begin{equation}
\begin{aligned}
\mu_r =& \mu \Bigg[ 1-\frac{\sigma+\kappa}{6D\pi}\frac{\gamma_0}{\mu}\ln\frac{D\Lambda^2}{\gamma_0}
-\frac{\sigma+\kappa}{12D\pi}\ln\frac{D\Lambda^2}{\mu}-\frac{\sigma+\kappa}{24D\pi} \\
&- \frac{\sigma+\kappa}{12D\pi} \left( 1+\frac{\sqrt{3}\nu_0}{2\mu}+\frac{\sqrt{3}\nu_0}{2\gamma_0}
+\frac{2\sqrt{3}\nu_0\gamma_0}{\mu^2} \right) \Bigg] \ , \\
D^o_r =& D-\frac{\sigma+\kappa}{48\pi}-\frac{\sigma+\kappa}{24\pi} \left( \frac{\gamma_0}{\mu}
-\frac{2\sqrt{3}\nu_0}{3\mu}+\frac{\sqrt{3}\nu_0}{3\gamma_0} \right) .
\end{aligned}
\end{equation}
For the $\psi_\pm$ modes, the renormalized parameters are
\begin{equation}
\begin{aligned}
\gamma_r =& \gamma_0 +\frac{\sigma+\kappa}{2\pi} \bigg[ \frac{\mathrm{Re}M_1}{2}
\ln\frac{D\Lambda^2}{\gamma_0} + \left( \mathrm{Re}M_2+\mathrm{Re}M_3\right)\ln\Lambda
-\theta\mathrm{Im}M_2+\eta\mathrm{Im}M_3 \\
&-\frac14 \left[ \mathrm{Re}M_2\ln \frac{\gamma_0^2+\nu_0^2}{D^2}+\mathrm{Re}M_3
\ln \frac{(\mu+\gamma_0)^2+\nu_0^2}{4D^2} \right] \\
&+\frac{D}{2\nu_0} \left( 1+\frac{\gamma_0^2}{\nu_0^2} \right)^{\!-1/2} 
\left(\mathrm{Re}M_4\cos\theta-\mathrm{Im}M_4\sin\theta \right) \\
&+\frac{D}{\nu_0} \left( 1+\frac{(\mu+\gamma_0)^2}{\nu_0^2} \right)^{\!-1/2}
\left(\mathrm{Re}M_5\cos\eta + \mathrm{Im}M_5\sin\eta\right) \bigg] \ , \\
\nu_r =& \nu_0 +\frac{\sigma+\kappa}{2\pi} \bigg[ \frac{\mathrm{Im}M_1}{2} \ln \frac{D\Lambda^2}{\gamma_0}
+\left(\mathrm{Im}M_2+\mathrm{Im}M_3\right) \ln\Lambda+\theta\mathrm{Re}M_2 - \eta\mathrm{Re}M_3 \\
&-\frac14 \left[ \mathrm{Im}M_2\ln \frac{\gamma_0^2+\nu_0^2}{D^2} \mathrm{Im}M_3 
\ln \frac{(\mu+\gamma_0)^2+\nu_0^2}{4D^2} \right] \\
&+\frac{D}{2\nu_0} \left( 1+\frac{\gamma_0^2}{\nu_0^2} \right)^{\!-1/2}
\left(\mathrm{Re}M_4\sin\theta+\mathrm{Im}M_4\cos\theta\right) \\
&+\frac{D}{\nu_0} \left( 1+\frac{(\mu+\gamma_0)^2}{\nu_0^2} \right)^{\!-1/2}
\left(-\mathrm{Re}M_5\sin\eta + \mathrm{Im}M_5\cos\eta\right) \bigg] \ ,
\end{aligned}
\end{equation}
and the renormalized diffusivity is
\begin{equation}
\begin{aligned}
\mathrm{Re}D_r = D  + \frac{\kappa+\sigma}{2\pi} \bigg[ &-\frac{\mathrm{Re}Q_1}{8} \ln 
\frac{\nu_0^2+(\mu+\gamma_0)^2}{4\gamma_0^2} +\frac14 \left(\eta\mathrm{Im}Q_1 - \theta\mathrm{Im}P_1\right) \\
&+\frac{D}{4\nu_0}\left(1+\frac{\gamma_0^2}{\nu_0^2}\right)^{\!-1/2}\left(\mathrm{Re}P_2\cos\theta-\mathrm{Im}P_2\sin\theta\right)\\
&+\frac{D^2}{8\nu_0^2} \left( 1+\frac{\gamma_0^2}{\nu_0^2} \right)^{\!-1}
\left(\mathrm{Re}P_3\cos2\theta-\mathrm{Im}P_3\sin2\theta \right) \\
&+\frac{D}{2\nu_0} \left( 1+\frac{(\mu+\gamma_0)^2}{\nu_0^2}\right)^{\!-1/2}
\left(\mathrm{Re}Q_2\cos\eta+\mathrm{Im}Q_2\sin\eta\right) \\
&+\frac{D^2}{2\nu_0^2} \left( 1+\frac{(\mu+\gamma_0)^2}{\nu_0^2} \right)^{\!-1}
\left(\mathrm{Re}Q_3\cos2\eta+\mathrm{Im}Q_3\sin2\eta\right) \bigg] \ .
\end{aligned}
\end{equation}

\subsection{$d=3$:} For the $\psi_o$ mode, the renormalized parameters read
\begin{equation}
\begin{aligned}
\mu_r =& \mu \left[ 1 + \frac{\sigma+\kappa}{16D\pi}\sqrt{\frac{\mu}{D}}+\frac{\sigma+\kappa}{6D\pi}\sqrt{\frac{\gamma_0}{D}} \left( \!-\frac14+\frac{\gamma_0}{\mu}-\frac{\sqrt{3}\nu_0}{16\mu}-\frac{\sqrt{3}\nu_0}{16\gamma_0}-\frac{\sqrt{3}\nu_0\gamma_0}{2\mu^2} \right)\!\right] , \\
D^o_r =& D-\frac{\sigma+\kappa}{96\pi}\sqrt{\frac{\mu}{D}}-\frac{\sigma+\kappa}{48\pi}\sqrt{\frac{\gamma_0}{D}}\left( \frac{\sqrt{3}\nu_0}{6\gamma_0}+\frac{\gamma_0}{\mu}-\frac{5\sqrt{3}\nu_0}{6\mu} \right) .
\end{aligned}
\end{equation}
For the $\psi_\pm$ modes, if $\gamma_0 \geq 0$, the renormalized parameters are
\begin{equation}
\begin{aligned}
\gamma_r =& \gamma_0 +\frac{\sigma+\kappa}{4\pi}\sqrt{\frac{\nu_0}{D}} \ \bigg[ -\mathrm{Re}M_1\sqrt{\frac{\gamma_0}{\nu_0}}-\left(1+\frac{\gamma_0^2}{\nu_0^2}\right)^{1/4}\left(\mathrm{Re}M_2\cos\frac{\theta}{2}+\mathrm{Im}M_2\sin\frac{\theta}{2}\right) \\
&- \frac{\sqrt{2}}{2} \left( 1+\frac{(\mu+\gamma_0)^2}{\nu_0^2} \right)^{\!1/4} 
\left(\mathrm{Re}M_3\cos\frac{\eta}{2}-\mathrm{Im}M_3\sin\frac{\eta}{2}\right) \\
&+ \frac{D}{2\nu_0} \left( 1+\frac{\gamma_0^2}{\nu_0^2} \right)^{\!-1/4}
\left(\mathrm{Re}M_4\cos\frac{\theta}{2}-\mathrm{Im}M_4\sin\frac{\theta}{2}\right) \\
&+ \frac{\sqrt{2}D}{2\nu_0 }\left( 1+\frac{(\mu+\gamma_0)^2}{\nu_0^2} \right)^{\!-1/4}
\left(\mathrm{Re}M_5\cos\frac{\eta}{2}+\mathrm{Im}M_5\sin\frac{\eta}{2}\right) \bigg] \ , \\
\nu_r =& \nu_0 +\frac{\sigma+\kappa}{4\pi}\sqrt{\frac{\nu_0}{D}} \ \bigg[ -\mathrm{Im}M_1
\sqrt{\frac{\gamma_0}{\nu_0}} - \left( 1+\frac{\gamma_0^2}{\nu_0^2} \right)^{\!1/4}
\left(-\mathrm{Re}M_2\sin\frac{\theta}{2}+\mathrm{Im}M_2\cos\frac{\theta}{2}\right) \\
&- \frac{\sqrt{2}}{2} \left( 1+\frac{(\mu+\gamma_0)^2}{\nu_0^2} \right)^{\!1/4}
\left(\mathrm{Re}M_3\sin\frac{\eta}{2}+\mathrm{Im}M_3\cos\frac{\eta}{2}\right) \\
&+ \frac{D}{2\nu_0} \left( 1+\frac{\gamma_0^2}{\nu_0^2} \right)^{\!-1/4}
\left(\mathrm{Re}M_4\sin\frac{\theta}{2}+\mathrm{Im}M_4\cos\frac{\theta}{2}\right) \\
&+ \frac{\sqrt{2}D}{2\nu_0} \left( 1+\frac{(\mu+\gamma_0)^2}{\nu_0^2} \right)^{\!-1/4}
\left(-\mathrm{Re}M_5\sin\frac{\eta}{2}+\mathrm{Im}M_5\cos\frac{\eta}{2}\right) \bigg] \ .
\end{aligned}
\end{equation}
When $\gamma_0 < 0$ and the system is rendered unstable, the renormalized parameters become
\begin{equation}
\begin{aligned}
\mu_r =& \left[ 1 + \frac{\sigma+\kappa}{16D\pi}\sqrt{\frac{\mu}{D}}+\frac{\sigma+\kappa}{6D\pi}\sqrt{\frac{\gamma_0}{D}} \left( -\frac14+\frac{\gamma_0}{\mu}-\frac{\sqrt{3}\nu_0}{16\mu}
-\frac{\sqrt{3}\nu_0}{16\gamma_0}-\frac{\sqrt{3}\nu_0\gamma_0}{2\mu^2} \right) \right] , \\
D^o_r =& D-\frac{\sigma+\kappa}{96\pi}\sqrt{\frac{\mu}{D}}-\frac{\sigma+\kappa}{48\pi}\sqrt{\frac{\gamma_0}{D}}
\left (\frac{\sqrt{3}\nu_0}{6\gamma_0}+\frac{\gamma_0}{\mu}-\frac{5\sqrt{3}\nu_0}{6\mu} \right) .
\end{aligned}
\end{equation}
For the $\psi_\pm$ modes, for $\gamma_0 \geq 0$, the renormalized parameters are
\begin{equation}
\begin{aligned}
\gamma_r =& \gamma_0 +\frac{\sigma+\kappa}{4\pi}\sqrt{\frac{\nu_0}{D}} \ \bigg[ -\mathrm{Im}M_1
\sqrt{\frac{|\gamma_0|}{\nu_0}}-\left( 1+\frac{\gamma_0^2}{\nu_0^2} \right)^{\!1/4}
\left(\mathrm{Re}M_2\cos\frac{\theta}{2}+\mathrm{Im}M_2\sin\frac{\theta}{2}\right) \\
&- \frac{\sqrt{2}}{2} \left( 1+\frac{(\mu+\gamma_0)^2}{\nu_0^2} \right)^{\!1/4}
\left(\mathrm{Re}M_3\cos\frac{\eta}{2}-\mathrm{Im}M_3\sin\frac{\eta}{2}\right) \\
&+ \frac{D}{2\nu_0} \left( 1+\frac{\gamma_0^2}{\nu_0^2} \right)^{\!-1/4}
\left(\mathrm{Re}M_4\cos\frac{\theta}{2}-\mathrm{Im}M_4\sin\frac{\theta}{2}\right) \\
&+ \frac{\sqrt{2}D}{2\nu_0} \left( 1+\frac{(\mu+\gamma_0)^2}{\nu_0^2} \right)^{\!-1/4}
\left(\mathrm{Re}M_5\cos\frac{\eta}{2}+\mathrm{Im}M_5\sin\frac{\eta}{2}\right) \bigg] \ , \\
\nu_r =& \nu_0 +\frac{\sigma+\kappa}{4\pi}\sqrt{\frac{\nu_0}{D}} \ \bigg[ \mathrm{Re}M_1
\sqrt{\frac{|\gamma_0|}{\nu_0}} - \left( 1+\frac{\gamma_0^2}{\nu_0^2} \right)^{\!1/4}
\left(-\mathrm{Re}M_2\sin\frac{\theta}{2}+\mathrm{Im}M_2\cos\frac{\theta}{2}\right) \\
&- \frac{\sqrt{2}}{2} \left (1+\frac{(\mu+\gamma_0)^2}{\nu_0^2} \right)^{\!1/4}
\left(\mathrm{Re}M_3\sin\frac{\eta}{2}+\mathrm{Im}M_3\cos\frac{\eta}{2}\right) \\
&+ \frac{D}{2\nu_0} \left( 1+\frac{\gamma_0^2}{\nu_0^2} \right)^{\!-1/4}
\left(\mathrm{Re}M_4\sin\frac{\theta}{2}+\mathrm{Im}M_4\cos\frac{\theta}{2}\right) \\
&+ \frac{\sqrt{2}D}{2\nu_0} \left( 1+\frac{(\mu+\gamma_0)^2}{\nu_0^2} \right)^{\!-1/4}
\left(-\mathrm{Re}M_5\sin\frac{\eta}{2}+\mathrm{Im}M_5\cos\frac{\eta}{2}\right) \bigg] \ .
\end{aligned}
\end{equation}
The renormalized diffusivity is not affected by the sign of $\gamma_0$,
\begin{equation}
\begin{aligned}
\mathrm{Re}D_r = D &+\frac{\kappa+\sigma}{3\pi}\sqrt{\frac{\nu_0}{D}} \ \bigg[ -\frac14 \left( 1+
\frac{\gamma_0^2}{\nu_0^2} \right)^{\!1/4} \mathrm{Im}P_1\sin\frac{\theta}{2} + \frac14 \mathrm{ReQ_1}
\sqrt{\frac{\gamma_0}{\nu_0}} \\
&+\frac{D}{8\nu_0} \left( 1+\frac{\gamma_0^2}{\nu_0^2} \right)^{\!-1/4}
\left(\mathrm{Re}P_2\cos\frac{\theta}{2}-\mathrm{Im}P_2\sin\frac{\theta}{2}\right) \\
&+\frac{D^2}{32\nu_0^2} \left( 1+\frac{\gamma_0^2}{\nu_0^2} \right)^{\!-3/4}
\left(\mathrm{Re}P_3\cos\frac{3\theta}{2}-\mathrm{Im}P_3\sin\frac{3\theta}{2}\right) \\
&-\frac{\sqrt{2}}{8} \left( 1+\frac{(\mu+\gamma_0)^2}{\nu_0^2} \right)^{\!1/4}
\left(\mathrm{Re}Q_1\cos\frac{\eta}{2}-\mathrm{Im}Q_1\sin\frac{\eta}{2}\right) \\
&+\frac{\sqrt{2}D}{8\nu_0} \left( 1+\frac{(\mu+\gamma_0)^2}{\nu_0^2} \right)^{\!-1/4}
\left(\mathrm{Re}Q_2\cos\frac{\eta}{2}+\mathrm{Im}Q_2\sin\frac{\eta}{2}\right) \\
&+\frac{\sqrt{2}D^2}{16\nu_0^2} \left( 1+\frac{(\mu+\gamma_0)^2}{\nu_0^2} \right)^{\!-3/4}
\left(\mathrm{Re}Q_3\cos\frac{3\eta}{2}+\mathrm{Im}Q_3\sin\frac{3\eta}{2}\right) \bigg] \ .
\end{aligned}
\end{equation}

\nocite{*}
\newcommand{\newblock}{}
\bibliographystyle{unsrt}
\bibliography{ref}
\end{document}